# High-Resolution Poverty Maps in Sub-Saharan Africa

Kamwoo Lee and Jeanine Braithwaite

*Up-to-date poverty maps are an important tool for policy makers, but until now, have been prohibitively expensive to produce. We propose a generalizable prediction methodology to produce poverty maps at the village level using geospatial data and machine learning algorithms. We tested the proposed method for 25 Sub-Saharan African countries and validated them against survey data. The proposed method can increase the validity of both single country and cross-country estimations leading to higher precision in poverty maps of 44 Sub-Saharan African countries than previously available. More importantly, our cross-country estimation enables the creation of poverty maps when it is not practical or cost-effective to field new national household surveys, as is the case with many low- and middle-income countries.*

# Introduction

One of the most pressing problems for global development policymakers is the lack of timely data on who is poor and where the poor are located, especially on the village level. Poverty maps provide this information and are usually constructed by imputing from household surveys to population census tracts, which must typically be done in situ at a country's statistical agency because of confidentiality concerns about census data. Survey data are expensive and difficult to obtain and require long time frames for cleaning and analysis. Furthermore, face to face interviews for surveys are sometimes impractical in times such as civil conflicts and global pandemics.

We develop a generalizable model for creating village-level[1] poverty maps for Sub-Saharan African (SSA) countries using readily available data sources and machine learning (ML) techniques to map wealth levels on the 1 square mile (1.6 x 1.6 km$^2$) level. The model works for both single country and cross-country estimation, which enables the creation of poverty maps when it is not practical or cost-effective to field new national household surveys. In the majority of SSA countries, we don't know precisely where the poor are located. In order to move people out of poverty, public policymakers need to know where they are located to place anti-poverty interventions in the right places. Our approach enables policymakers to use up-to-date information to map out areas of SSA countries to locate the poor to a higher degree of precision than previously available.

In the following sections, we cover the literature about conventional poverty maps and previous studies that have used ML estimation techniques to produce poverty maps. Next, we explain the methodology that we propose and present our results for 25 SSA countries - Angola, Benin, Burkina-Faso, Burundi, Cameroon, Chad, Ethiopia, Ghana, Guinea, Kenya, Liberia, Madagascar, Malawi, Mali, Mozambique, Nigeria, Rwanda, Senegal, Sierra-Leone, South Africa, Tanzania, Togo, Uganda, Zambia, and Zimbabwe, including our cross-checks against survey data for validation. Then, we apply our method to the remaining SSA countries which lack survey data. We also speculate on how our poverty mapping technique could be extended to other regions of the world. Lastly, we draw conclusions and implications of our method for real-world policy makers.

## Poverty Mapping

Literature on poverty and poverty measurement is vast. The first survey on living standards was done in England more than 100 years ago (Rowntree 1901) and international organizations have produced both complementary and competing definitions of poverty for the past thirty years (World Bank 1990; UN 1990). The literature just by the World Bank and UN authors, let alone academics, on poverty is too extensive to be fully cited here[2]. We acknowledge this immense literature while leaving the debates over the poverty definition of poverty and poverty measurement to other fora.

Poverty mapping is the production of a geospatial dataset of disaggregated, small area estimates of poverty or welfare indicators for a country or countries. When countries and international agencies began measuring and estimating poverty, these estimates were based on household surveys that were representative only on the national or provincial level. In the mid to late 1990s, analysts began combining household survey data with population censuses to generate poverty maps to produce disaggregated estimates for poverty on the village or small area level (Hentschel and Lanjouw 1998).

While much of the initial impetus for poverty mapping was intertwined with the emerging desire to target social protection benefits, policy makers realized that poverty maps were useful for many types of policy decisions, including where to site clinics, schools, and other social services and infrastructure (Bedi, Coudouel, and Simler 2007). The basis for an empirical technique for poverty mapping was quickly established (Elbers, Lanjouw, and Lanjouw 2003) and extended to other techniques to predict poverty in small area estimations (Molina and Rao 2010; Chambers and Tzavidis 2006; Tzavidis et al. 2010).

However, these poverty mapping methods all required the availability of a population census, which is typically conducted once a decade, and at least one household survey. Household surveys and population censuses are expensive, and emerging economies may not be able to afford frequently updating them. Additionally, the precision of poverty maps was debated at the World Bank in the 2000s (Banerjee et al. 2006; Lanjouw and Ravallion 2006). Now, new technology and computing power make it possible to produce poverty/wealth maps in finer granularity cost-efficiently.

## *Machine Learning Methods*

In recent years, there have been significant advances in estimating socioeconomic indicators with geospatial information and machine learning. Geospatial information, such as nighttime lights, day-time satellite imagery, and crowd-sourced map data provide a low-cost means of creating granular poverty maps. The geospatial information not only shows patterns and disparities of poverty in a country but is also one of the most accurate predictors of sub-national and even community-level wealth standing.

Machine learning (ML), which is an application of artificial intelligence, is widely used to effectively and efficiently find patterns of relationships between input data and output labels. ML's popularity is largely due to the fact that it takes any form of data such as numbers, texts, and images, and generalizes the relationships found in the existing data to unforeseen instances. ML algorithms have been proven to have better performance than humans in understanding patterns of what we used to believe only humans could understand, such as object detection and facial recognition. Now ML techniques can be used to recognize which villages are impoverished.

We categorize these ML techniques that use geospatial information for wealth level estimation into two groups. We call the first group a feature-based prediction model (Zhao and Kusumaputri 2016; Lee et al. 2017; Tingzon et al. 2019), which employs quantifiable geospatial features, such as the number of buildings in a region, total area of buildings, length of roads, number of junctions as well as the distances to the closest school, hospital, market and other types of locations. Since this kind of geospatial information is closely related to the economic activities in a region, a prediction model can be trained to learn the best features that explain the wealth level of an area. The second group (Jean et al. 2016; Head et al. 2017; Babenko et al. 2017; Engstrom et al. 2017; Heitmann and Buri 2019; Tingzon et al. 2019; Yeh et al. 2020) - we call an image-based prediction model - which works to capture geospatial characteristics in satellite images or aerial pictures. This type of model can recognize qualitative characteristics such as types of buildings, shapes of roads, and man-made structures in the pictures and how they relate to the wealth level of a region.

Each type of model has its own limitations. The limitation of the feature-based model is that it cannot capture qualitative differences in the geospatial features. For instance, agricultural buildings like hay barns or cow barns are distinct in nature from high-story buildings at the city center even though they have the same area, while a single-track road in the mountain is fundamentally different from an 8-lane highway with the same length. The feature-based model does not make that distinction. At the same time, we cannot incorporate out-of-image information with the image-based model. In this type of model, there is no information about the distance to the nearest hospital, school, or market that are outside of an image grid cell, which are vital indicators to predict wealth levels. To overcome the first limitation, Zhao et al. (2019) proposed a model that utilizes both quantifiable geospatial features and satellite images in 10km x 10km grid cells.

There has been a parallel effort to utilize cell phone data in combination with other data to estimate poverty (Hernandez et al. 2017; Lee et al. 2017; Steele et al. 2017; Heitmann and Buri 2019). However, as Heitmann and Buri (2019) point out, cell detail records (CDR) are hard to rely on for regular poverty measurement as they are not publicly available. Moreover, and most importantly, all of the previous studies, except for Yeh et al. (2020), are single country estimations where the prediction model has been applied to a specific country that has a recent household survey with GPS coordinates. These studies are difficult to generalize because only a few countries meet this condition.

## Proposed Method

We create a cross-country prediction model with publicly available data using both feature-based and image-based machine learning methods to solve the limitations aforementioned. First, we derive from the Demographic and Health Survey (DHS) the International Wealth Index (IWI) for 25 countries, which is a strictly comparable asset-based index that can be used for low- and middle-income countries. We have also devised a training data refining mechanism by employing both the feature-based and image-based models in a way that they improve each other's result while accounting for quantitative and qualitative differences in geospatial features among different places in different countries.

*Wealth Index and Cross-country Estimation*

For single country estimations, one could use the DHS Wealth Index (DHS WI) as the dependent variable of a prediction model. However, the DHS WI is a survey-specific measure of the relative economic standing of a household at a particular point in time and country. The scores can only be used within a specific survey and cannot be compared with other surveys for different time and countries (Figure 1). In fact, our initial attempts to predict the DHS WI across countries showed unsatisfactory results. Cross-country estimations require comparable wealth measurements.

There have been attempts to make the DHS WI comparable across surveys. Rutstein and Stavteig (2014) proposed the DHS Comparative Wealth Index (CWI), a methodology to calculate a linear displacement of the DHS WI factor scores by regressing eight anchoring points against a baseline survey. To extend and complement the CWI approach, Staveteig and Mallick (2014) proposed the Harmonized Wealth Index (HWI), which is constructed by harmonizing household asset categories that are present in all surveys covered. In an effort to create a metric that can be used for all low- and middle-income countries, Smits and Steendijk (2015) proposed the International Wealth Index (IWI) and developed a universal set of asset weights based on 12 common assets and asset categories. We decided to use the IWI as the dependent variable since our focus is cross-country comparisons, not an intertemporal analysis, and the IWI provides a better way to cover a large set of countries at a fixed time frame.

We measure the IWI from 34 DHS surveys that took place in 25 countries within the past 5 years. From the household section in each DHS dataset, we calculate the IWI based on whether each household has the following ten assets: five consumer

**Figure 1**. Different Wealth Levels among Countries

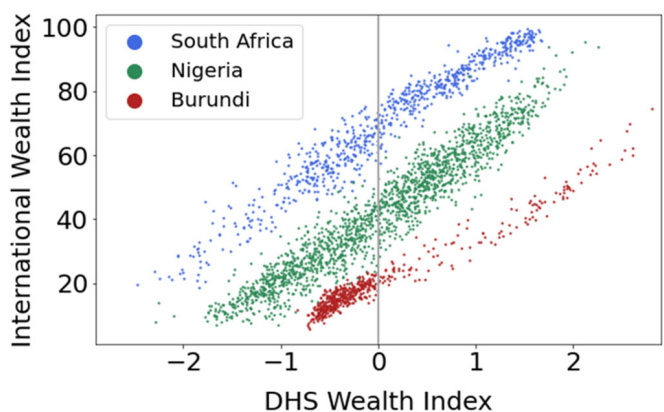

*Source:* DHS data and authors' calculation of IWI.
*Note:* The same DHS WI can have significantly different IWI depending on the country

durables (TV, fridge, phone, bike, and car)[3], access to two public services (water and electricity) and three housing characteristics (number of sleeping rooms, quality of floor material and of toilet facility). The derived IWI is highly correlated to the original DHS Wealth Index with Pearson's correlation coefficients ranging from 0.93 to 0.99.

## Geospatial Datasets

For the independent variables of our prediction model, we utilize OpenStreetMap data, the Visible Infrared Imaging Radiometer Suite Day/Night Band (VIIRS DNB) nighttime lights dataset, day-time satellite images, and the High-Resolution Settlement Layer (HRSL) datasets for the 25 model countries. We extract from OpenStreetMap, which is a crowd sourced, free, and editable map of the whole world, the total length of roads[4], distance to the closest road, number of junctions[5], distance to the closest junctions, total building area, and the number of buildings for each 1 square-mile populated area. We also collect the number of and distance to 24 locations of interest such as schools, hospitals, and markets[6].

The VIIRS DNB dataset has 15 arcsec (about 463m at the equator) resolution nighttime luminosity. We compute six aggregate summary statistics[7] of the night light intensities within 1 square mile (1.6x1.6km$^2$), 5x5km$^2$, and 10x10km$^2$ areas around each populated place. We download the satellite images through the Google Static Map API, and each of the 640x640-pixel images with zoom-16 resolution[8] covers roughly 1 square mile. We also extract populations within 1.6x1.6km$^2$, 5x5km$^2$, and 10x10km$^2$ areas around each populated place from the HRSL dataset, which provides population estimates at 1 arc-second (approximately 30m) resolution based on image recognition methods and satellite imagery (Facebook & Columbia University 2016). The HRSL is not available in Somalia, Sudan, and South Sudan, so we used the gridded population dataset (100m resolution) from WorldPop in place of the HRSL for these countries.

To identify populated places, we mainly use two datasets - the settlement data from the United Nations Office for the Coordination of Humanitarian Affairs (UN OCHA) and OpenStreetMap (OSM)[9]. Interestingly, the two datasets sometimes list quite different populated places. For some countries, one dataset has significantly more places than the other, or the locations are very different[10]. For example, Uganda's OSM has around 10,000 places but UN OCHA data has only half as many. Ethiopia's OSM populated places barely overlap with UN OCHA settlement places. Thus, we collect all places in both datasets unless they are less than 1km apart. The benefit of the UN OCHA dataset is that it contains place names without omission and the administrative regions they belong to. On the other hand, the OSM data are updated more frequently. In order not to overlook numerous unnamed places and many smaller communities within big cities, we extracted places where population

**Figure 2**. Narrowing Locations of a DHS Cluster

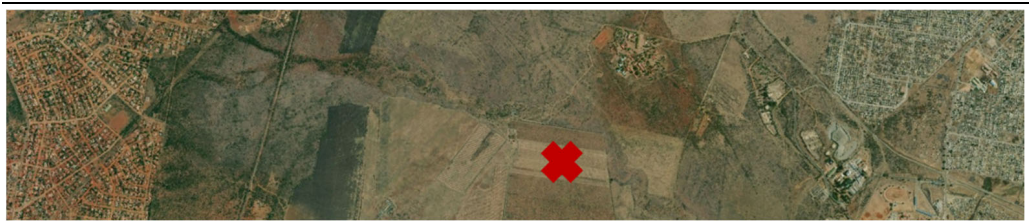

*Source:* Google Static Map.
*Note:* The red X mark in the middle is the GPS coordinates of a DHS cluster. The village on the left is predicted to be very poor while the right village to be very rich. Judging by the survey data that this place is extremely poor, we can decide that the left one is a closer candidate for the actual DHS cluster. We narrow locations of DHS clusters and average them only for ML training purposes, and we don't single out a DHS location candidate nor generate the narrowed locations in any data form.

is larger than 100 within an area of 1 square mile but are not listed in UN OCHA and OSM datasets.

We also use the identified populated places to narrow the possible locations of the DHS clusters. The GPS coordinates that are attached to DHS responses are randomly displaced to maintain respondent confidentiality. The maximum displacement is 2km for urban areas and 5km for rural areas (1% further displaced up to 10km). When estimating wealth index for DHS clusters, we initially take the average of the wealth index in populated places within 2km radius for urban clusters and 5km for rural clusters. In case there is no populated places around the cluster, which can happen when the DHS clusters are extremely rural, and neither OSM nor UN OCHA data list them, we use the HRSL dataset to select most populated places in the evenly divided 4 areas around the cluster and then average the wealth index of the 4 places. After the initial prediction of the model, we can further narrow down the potential DHS cluster locations by splitting the potential locations into three groups (richer, middle, and poorer) with tercile thresholds. We then average the geospatial features of the locations within the segment that the DHS IWI belongs to (Figure 2).

## Machine Learning Models

The overall machine learning process is illustrated in Figure 3. We use both feature-based and image-based models to improve each other's performance by providing better training data. This refining process can be used for either the cross-country estimation model or the single country estimation model. For cross-country estimation, we leave out one country and train our model on 24 other countries' data. After the training, we validate the prediction with the 25th country's data. We repeat this 25 times. For single country estimation, we train each country by holding out 1/5 of the data of a single country for validation and training on the remaining 4/5.

**Figure 3**. Refining Mechanism

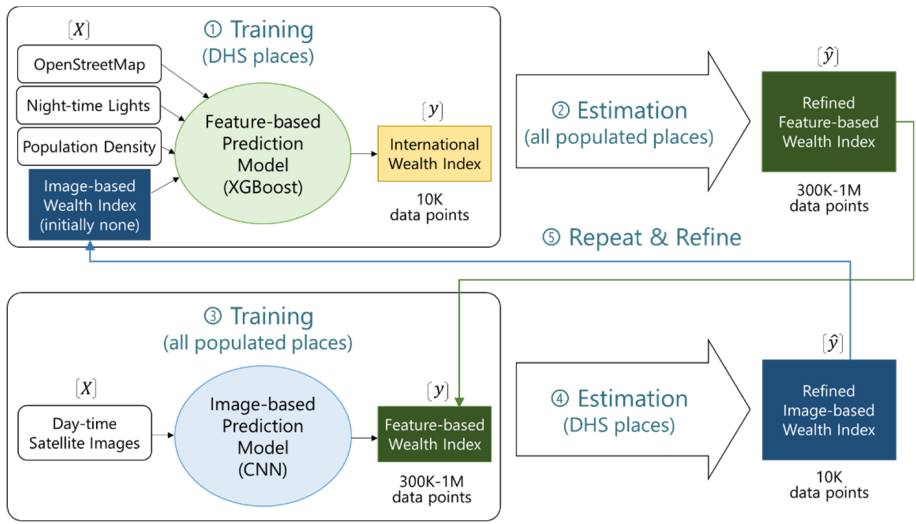

*Source:* Authors' illustration.
*Note:* The two types of models improve their prediction performance by taking better training data from the other type model through iteration. The first pass of the iteration for the 25 countries takes around 80 hours on a workstation[12], then 30 hours from the second iteration.

First, we train an XGBoost (eXtreme Gradient Boosting) algorithm, which is one of the most powerful prediction models at the moment, on the OSM features, night-time luminosity, and population density as input variables and the IWI as an output variable. There are around 14,000 DHS clusters in 25 countries and around 200-1000 clusters in a single country. An XGBoost model can have many configurations, also known as hyperparameters, that affect training performance. We choose[11] the configuration that performs best for the training dataset (24 countries data for cross-country estimation and 4/5 of data for single country estimation).

Second, we estimate wealth levels for all populated places in the 25 countries using the trained XGBoost model. There are 662,061 populated places in total that are listed in the UN OCHA and OSM datasets and partly extracted from the HRSL dataset. At this point, the estimation model is already a good predictor of the IWI. We choose the better of the cross-country and single country estimators for the next step (Figure 4).

Third, we train a convolutional neural network (CNN) model, which is the most popular deep learning algorithm for image recognition, on the day-time satellite images. CNN models are only effective when there are abundant labeled training datasets. Since the number of datapoints in surveys are not enough to train a CNN, previous researchers have resorted to using nighttime luminosity as a proxy for the wealth index. Although using nighttime luminosity for CNN training and applying transfer learning is a brilliant idea, nighttime luminosity is fundamentally limited in that it is hard to differentiate villages in rural areas with night lights. For instance,

**Figure 4**. Initial Results of Single country and Cross-country Estimations for Kenya

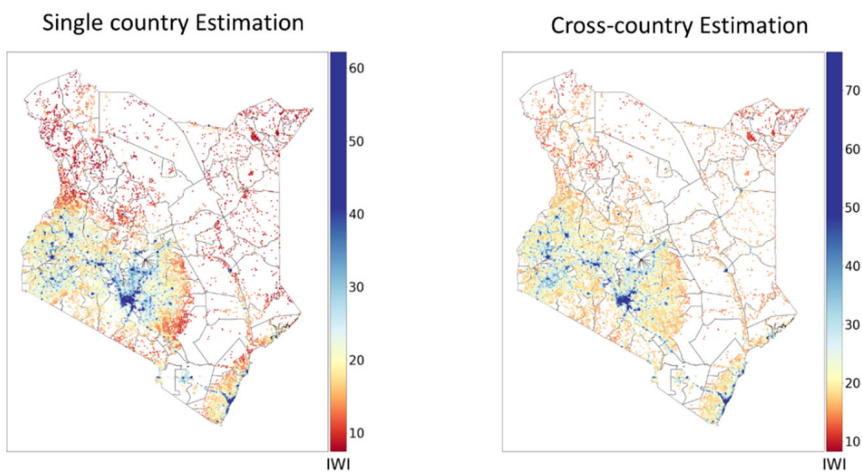

*Source:* Authors' calculation.
*Note:* 20,167 populated places in Kenya and estimated wealth levels. Initially, single country estimation (left) is more accurate than cross-country estimation (right). But later through the refining process cross-country estimation becomes more accurate.

6,466 out of 30,131 (21.5%) populated places in Senegal have zero luminosity within 1 square mile around them. Here, we propose to use the predicted wealth indices in the second step to train a customized CNN[13, 14]. The distribution of the predicted wealth indices is much closer to the ground truth wealth distribution than the nighttime luminosity and will generate much more accurate predictions in wealth level estimation. We are cautious in our method design to not let test data leak into the training process. We resample training data in each iteration. Every prediction used as a dependent variable is a cross-estimated one regardless of whether it is a cross-country model or single country model. Even in the hyperparameter tuning, we maintain separation of training data and testing data thoroughly. The output of the customized CNN model is a probability distribution of whether a 1 square mile area is rich, upper middle class, lower middle class or poor. We choose this classification over regression in order to capture multi-faceted aspects of qualitative geospatial features shown in the picture that might not be captured by a single number with the XGBoost model.

    Fourth, we estimate the four probabilities (rich, upper-middle class, lower-middle class, or poor) for all populated places using the CNN model. We then channel this featurized information into the XGBoost model back. The whole process creates a cycle where the models improve each other by providing better training data with every iteration. An additional benefit of this cycle is that it allows a human modeler to incrementally improve model structures. Both XGBoost and CNN algorithms have many hyperparameters, which have to be set before training. A modeler can try different configurations of the models in each iteration as long as

the prediction performance is improving. As for CNN training, we apply transfer learning[15] from the second iteration to speed up the process by using part of the model that has already been trained in the previous iteration and further learning on the refined training data.

# Results

## Validation

Following previous research, we measured the coefficient of determination (R-squared) between the estimated wealth index of the model and the observed wealth index in the recent 5-year DHS surveys (Figure 5). The R-squared can be interpreted as the proportion of the variance for the observed wealth index that is explained by the estimated wealth index. Although the R-squared doesn't precisely represent prediction accuracy, it conveys a relative performance in an intuitive way. Our result of single country estimation with the proposed models for 25 countries is 88.12% on average ranging from 73.03% to 94.91% outperforming all previous research results. Our cross-country estimation is also as high as the single country estimation results with the average of 85.6% (Table 1; Figure 6).

**Table 1**. Validation Results

| Country | R-squared (single country) | R-squared (cross-country) | Country | R-squared (single country) | R-squared (cross-country) |
|---|---|---|---|---|---|
| Angola | 90.82% | 87.82% | Mali | 89.84% | 86.99% |
| Benin | 85.80% | 81.51% | Mozambique | 94.91% | 93.04% |
| Burkina Faso | 81.42% | **84.84%** | Nigeria | 88.64% | 86.67% |
| Burundi | 88.35% | 83.01% | Rwanda | 87.41% | 86.32% |
| Cameroon | 92.54% | 87.42% | Senegal | 88.68% | 87.37% |
| Chad | 88.55% | 80.06% | Sierra Leone | 89.82% | **91.12%** |
| Ethiopia | 94.04% | 91.53% | South Africa | 73.03% | 60.09% |
| Ghana | 89.56% | 88.22% | Tanzania | 89.93% | 87.76% |
| Guinea | 94.08% | 91.66% | Togo | 89.34% | **91.60%** |
| Kenya | 74.39% | **79.98%** | Uganda | 89.02% | 86.12% |
| Liberia | 89.30% | 88.73% | Zambia | 87.94% | 83.03% |
| Madagascar | 89.29% | 84.92% | Zimbabwe | 92.10% | 89.18% |
| Malawi | 84.14% | 80.02% | **Average** | **88.12%** | **85.60%** |

*Source:* Authors' calculation.
*Note:* Single country estimations are validated by partitioning the data into 4 complementary subsets and rotating the subsets for 3 training sets and 1 validation set (4-fold cross-validation). Cross-country estimations are trained on 24 countries' data and validated on the 25th country's data. Notice that cross-country estimations are more accurate than single country estimations for Burkina Faso, Kenya, Sierra Leone, and Togo.

**Figure 5.** Validation of estimation for Sierra Leone

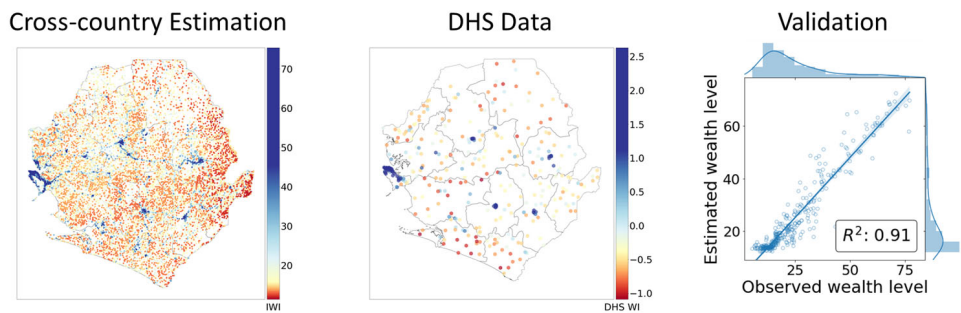

*Source:* Authors' calculation.
*Note:* Cross-country estimation of wealth levels for 13,040 populated places in Sierra Leone (left), 336 DHS clusters (middle), and validation plots (right). The R-squared for the estimation is 91.12%.

It is worth noting that the cross-country estimation surpasses the single country estimation in Burkina Faso, Kenya, Sierra Leone, and Togo. Normally, single country estimation is more accurate than cross-country estimation since there are many common geospatial elements that can translate into relative wealth levels inside a country. But cross-country estimation can sometimes give better estimates because of the greater number of data points to train a model, making a model more robust and stable. The fact that examples from outside the country can provide estimates of the wealth levels inside the country shows promising potential for cross-country estimation methods in general.

**Figure 6.** Single country and Cross-country Estimation Poverty Maps

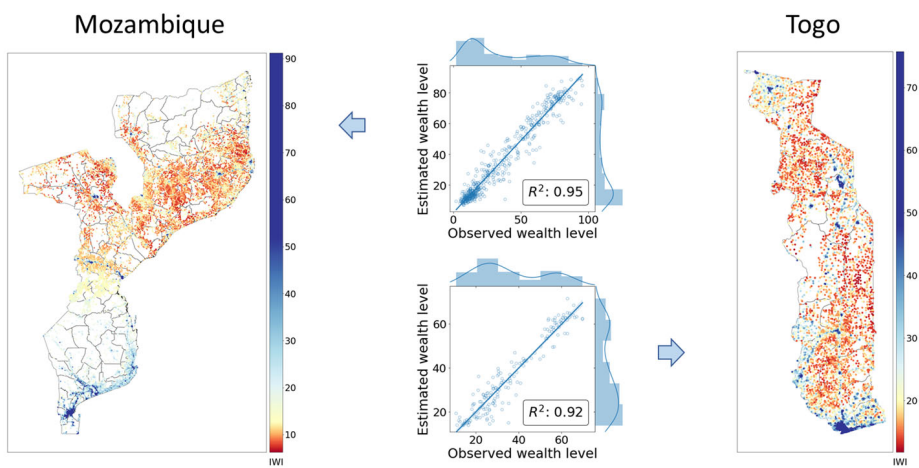

*Source:* Authors' calculation.
*Note:* 21,623 populated places in Mozambique (left, single country estimation) and 7,803 populated places in Togo (right, cross-country estimation). The R-square values for the estimations are 94.91% and 91.60%, respectively.

## Analysis of Results

The validation results improved over the iteration of our refining process (Figure 7). We analyzed each component of the process and its contribution to the overall improvement of both the single country estimation and the cross-country estimation. The refining process started with the feature-based model (iteration 0), and then narrowed the DHS cluster candidates using the previous estimates, which leads to better training data and higher R-squared (iteration 1). At this point, the single-country estimation for South Africa and the cross-country estimation for Kenya showed the biggest improvements (16.82% points and 15.43% points, respectively). In the next following refining iterations, an image-based model was trained with the previous estimates of the feature-based model, and the estimation of the image-based model was channeled into the next version of the feature-based model. This resulted in steep improvement in Nigeria's cross-country estimation (20.24% points) among others.

The most notable improvements are Malawi and South Africa, which showed biggest increases for both estimations (over 20% points for single country estimation and near 30% points for cross-country estimation). Although South Africa's cross-country estimation was less accurate than the other model countries, it still provides influential training data points for other countries. The essence of the refining process is to generate better training data in every iteration.

**Figure 7**. Improvement of Estimation over the Refining Process

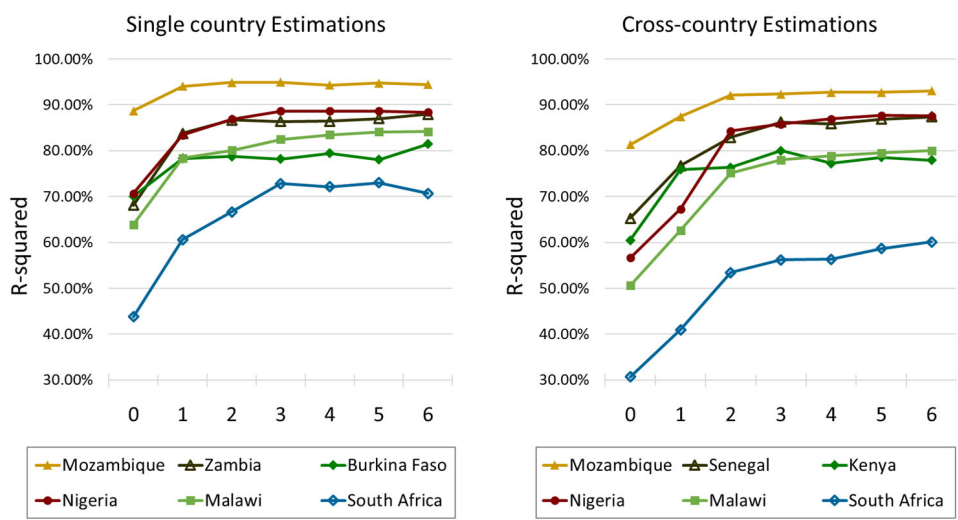

*Source:* Authors' calculation.
*Note:* Countries that showed big improvements in single country estimation (left) and cross-country estimation (right). The first two estimation models (iteration 0 and 1) are feature-based models and the following models (iteration 2, 3, 4, 5 and 6) are the combined models with feature-based and image-based estimations.

We used different validation schemes between single country and cross-country estimations. Single country estimation is a typical ML application, so we used a 5-fold cross-validation, which is a default setting in many ML libraries. Since cross-country estimation is a relatively new technique, we chose a more rigorous way to validate the result. By using leave one-group-out cross-validation, we essentially tried to find a counter-example that our method would fail. Using only 24 at a time and rotating through is the same as validating our method with a new country 25 times. After the 25 repeated experiments validation, we gained confidence that our methodology was ready for cross-country estimation.

**Figure 8**. SSA Countries with Predicted Wealth Levels at the Village Level

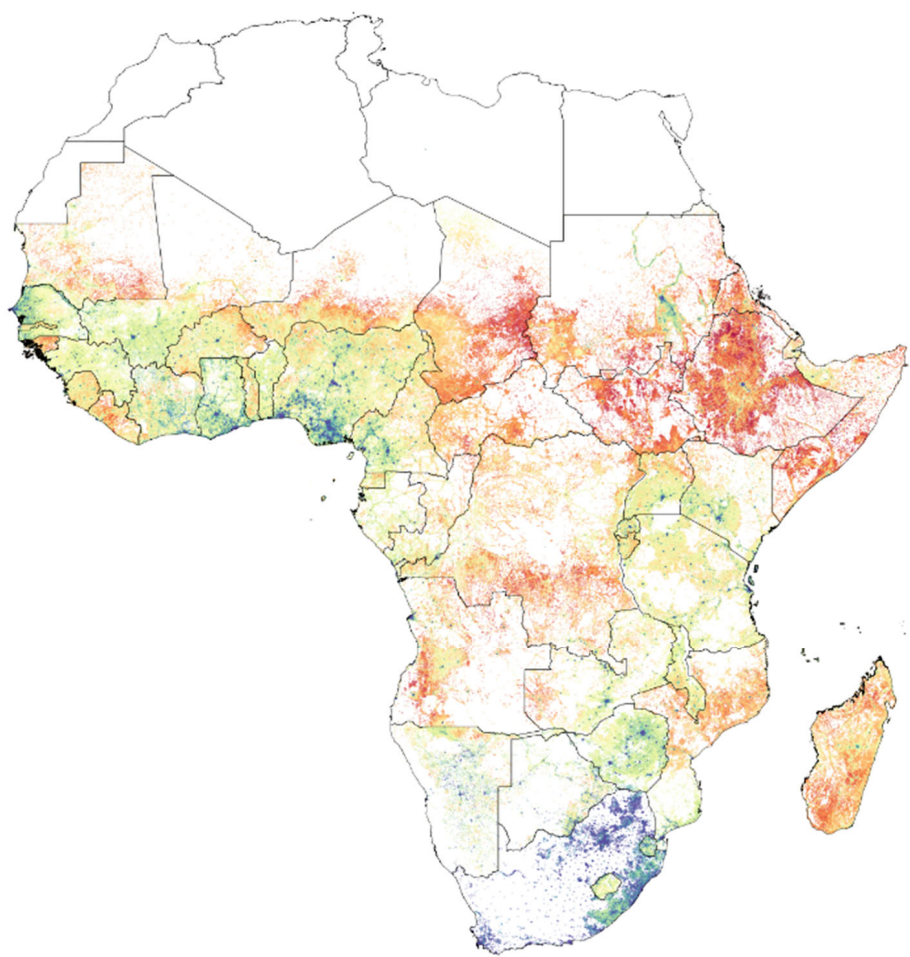

*Source:* Authors' calculation.
*Note:* Wealth level estimates for 929,295 populated places in 44 countries displayed on the same color scale. On this continental map, difference in wealth levels within a single country look small.

# Poverty Mapping for All SSA Countries and Beyond

We created poverty maps for the remaining SSA countries with the cross-country estimation model by regarding the whole SSA as a single territory (Figure 8). The model was trained and validated with pooled data from the 25 countries. The 25 model countries demonstrate multiple characteristic spectrums of the SSA countries. Senegal is the westmost country, and Madagascar is the furthest east. On the latitudinal range, we have Mali and South Africa. Nigeria is the most populous country in Africa, while Liberia's population is only 2% the size of Nigeria's population. Burundi's GDP per capita is $261.2 while South Africa's is $6,001.4 (World Bank 2019). When evaluated with 5-fold cross-validation while ignoring the country borders, the R-squared of our estimation is 91.7%. Among the remaining 24 SSA countries, we left out Cabo Verde, Comoros, Mauritius, São Tomé & Príncipe, and Seychelles since they are small islands the size of a city, which are not suited for this type of poverty maps (e.g., only 49 populated places in Seychelles). We generated not only pictures of maps, but also corresponding datasets including names of places, GPS coordinates, administrative hierarchy, population density, and wealth levels for full applications of our poverty maps.

We are also considering proper steps to maintain ongoing accuracy of the maps and the resulting data. Although the OpenStreetMap data and satellite images are updated frequently, we believe that it would be meaningful for policy impact to update our model and resulting maps bi-annually or annually. We should re-train our model every time a new DHS dataset is released. Thus, we will remove old DHS data and add new data on a rolling basis. Our predicted poverty maps also depend on the estimation of how many people live in which areas. As the estimations of populated places are revised, our prediction maps should be updated accordingly.

To extend to work beyond SSA, we first need to divide the world into regions and create a model for each region based on survey data. Outside of SSA, there are many low- and middle-income countries in the Middle East and North Africa (MENA),

**Table 2.** Regions where the Most Low- and Middle-income Countries are Located

| Region | [# of Countries] | [# of Countries with DHS] |
| --- | --- | --- |
| Sub-Saharan Africa | 49 | 25 |
| Middle East and North Africa | 21 | 1 |
| Latin America and Caribbean | 33 | 3 |
| Central, South, and Southeast Asia | 24 | 6 |

*Source:* World Bank, USAID
*Note:* [# of Countries] only includes sovereign states (excluding dependent territories and administrative subdivisions). [# of Countries with DHS] counts countries that have DHS studies in recent 5 years with corresponding GPS coordinates available as of August 2020.

Latin America and Caribbean (LAC), and Central, South & Southeast Asia (CSSA) regions (Table 2). We conjecture that CSSA can be the next approachable region after SSA since there are 6 DHS studies within the past 5 years in this region that will allow us to create cross-country estimates based on the 6 model countries. For MENA and LAC, there are not enough model countries. This dilemma introduces a new series of questions: Is cross-continent estimation possible? Are there common geospatial elements that can be applicable globally? If so, what data can we use and how? We will also have to experiment with other ML techniques as one of the drawbacks of XGBoost and other popular decision-tree-based algorithms is that they do not extrapolate well. Considering the importance of extrapolation in cross-continent estimations, we will look into other recent research studies in the ML community that are working to increase extrapolation performance of ML algorithms.

## Conclusions

We proposed a generalizable prediction methodology to produce poverty maps at the village level using geospatial data and machine learning algorithms. Along the way, we also achieved higher validity of single country estimations with our refining mechanism than previous studies. Our highest R-squared is 94.91% for Mozambique and our average R-squared is 88.62% for the 25 countries. This demonstrates that we created highly accurate poverty maps for the 25 countries. While each one of these maps has policy implications for its country, we want to emphasize the efficacy of the cross-country estimation, which is the main topic of this study.

We repeatedly tested our cross-country prediction model 25 times by assuming that there was no survey data in one of the countries, estimating the missing country's wealth levels from other 24 countries, and validating the results. The best and average R-squared values are 93.04% and 85.60% respectively. We believe that this level of validity is high enough to apply our model to other countries where there is no survey data with GPS coordinates. We thus predicted wealth levels for the remaining 19 SSA countries that lack survey data, completing village-level poverty maps in SSA. Our cross-country estimations can be updated easily and cheaply when new data become available, enabling policy makers to rapidly access up-to-date geospatial data about where the poor are located in their countries.

Poverty maps have long been desirable tools for policy makers to ensure that anti-poverty programs reach their targeted beneficiaries, and that services and infrastructure are sited appropriately. However, traditional poverty maps are expensive and rapidly become outdated. Using new public data sources and ML techniques, our methodology will bring the advantages of updated poverty maps to low- and medium-income countries affordably.

# Notes

Kamwoo Lee is a Data Scientist at the World Bank, working in the Data Analytics and Tools Unit (DECAT), and can be reached at klee16@worldbank.org. Jeanine Braithwaite is a Professor of Public Policy at the Frank Batten School of Leadership & Public Policy, UVA, and can be reached at jdb6bc@virginia.edu. This work was supported by the Center for Global Health at UVA.

1. In this paper, we loosely define "village-level", for brevity, as an area covered by a zoom-16, 640 px x 640 px (approximately 1 square mile) satellite image, which we refer to as high-resolution.
2. See Ravallion (2015) for a summary
3. The IWI allows up to three missing indicators. We dropped cheap/expensive utensil indicators because the IWI and the DHS wealth index have higher correlation for most countries without the two indicators.
4. We define roads as ways in OpenStreetMap that have one of the following tags: 'primary', 'primary_link', 'secondary', 'secondary_link', 'tertiary', 'tertiary_link', 'trunk', 'trunk_link', 'motorway'. We also collected the surface types of roads to distinguish paved and unpaved roads.
5. Our definition of junction is a node in OpenStreetMap where two or more ways cross, or three or more ways begin. This definition includes nodes that don't count as a typical road intersection, such as a road forking to two parallel trunk roads and a link to an overpass. We count them as junctions since they reflect a similar concept of man-made structure complexity to a road intersection.
6. 24 points of interest: bar, cafe, fast food, pub, college, kindergarten, library, school, university, bus station, atm, bank, clinic, dentist, hospital, pharmacy, veterinary, cinema, community centre, courthouse, embassy, marketplace, police, townhall.
7. Max/mean/median luminosity, ratio of zero luminosity, average of upper/lower third luminosity.
8. The meter per pixel resolution is calculated by $156{,}543 \times \cos(\text{latitude} \times \pi \div 180) \div 2^{\text{zoom}}$
9. We collected all nodes in OpenStreetMap that have one of the following tags: 'city', 'town', 'village', 'hamlet', or 'isolated_dwelling'.
10. The number of populated places in a country may not be proportional to the population size of the country since both OCHA settlement and OSM data are updated by different organizations or people.
11. We used Bayesian optimization for tuning hyperparameters. Our hyperparameter space is as follows: 'learning_rate' (0.005, 0.03), 'n_estimators' (200, 300), 'max_depth' (3, 10), 'min_child_weight' (2, 20), 'subsample' (0.2, 1.0), 'colsample_bytree' (0.2, 1.0), 'lambda' (0, 0.1)
12. Workstation with dual 8-core Xeon processors, 128GB RAM, and NVIDIA TITAN RTX graphics processing unit.
13. For the detailed configuration, our customized CNN has input of 640x640x3 size (random rotation, random horizontal/vertical flip during training), 7 convolutional layers (# channels: 16, 32, 64, 128, 256, 512, 1024 respectively with batch normalization and ReLU activation on each layer), 5 fully-connected layers (# nodes: 25600, 4096, 1024, 256, 64 respectively with 50% dropout and ReLU activation on each layer except for the last layer), and output of 4 dimensions (poor, lower middle, upper middle, rich). We train it with a batch size of 64 and Adam optimizer (learning rate: initially 1e-4, reducing by a factor of 10 on a validation loss plateau).

14. It is possible to use a pre-trained model such as VGGNet and Inception. However, our input image size (640 x 640 pixels) is much bigger than the input images of any pre-trained models. We tried using the VGG-16 model by replacing the fully-connected top layers with fully-convolutional layers to adapt a pre-trained model for larger images, but the replaced fully-convolutional layers became too big to take advantage of the trained lower layers. The number of trainable parameters and the total size of our customized CNN is much smaller but as effective as VGG-16 in estimating wealth levels.

15. Typically, transfer learning is applied by taking lower layers (convolutional layers) of a pre-trained model and adding new upper layers (fully connected layers) that need to be trained on new data. By doing so, we can retain the pre-trained model's ability to discern low-level image information such as edges, blobs and other trivial geometries. We followed this approach with different initial learning rates of 1e-6 and 1e-4 for lower and upper layers respectively.

Appendix A

# High-Resolution Poverty Maps in Sub-Saharan Africa

## Kamwoo Lee and Jeanine Braithwaite

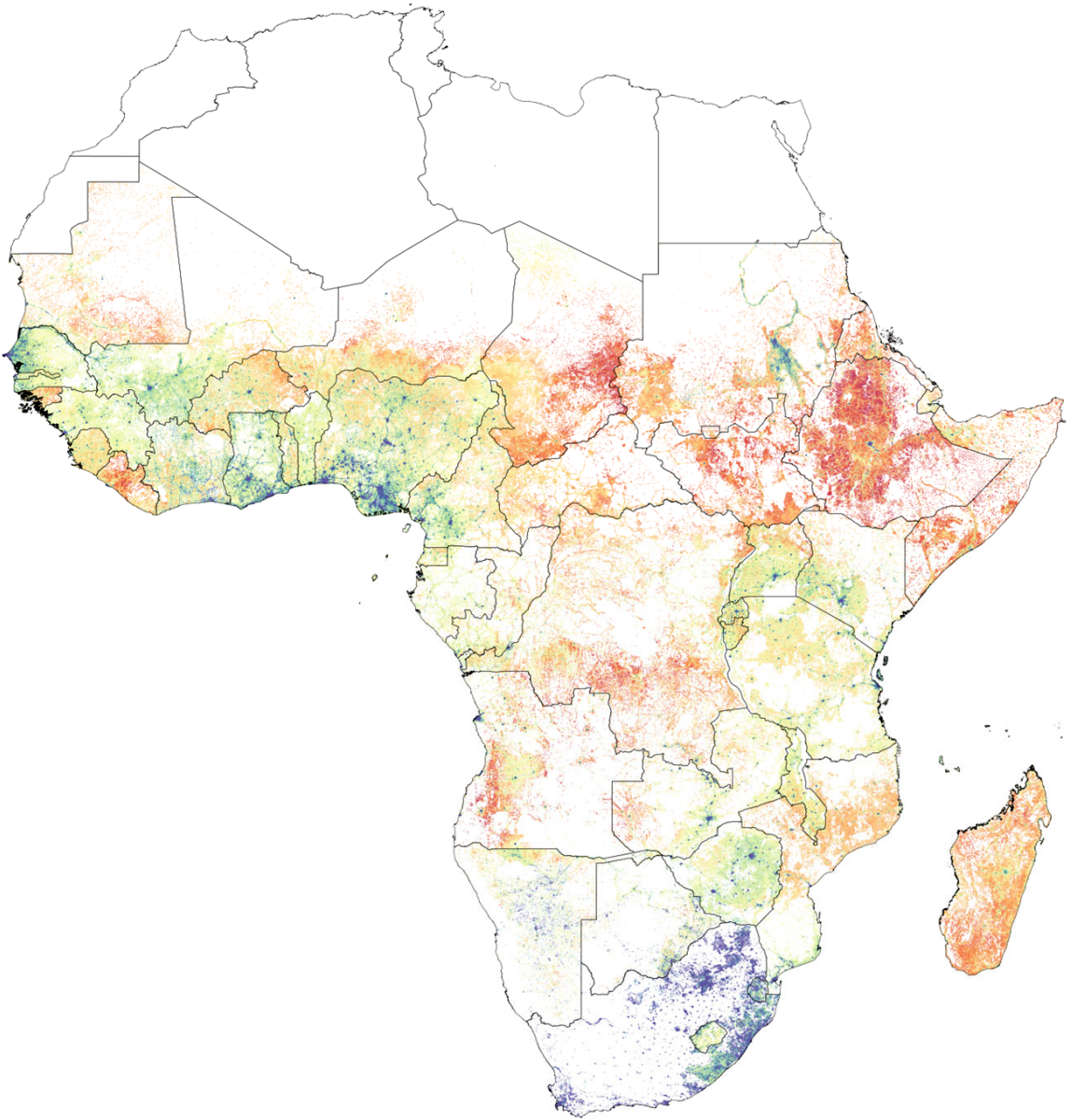

# High-Resolution Poverty Map
# Angola

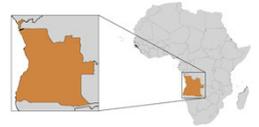

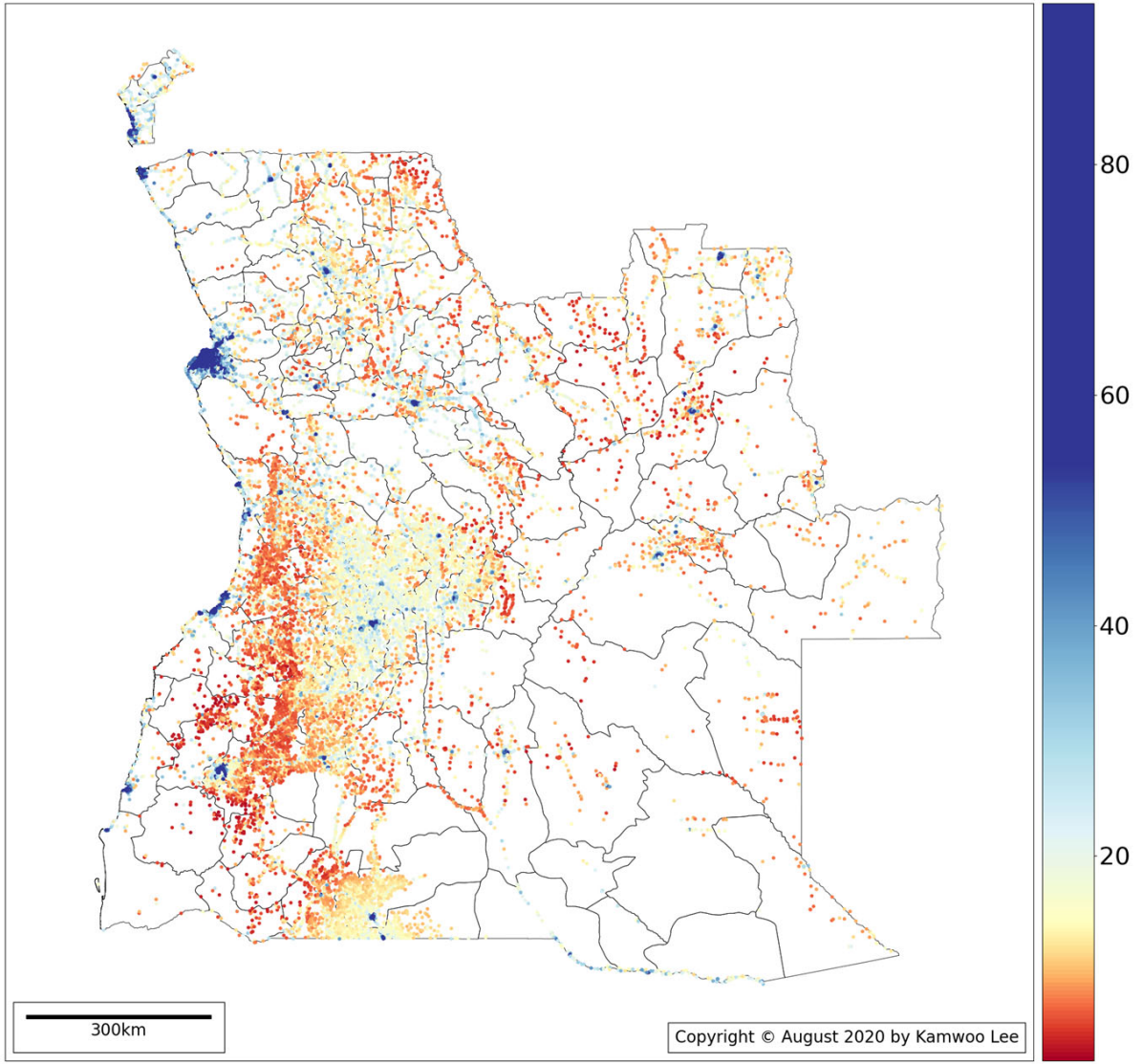

### Estimated Wealth Level (August 2020)

This map displays estimated average wealth levels of households in 1 square-mile populated areas. The wealth level was estimated on the 0-100 International Wealth Index scale (color code: red-poor, yellow-median, blue-rich) using machine learning methods with geospatial information including OpenStreetMap, daytime satellite images, nighttime luminosity, and High-Resolution Population Densities. The estimation was validated with 2015-2016 Standard DHS.

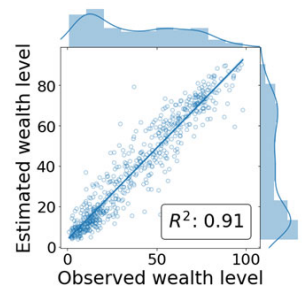

# High-Resolution Poverty Map
# Benin

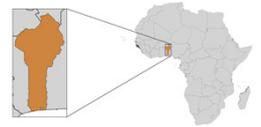

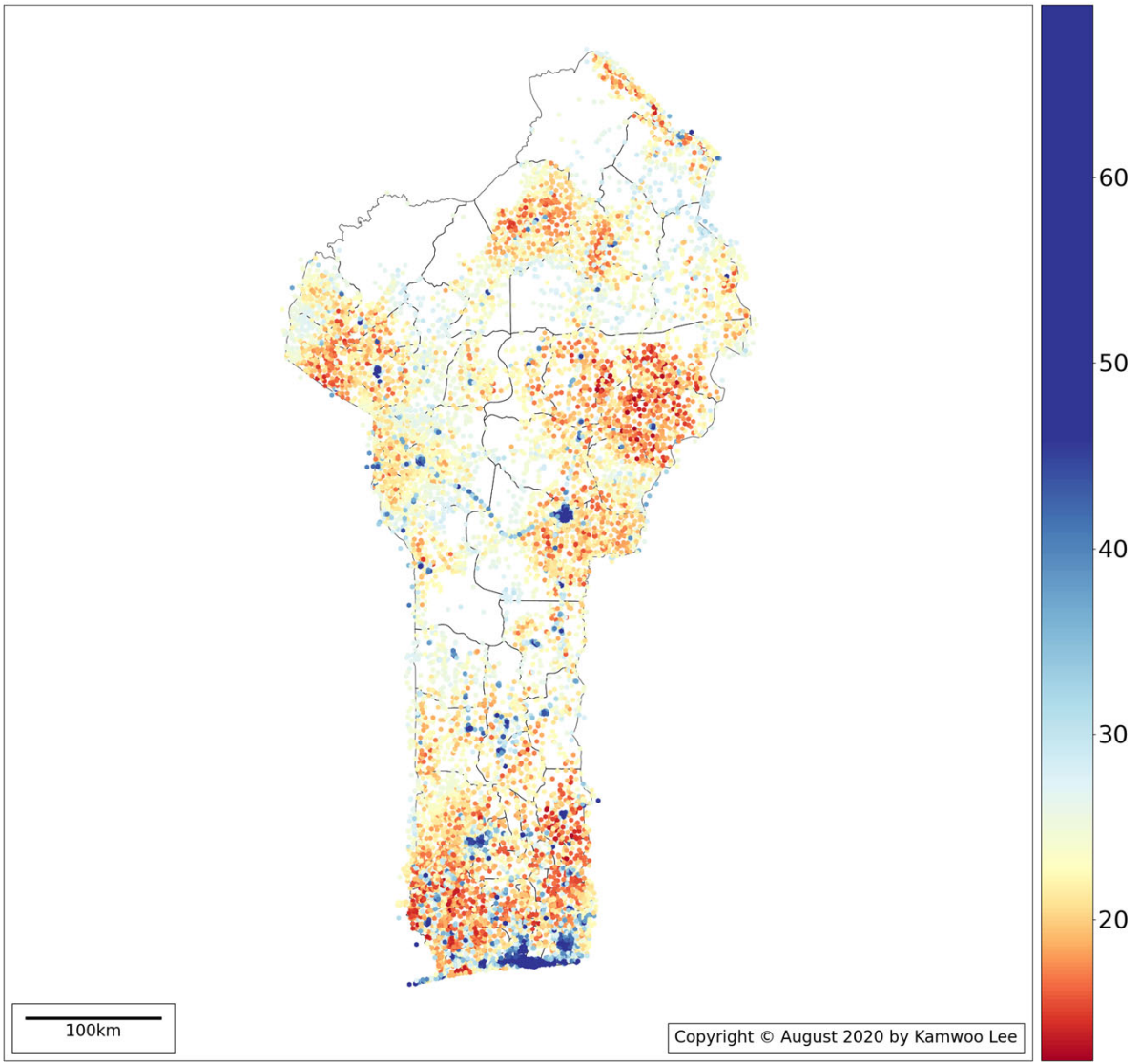

### Estimated Wealth Level (August 2020)
This map displays estimated average wealth levels of households in 1 square-mile populated areas. The wealth level was estimated on the 0-100 International Wealth Index scale (color code: red-poor, yellow-median, blue-rich) using machine learning methods with geospatial information including OpenStreetMap, daytime satellite images, nighttime luminosity, and High-Resolution Population Densities. The estimation was validated with 2017-2018 Standard DHS.

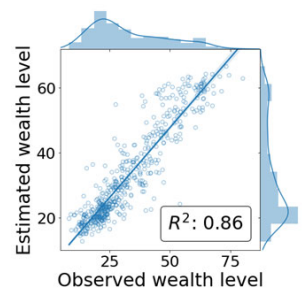

# High-Resolution Poverty Map
# Botswana

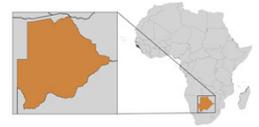

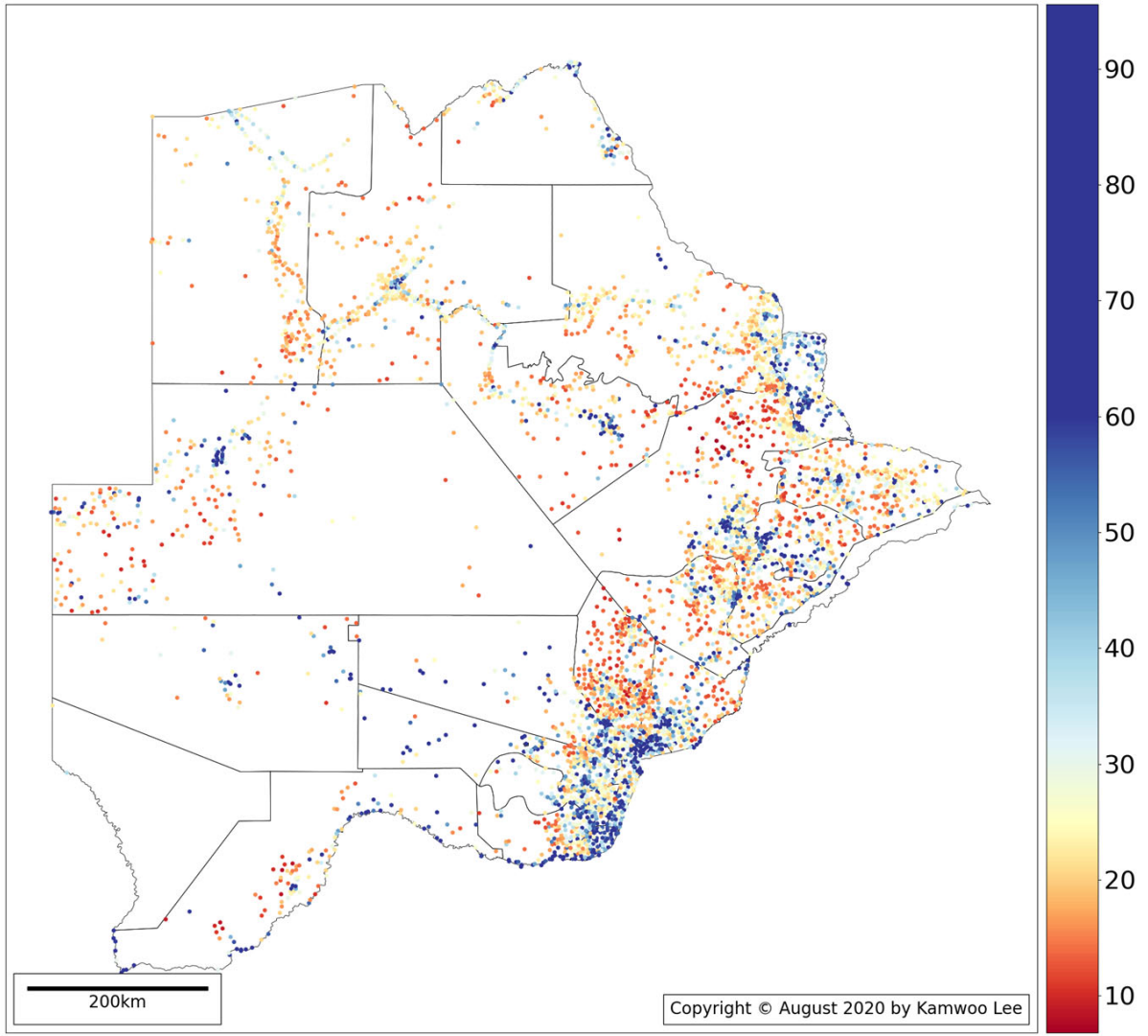

### Estimated Wealth Level (August 2020)
This map displays estimated average wealth levels of households in 1 square-mile populated areas. The wealth level was estimated on the 0-100 International Wealth Index scale (color code: red-poor, yellow-median, blue-rich) using machine learning methods with geospatial information including OpenStreetMap, daytime satellite images, nighttime luminosity, and High-Resolution Population Densities. This is a cross-country estimation that is validated with the DHS data from 25 SSA countries ($R^2$: 0.91).

# High-Resolution Poverty Map
# Burkina Faso

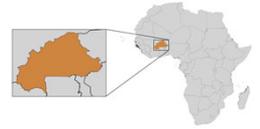

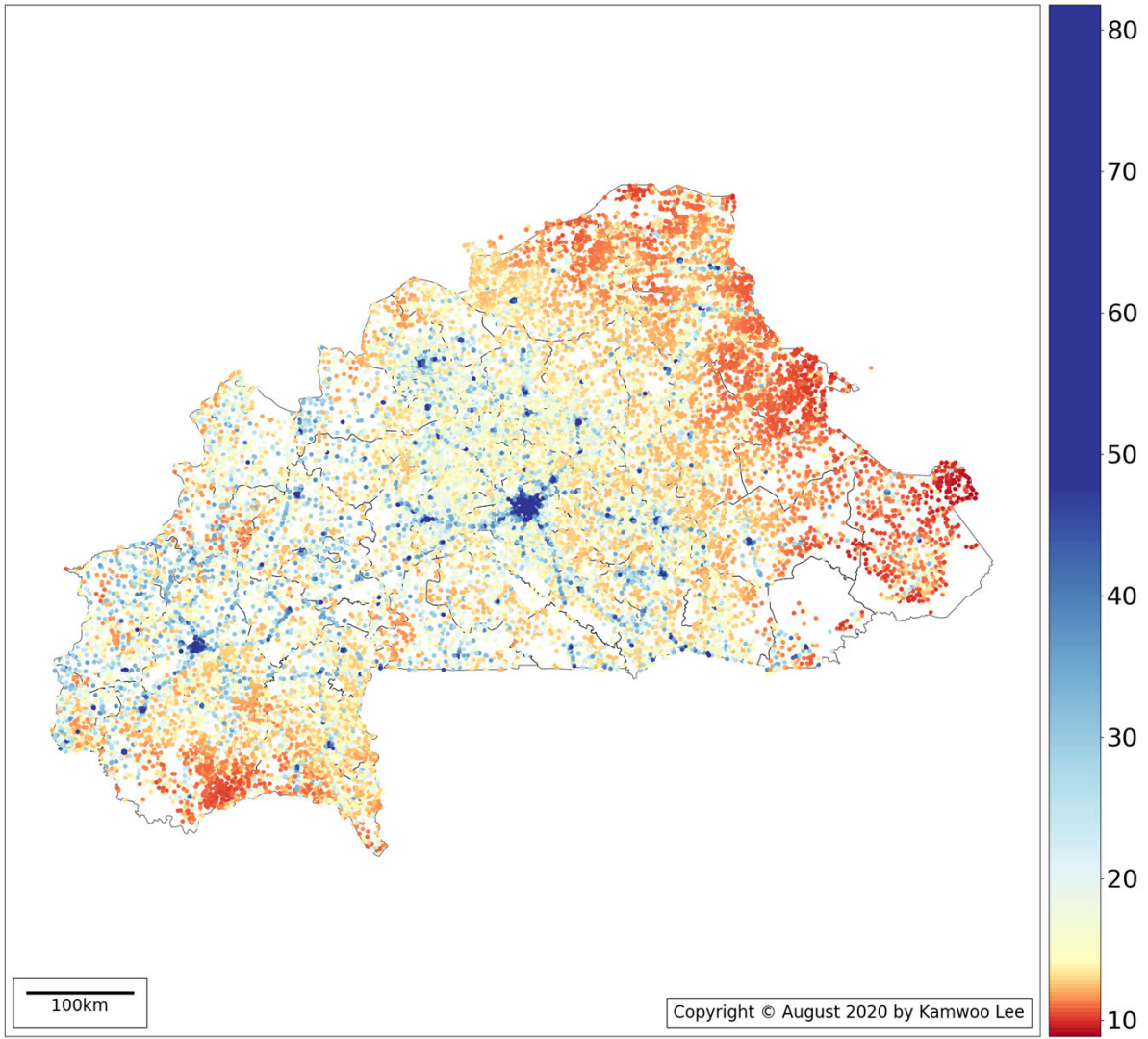

### Estimated Wealth Level (August 2020)
This map displays estimated average wealth levels of households in 1 square-mile populated areas. The wealth level was estimated on the 0-100 International Wealth Index scale (color code: red-poor, yellow-median, blue-rich) using machine learning methods with geospatial information including OpenStreetMap, daytime satellite images, nighttime luminosity, and High-Resolution Population Densities. The estimation was validated with 2017-2018 MIS.

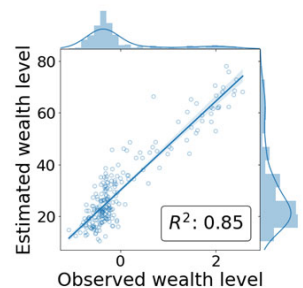

# High-Resolution Poverty Map
# Burundi

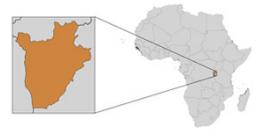

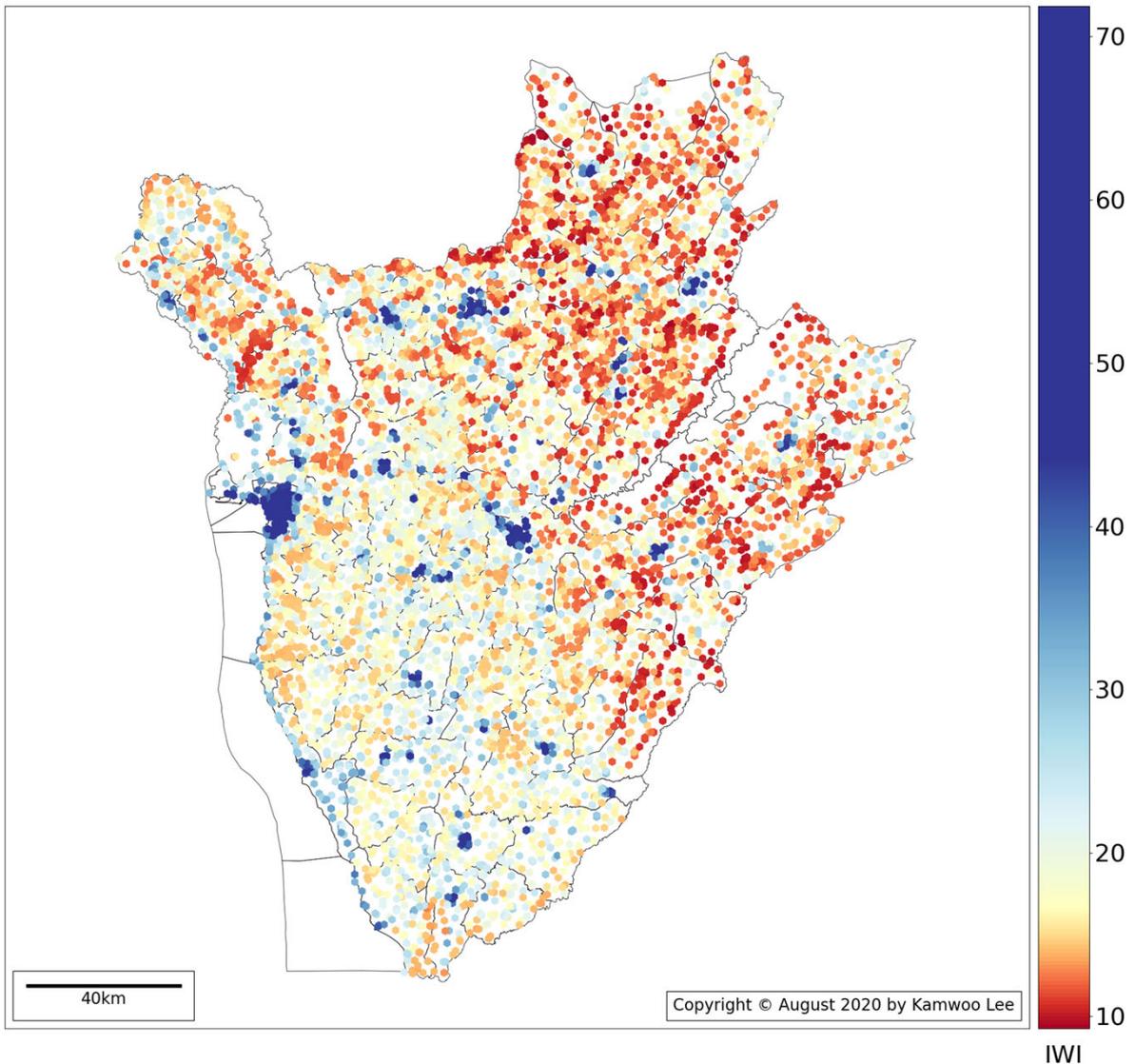

### Estimated Wealth Level (August 2020)
This map displays estimated average wealth levels of households in 1 square-mile populated areas. The wealth level was estimated on the 0-100 International Wealth Index scale (color code: red-poor, yellow-median, blue-rich) using machine learning methods with geospatial information including OpenStreetMap, daytime satellite images, nighttime luminosity, and High-Resolution Population Densities. The estimation was validated with 2016-2017 Standard DHS.

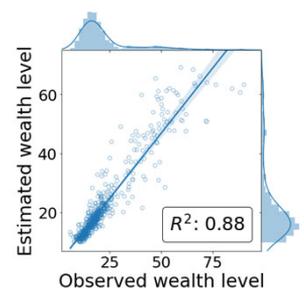

# High-Resolution Poverty Map
# Cameroon

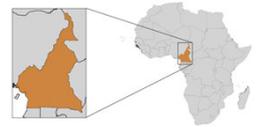

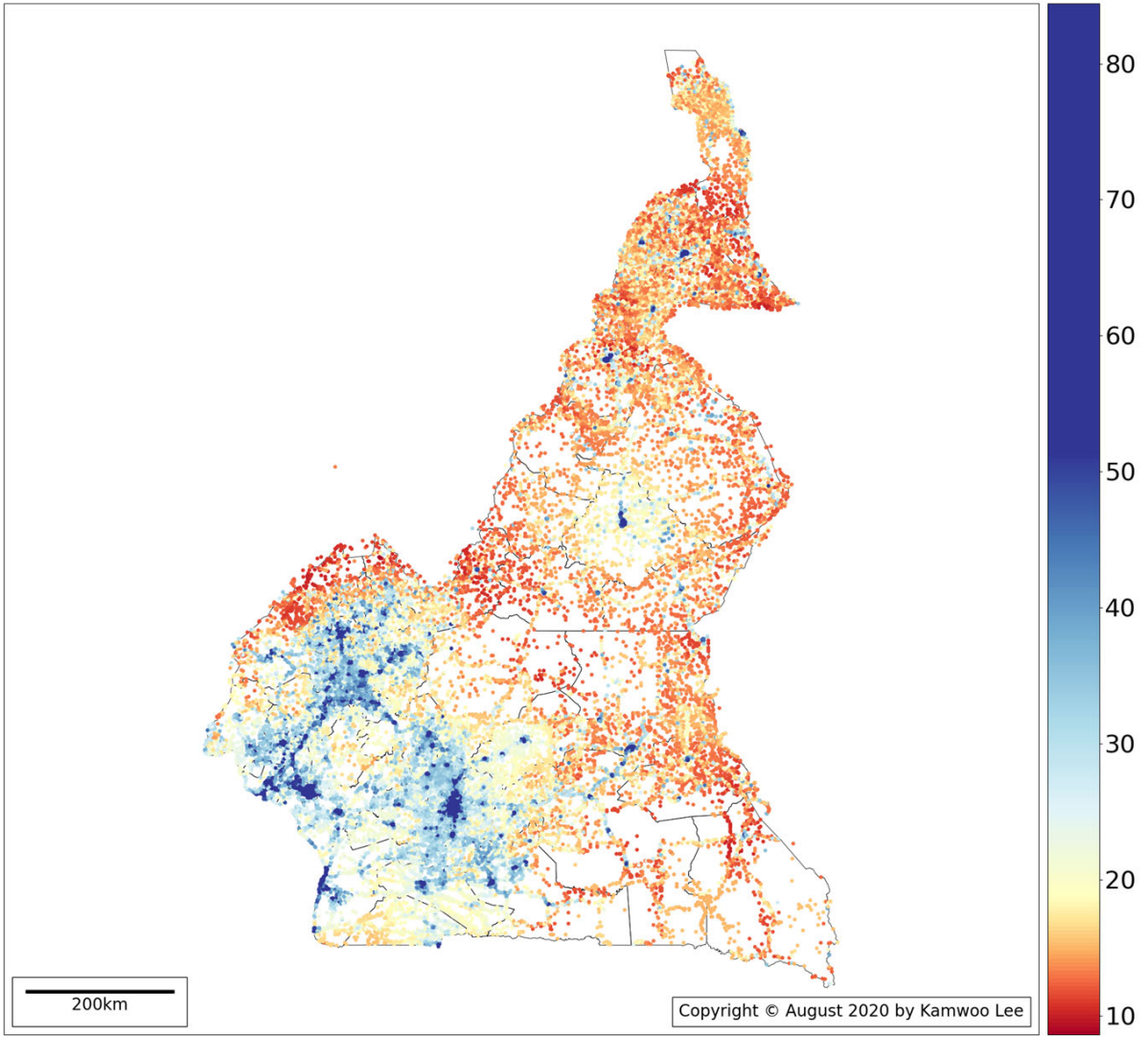

### Estimated Wealth Level (August 2020)
This map displays estimated average wealth levels of households in 1 square-mile populated areas. The wealth level was estimated on the 0-100 International Wealth Index scale (color code: red-poor, yellow-median, blue-rich) using machine learning methods with geospatial information including OpenStreetMap, daytime satellite images, nighttime luminosity, and High-Resolution Population Densities. The estimation was validated with 2018 Standard DHS.

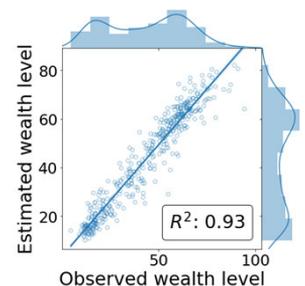

# High-Resolution Poverty Map
# Central African Republic

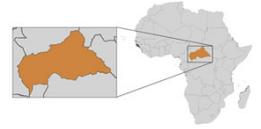

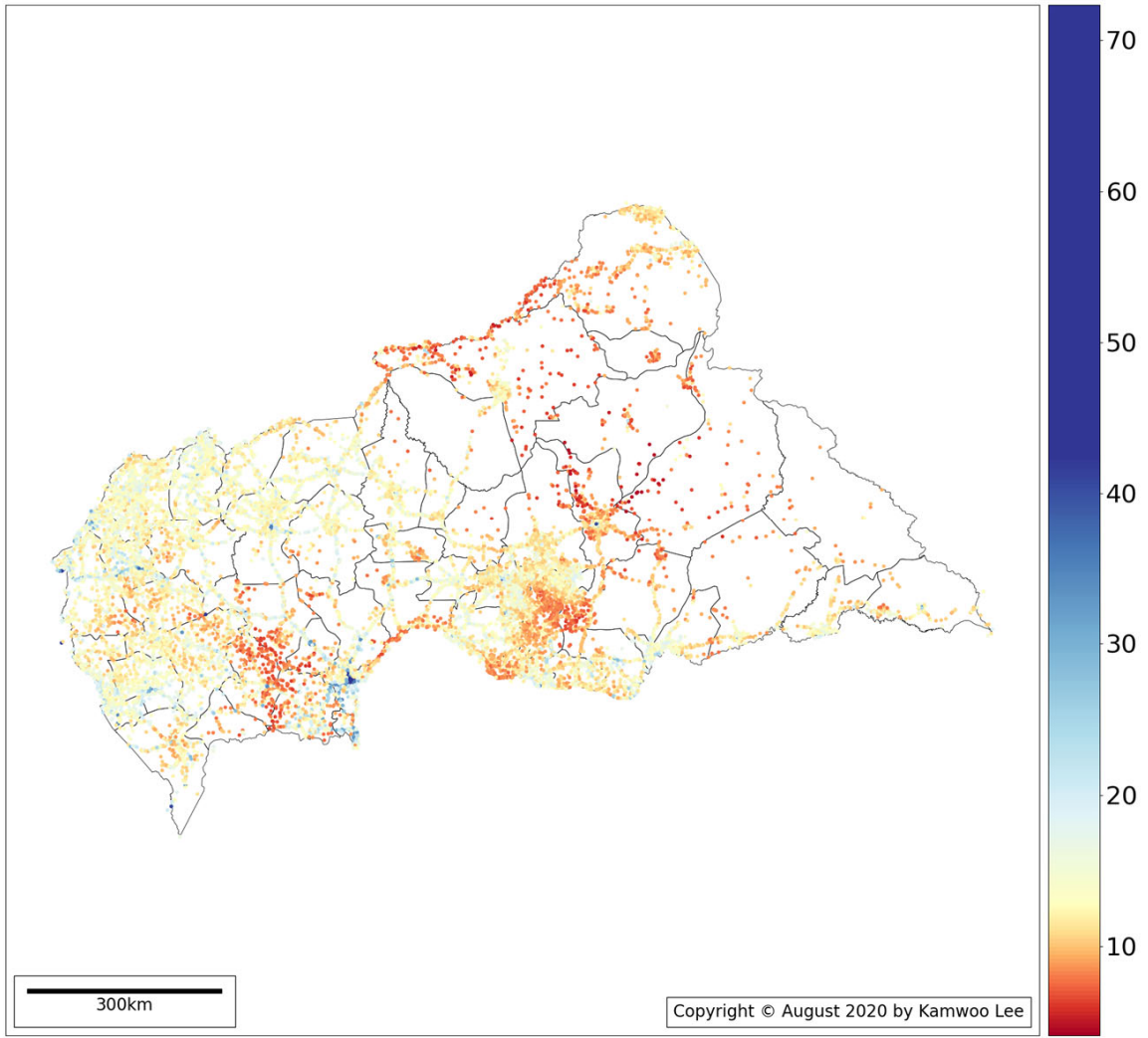

**Estimated Wealth Level (August 2020)**
This map displays estimated average wealth levels of households in 1 square-mile populated areas. The wealth level was estimated on the 0-100 International Wealth Index scale (color code: red-poor, yellow-median, blue-rich) using machine learning methods with geospatial information including OpenStreetMap, daytime satellite images, nighttime luminosity, and High-Resolution Population Densities. This is a cross-country estimation that is validated with the DHS data from 25 SSA countries ($R^2$: 0.91).

# High-Resolution Poverty Map
# Chad

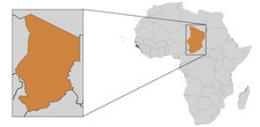

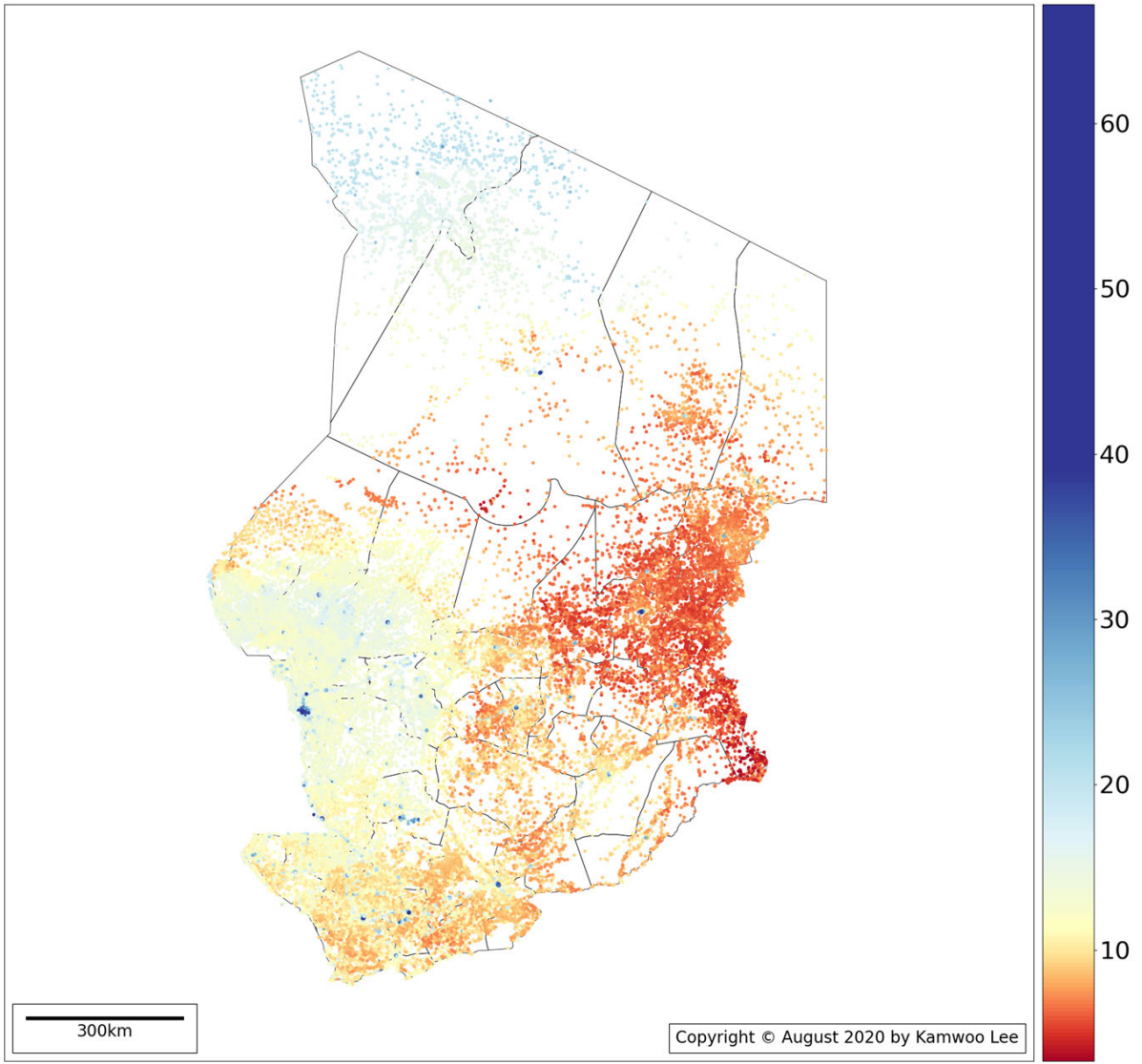

### Estimated Wealth Level (August 2020)

This map displays estimated average wealth levels of households in 1 square-mile populated areas. The wealth level was estimated on the 0-100 International Wealth Index scale (color code: red-poor, yellow-median, blue-rich) using machine learning methods with geospatial information including OpenStreetMap, daytime satellite images, nighttime luminosity, and High-Resolution Population Densities. The estimation was validated with 2014-2015 Standard DHS.

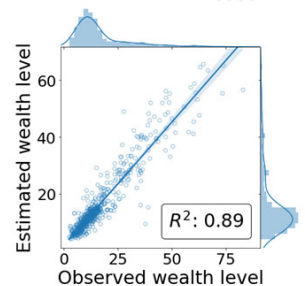

# High-Resolution Poverty Map
# Congo, Democratic Republic of the

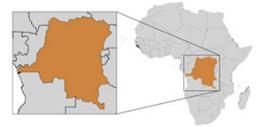

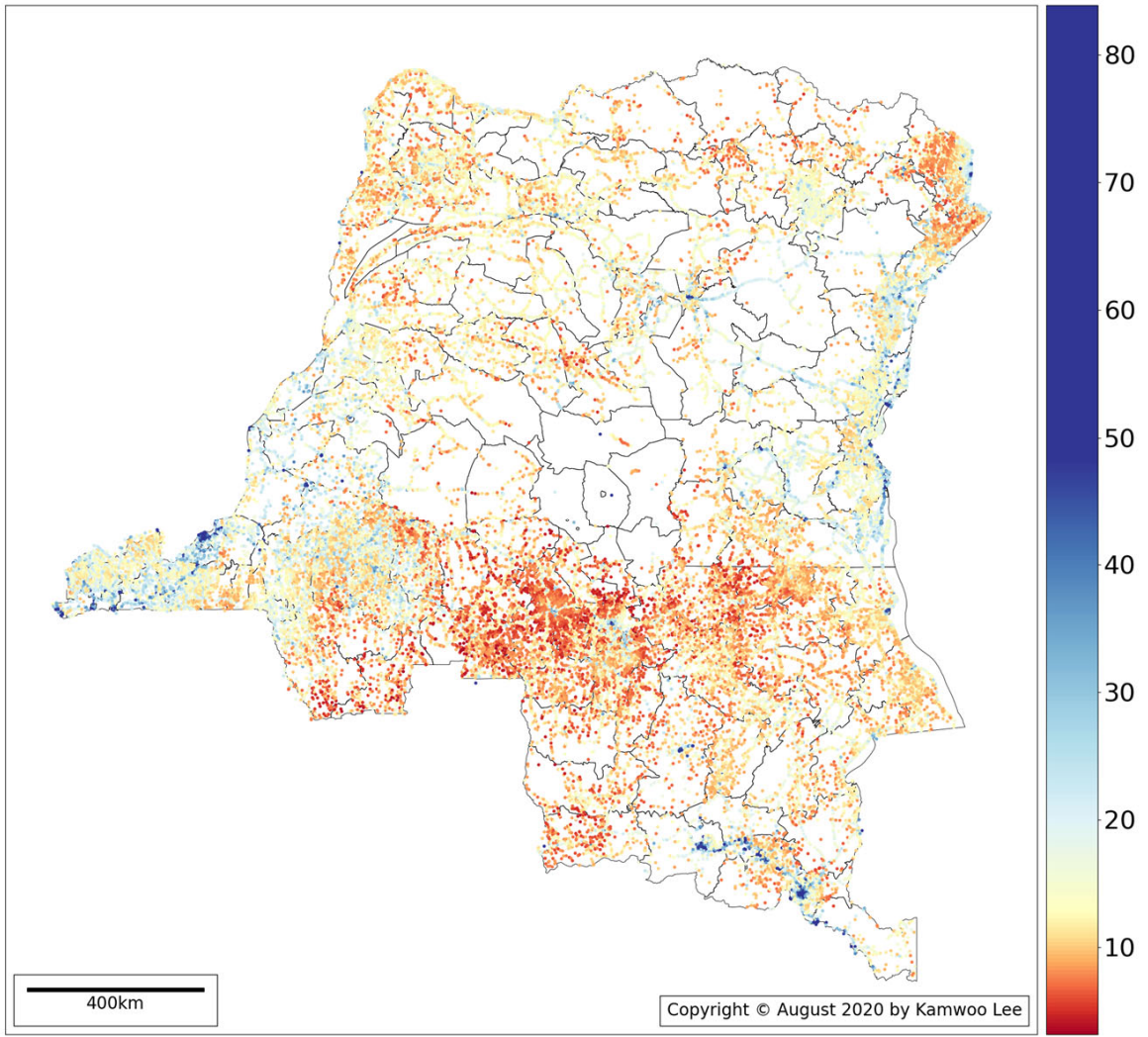

### Estimated Wealth Level (August 2020)
This map displays estimated average wealth levels of households in 1 square-mile populated areas. The wealth level was estimated on the 0-100 International Wealth Index scale (color code: red-poor, yellow-median, blue-rich) using machine learning methods with geospatial information including OpenStreetMap, daytime satellite images, nighttime luminosity, and High-Resolution Population Densities. This is a cross-country estimation that is validated with the DHS data from 25 SSA countries ($R^2$: 0.91).

# High-Resolution Poverty Map

# Congo, Republic of the

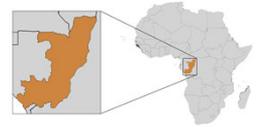

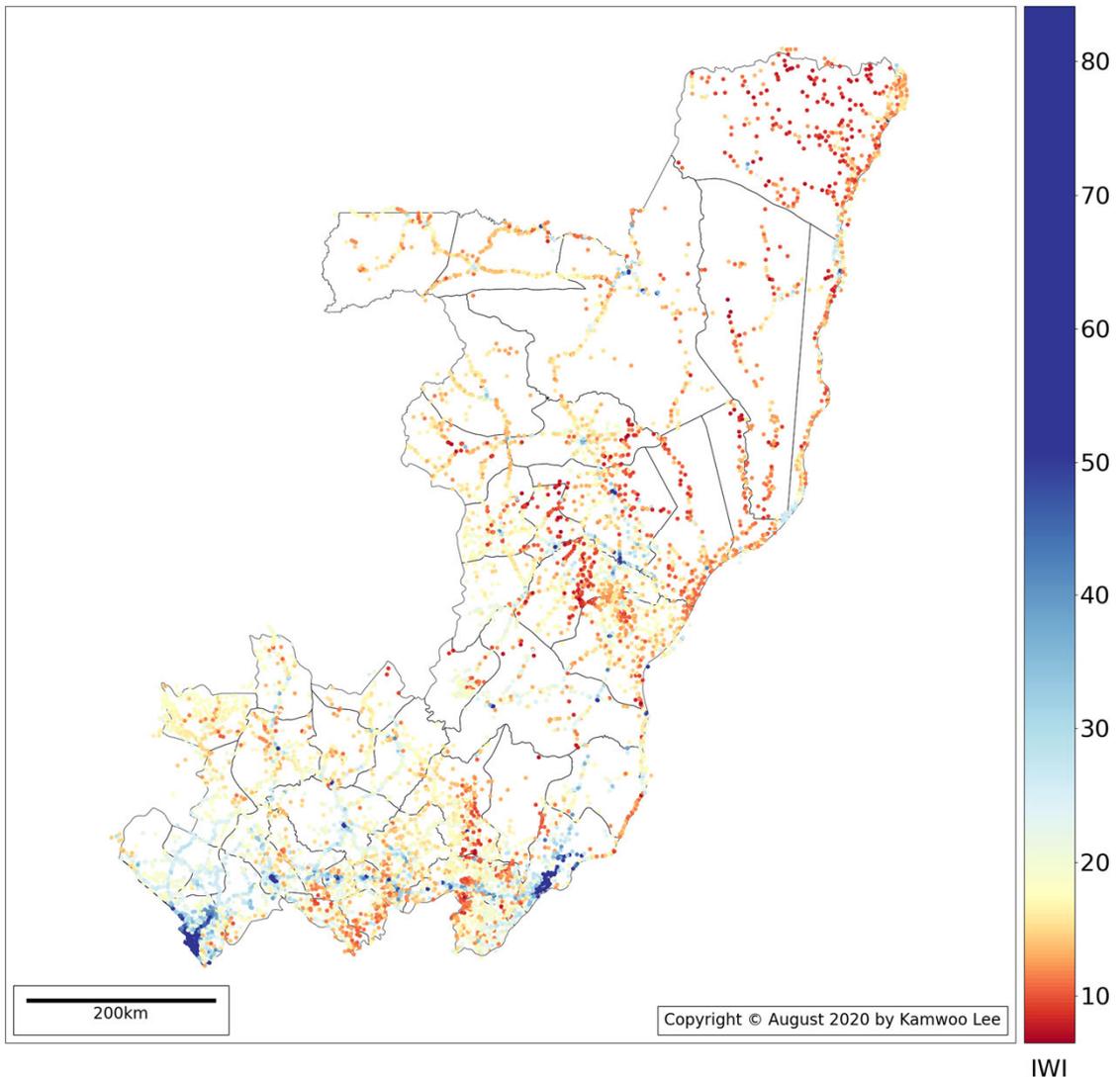

**Estimated Wealth Level (August 2020)**
This map displays estimated average wealth levels of households in 1 square-mile populated areas. The wealth level was estimated on the 0-100 International Wealth Index scale (color code: red-poor, yellow-median, blue-rich) using machine learning methods with geospatial information including OpenStreetMap, daytime satellite images, nighttime luminosity, and High-Resolution Population Densities. This is a cross-country estimation that is validated with the DHS data from 25 SSA countries ($R^2$: 0.91).

# High-Resolution Poverty Map
# Côte d'Ivoire

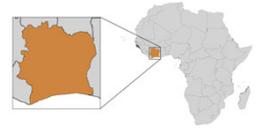

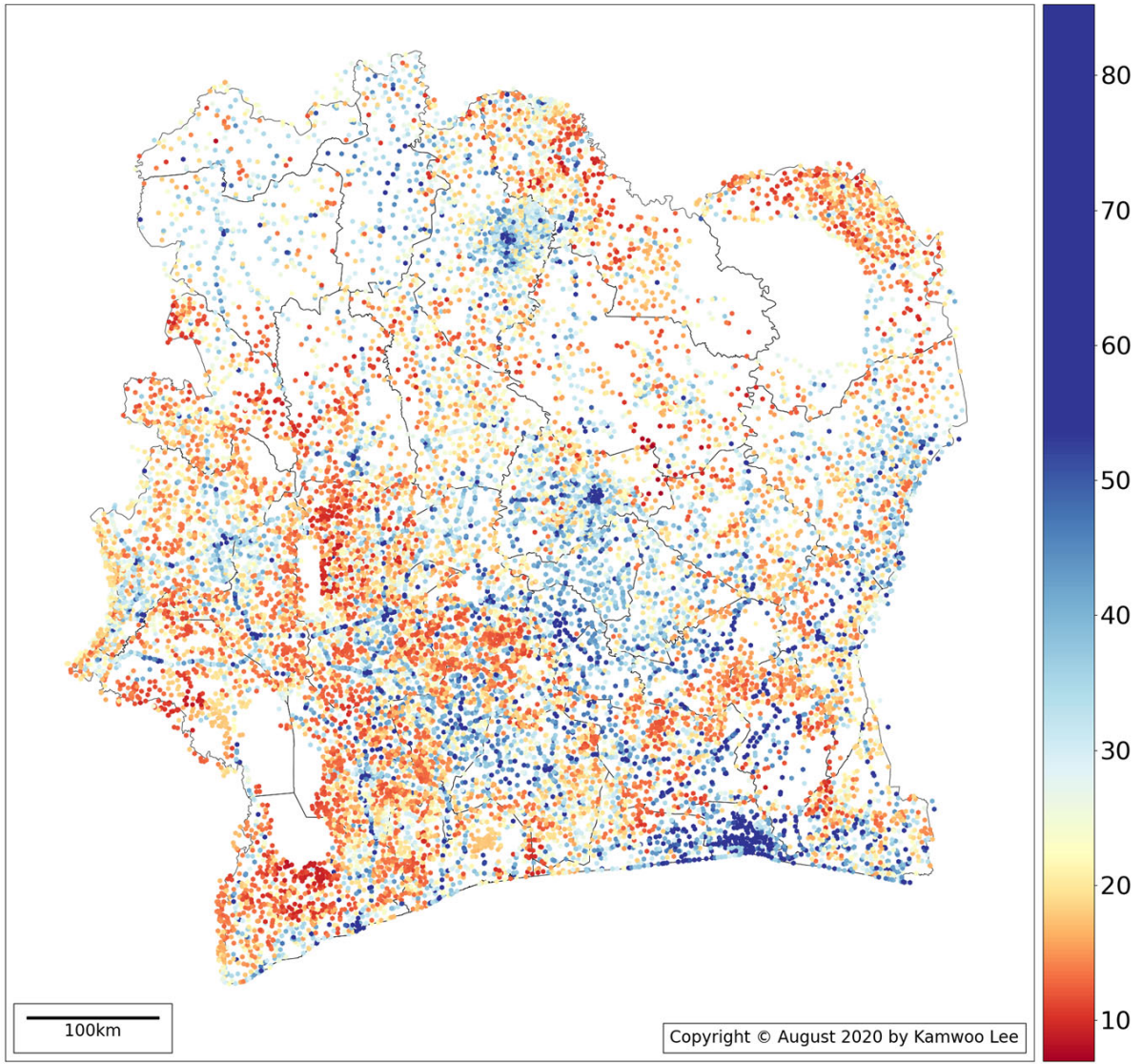

**Estimated Wealth Level (August 2020)**
This map displays estimated average wealth levels of households in 1 square-mile populated areas. The wealth level was estimated on the 0-100 International Wealth Index scale (color code: red-poor, yellow-median, blue-rich) using machine learning methods with geospatial information including OpenStreetMap, daytime satellite images, nighttime luminosity, and High-Resolution Population Densities. This is a cross-country estimation that is validated with the DHS data from 25 SSA countries ($R^2$: 0.91).

# High-Resolution Poverty Map
# Djibouti

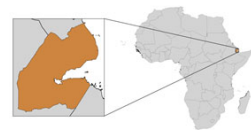

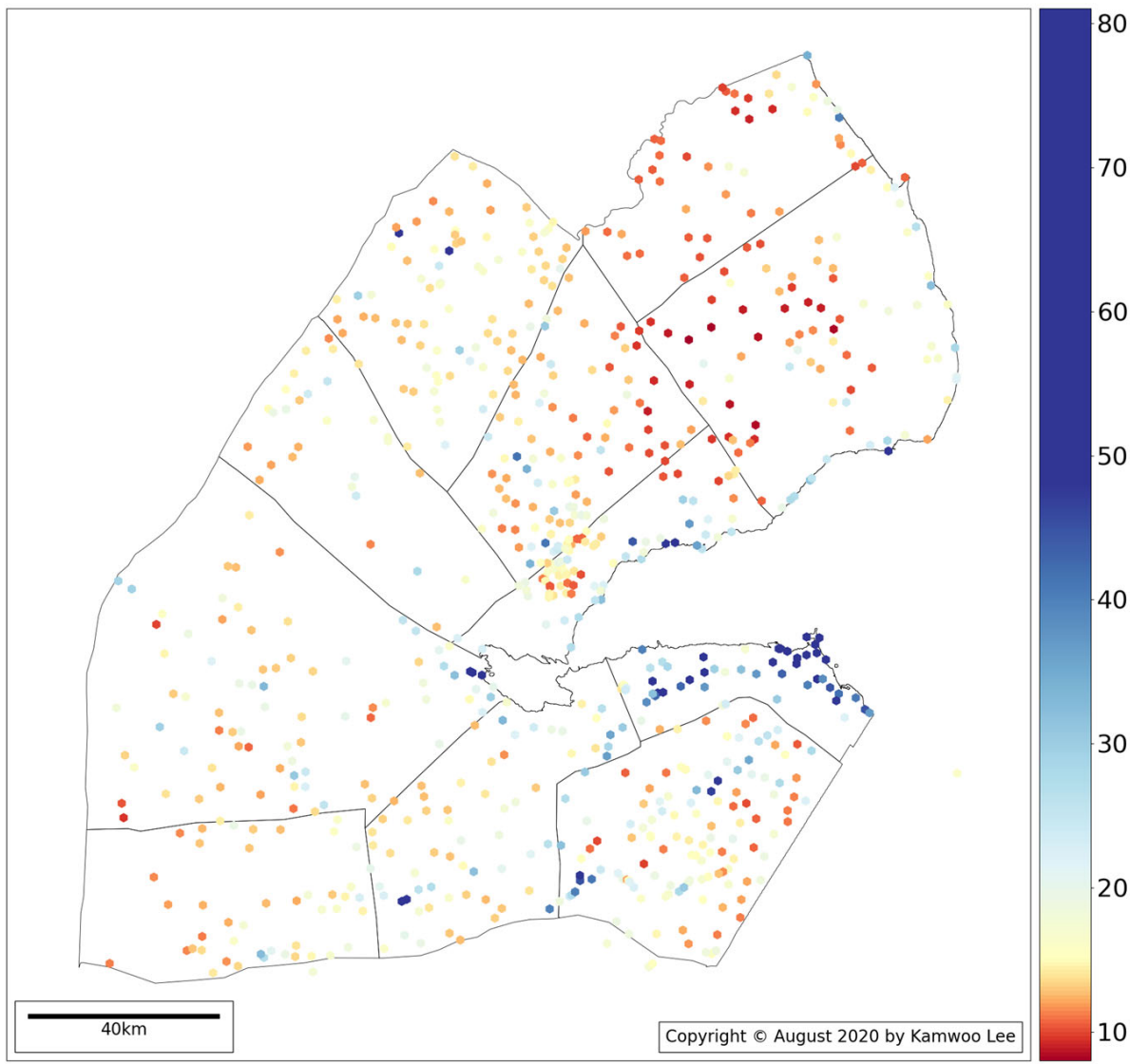

### Estimated Wealth Level (August 2020)
This map displays estimated average wealth levels of households in 1 square-mile populated areas. The wealth level was estimated on the 0-100 International Wealth Index scale (color code: red-poor, yellow-median, blue-rich) using machine learning methods with geospatial information including OpenStreetMap, daytime satellite images, nighttime luminosity, and High-Resolution Population Densities. This is a cross-country estimation that is validated with the DHS data from 25 SSA countries ($R^2$: 0.91).

# High-Resolution Poverty Map
# Equatorial Guinea

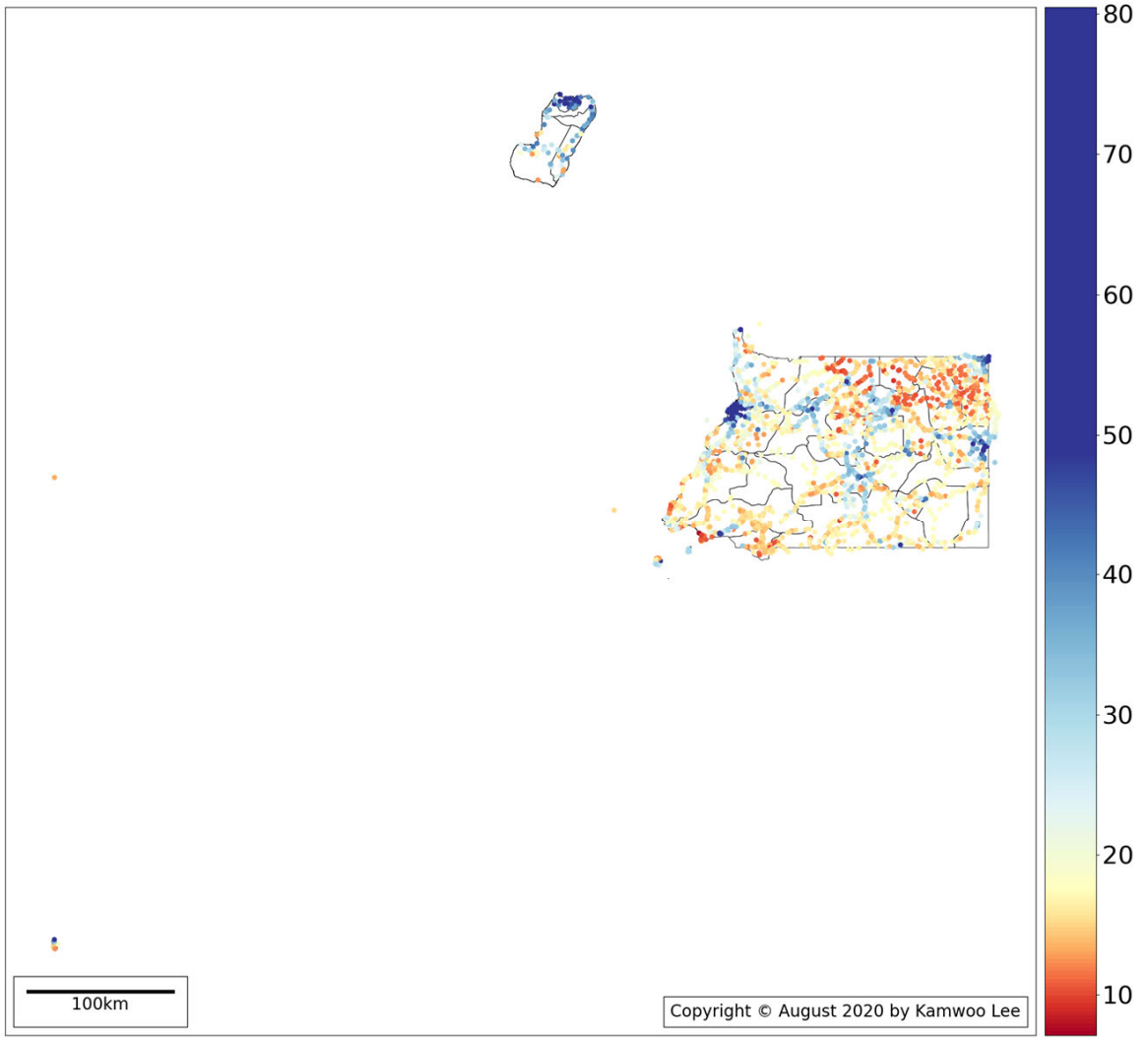

### Estimated Wealth Level (August 2020)
This map displays estimated average wealth levels of households in 1 square-mile populated areas. The wealth level was estimated on the 0-100 International Wealth Index scale (color code: red-poor, yellow-median, blue-rich) using machine learning methods with geospatial information including OpenStreetMap, daytime satellite images, nighttime luminosity, and High-Resolution Population Densities. This is a cross-country estimation that is validated with the DHS data from 25 SSA countries ($R^2$: 0.91).

# High-Resolution Poverty Map
# Eritrea

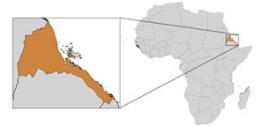

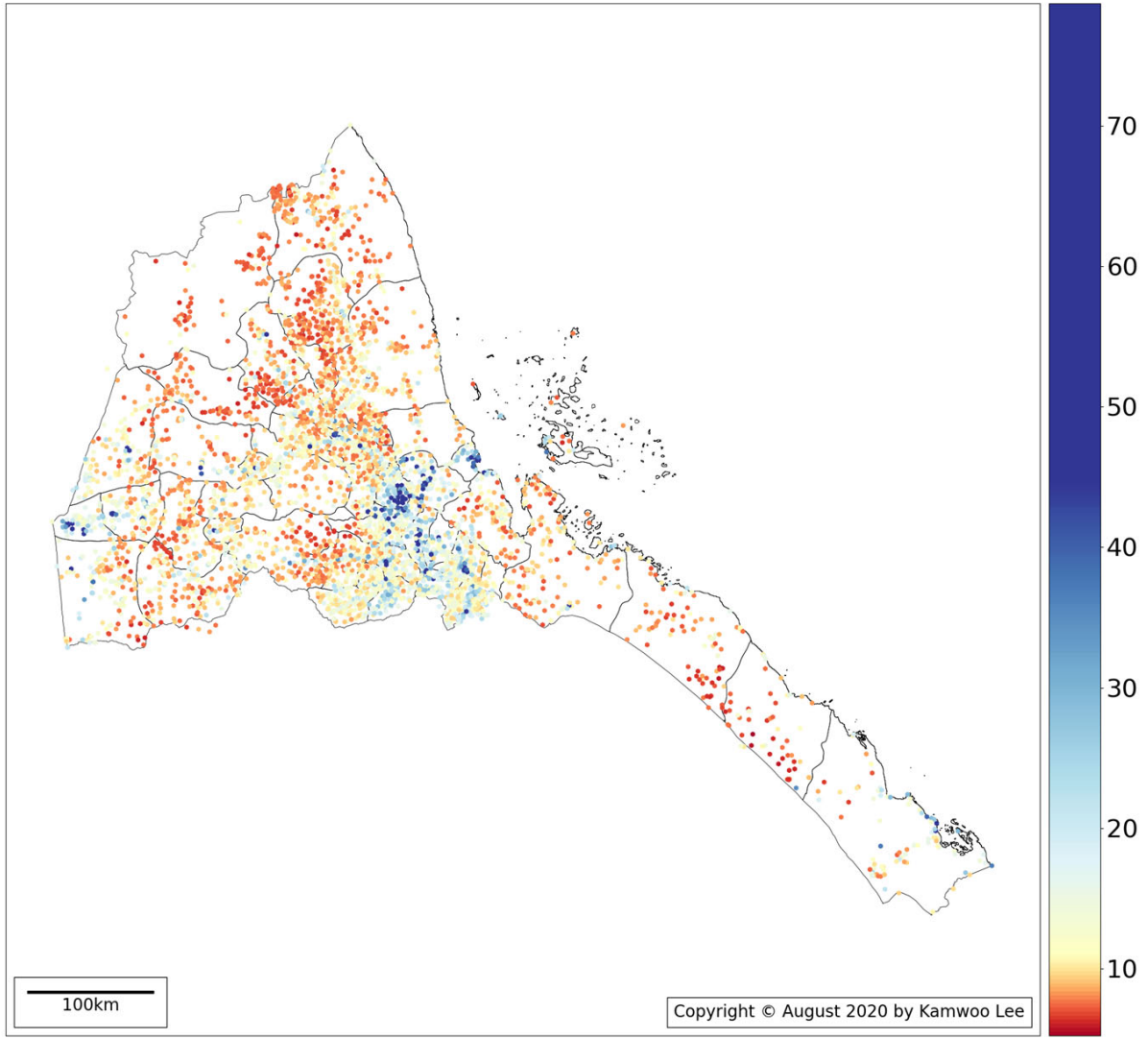

### Estimated Wealth Level (August 2020)
This map displays estimated average wealth levels of households in 1 square-mile populated areas. The wealth level was estimated on the 0-100 International Wealth Index scale (color code: red-poor, yellow-median, blue-rich) using machine learning methods with geospatial information including OpenStreetMap, daytime satellite images, nighttime luminosity, and High-Resolution Population Densities. This is a cross-country estimation that is validated with the DHS data from 25 SSA countries ($R^2$: 0.91).

# High-Resolution Poverty Map
# Eswatini

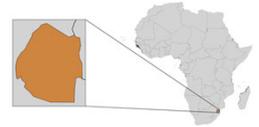

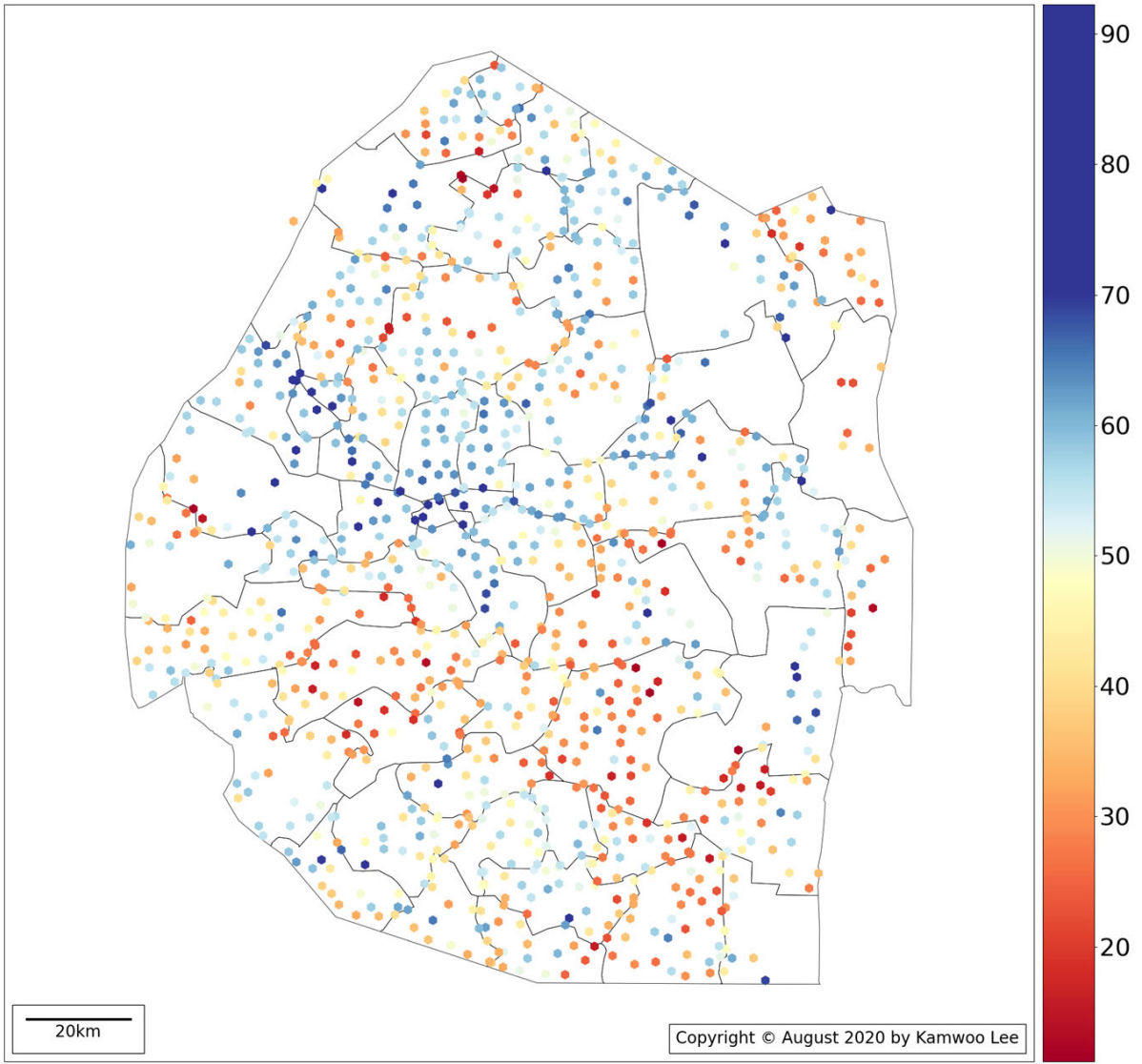

### Estimated Wealth Level (August 2020)
This map displays estimated average wealth levels of households in 1 square-mile populated areas. The wealth level was estimated on the 0-100 International Wealth Index scale (color code: red-poor, yellow-median, blue-rich) using machine learning methods with geospatial information including OpenStreetMap, daytime satellite images, nighttime luminosity, and High-Resolution Population Densities. This is a cross-country estimation that is validated with the DHS data from 25 SSA countries ($R^2$: 0.91).

# High-Resolution Poverty Map
# Ethiopia

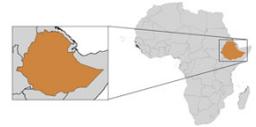

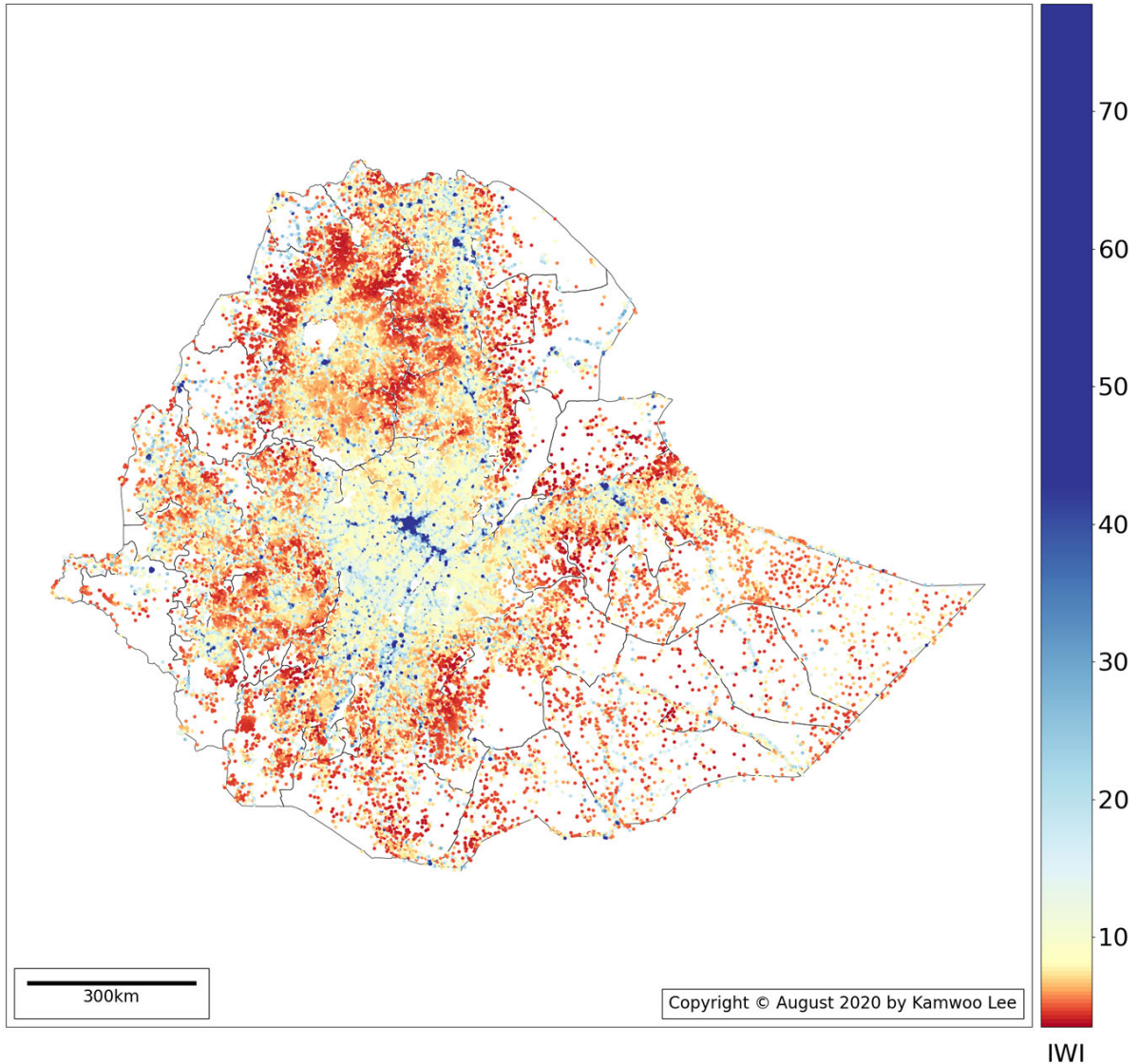

### Estimated Wealth Level (August 2020)
This map displays estimated average wealth levels of households in 1 square-mile populated areas. The wealth level was estimated on the 0-100 International Wealth Index scale (color code: red-poor, yellow-median, blue-rich) using machine learning methods with geospatial information including OpenStreetMap, daytime satellite images, nighttime luminosity, and High-Resolution Population Densities. The estimation was validated with 2016 Standard DHS.

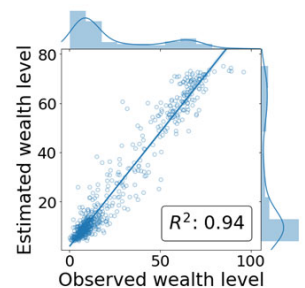

# High-Resolution Poverty Map
# Gabon

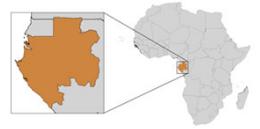

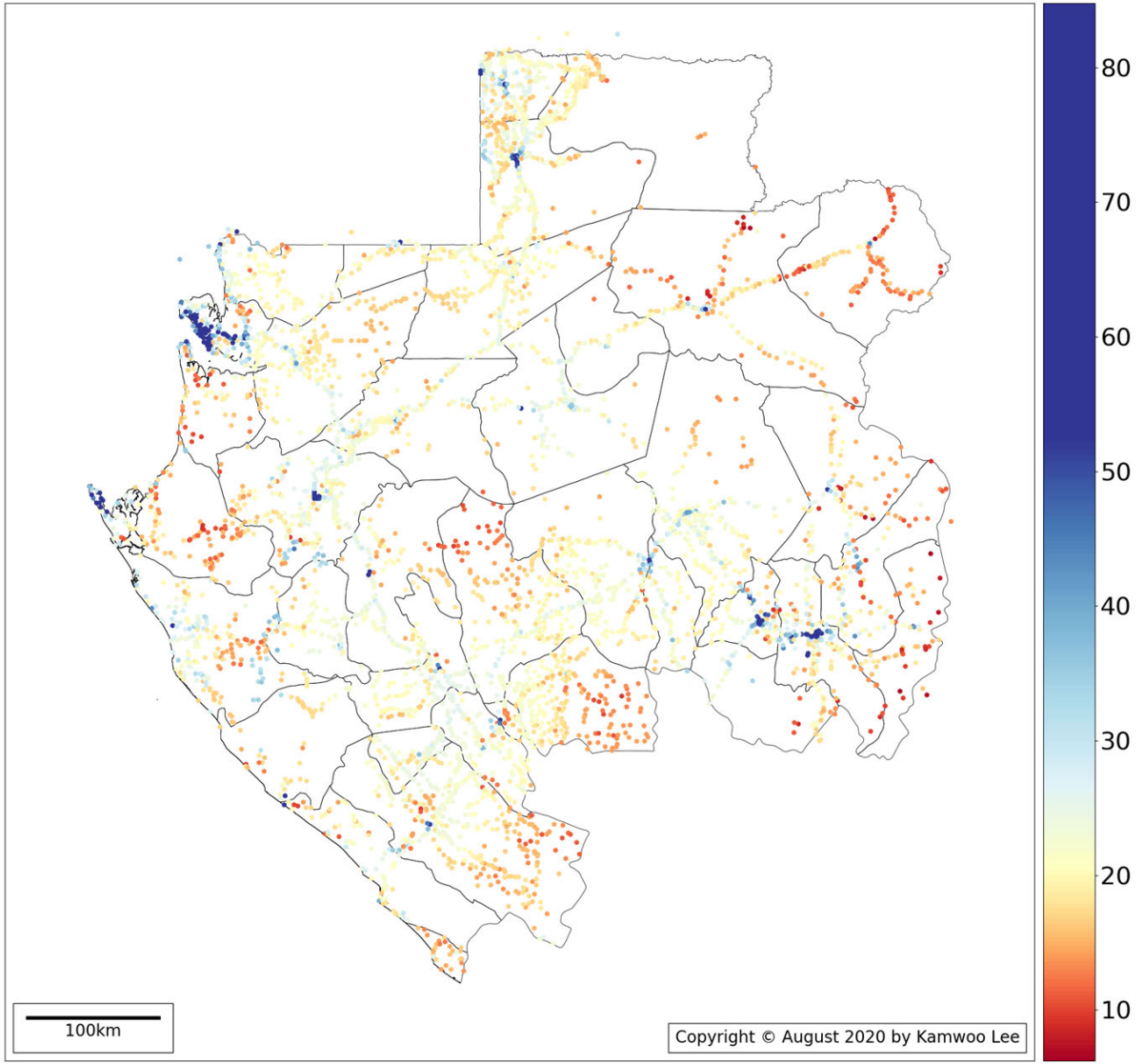

### Estimated Wealth Level (August 2020)
This map displays estimated average wealth levels of households in 1 square-mile populated areas. The wealth level was estimated on the 0-100 International Wealth Index scale (color code: red-poor, yellow-median, blue-rich) using machine learning methods with geospatial information including OpenStreetMap, daytime satellite images, nighttime luminosity, and High-Resolution Population Densities. This is a cross-country estimation that is validated with the DHS data from 25 SSA countries ($R^2$: 0.91).

# High-Resolution Poverty Map
## Gambia

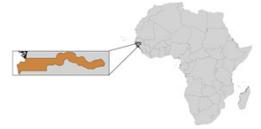

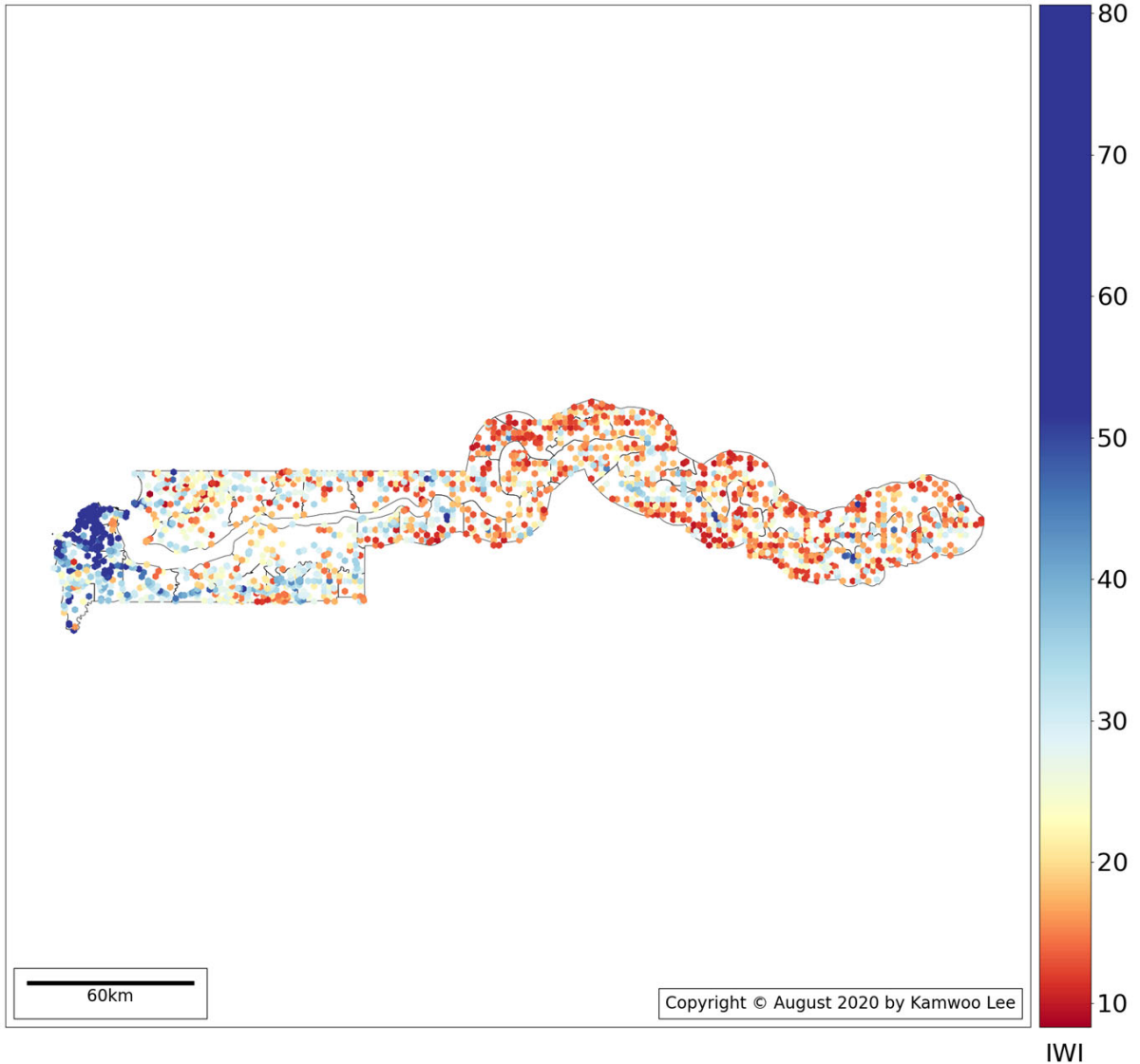

**Estimated Wealth Level (August 2020)**
This map displays estimated average wealth levels of households in 1 square-mile populated areas. The wealth level was estimated on the 0-100 International Wealth Index scale (color code: red-poor, yellow-median, blue-rich) using machine learning methods with geospatial information including OpenStreetMap, daytime satellite images, nighttime luminosity, and High-Resolution Population Densities. This is a cross-country estimation that is validated with the DHS data from 25 SSA countries ($R^2$: 0.91).

# High-Resolution Poverty Map
# Ghana

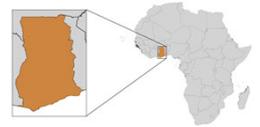

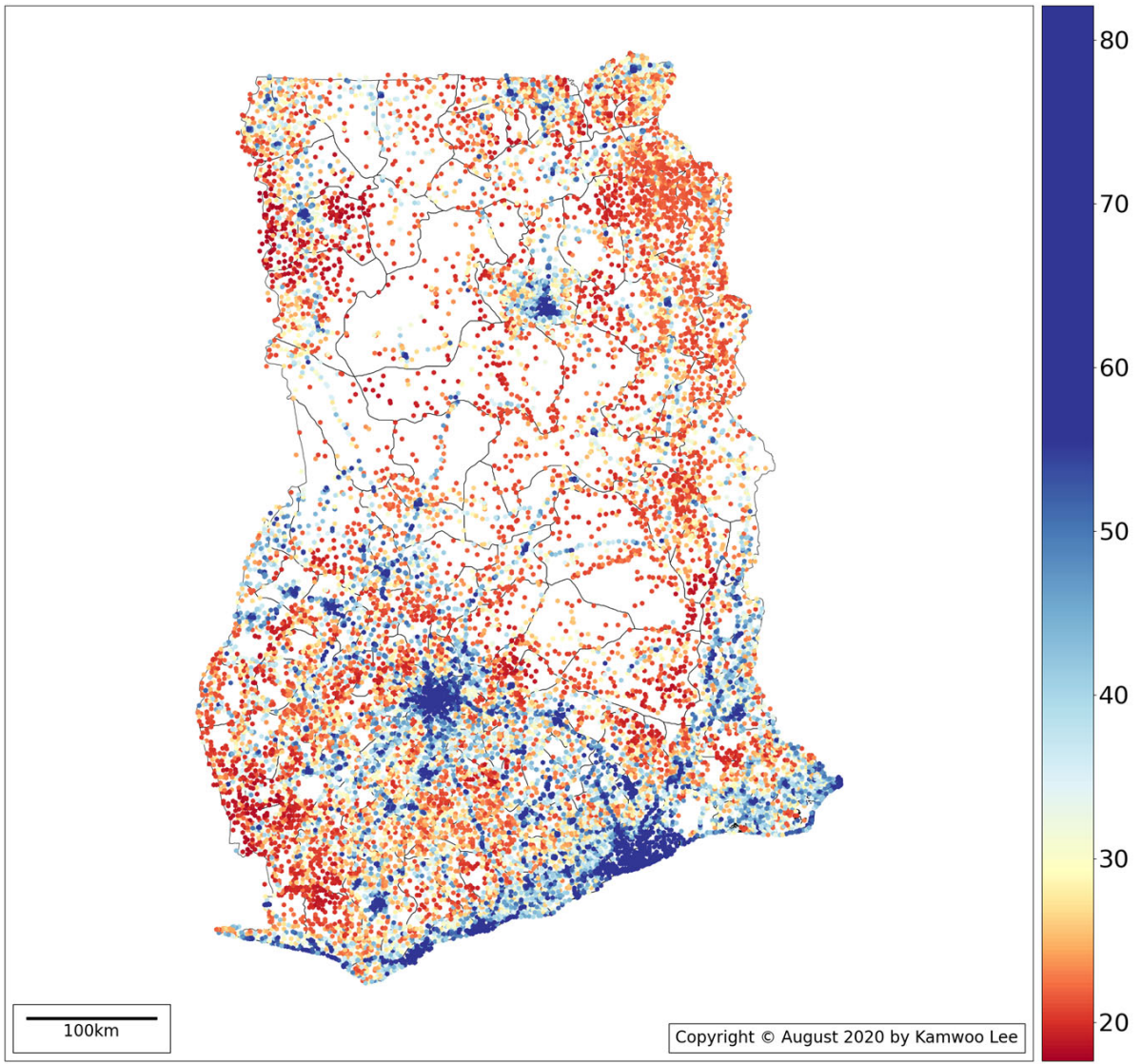

### Estimated Wealth Level (August 2020)
This map displays estimated average wealth levels of households in 1 square-mile populated areas. The wealth level was estimated on the 0-100 International Wealth Index scale (color code: red-poor, yellow-median, blue-rich) using machine learning methods with geospatial information including OpenStreetMap, daytime satellite images, nighttime luminosity, and High-Resolution Population Densities. The estimation was validated with 2016 and 2019 MIS.

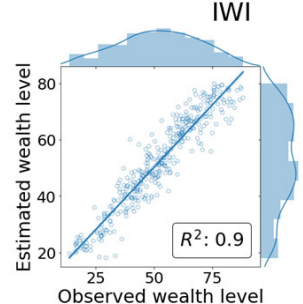

# High-Resolution Poverty Map
## Guinea

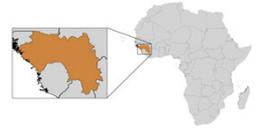

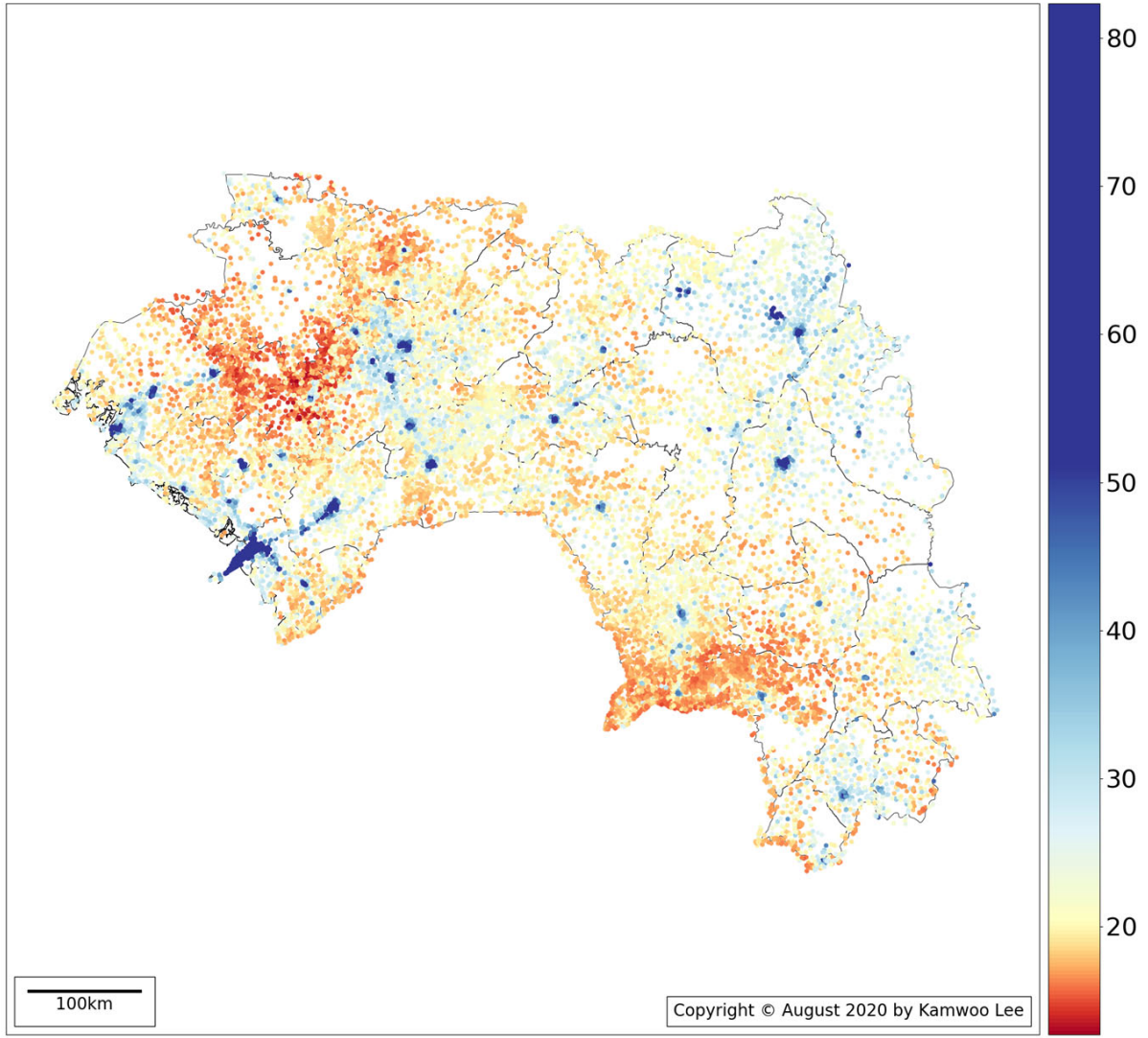

### Estimated Wealth Level (August 2020)

This map displays estimated average wealth levels of households in 1 square-mile populated areas. The wealth level was estimated on the 0-100 International Wealth Index scale (color code: red-poor, yellow-median, blue-rich) using machine learning methods with geospatial information including OpenStreetMap, daytime satellite images, nighttime luminosity, and High-Resolution Population Densities. The estimation was validated with 2018 Standard DHS.

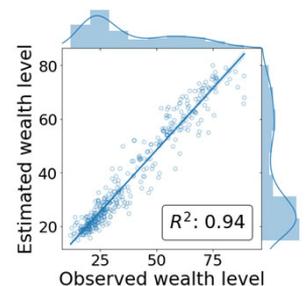

# High-Resolution Poverty Map
# Guinea-Bissau

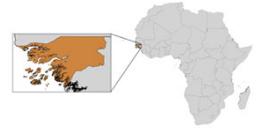

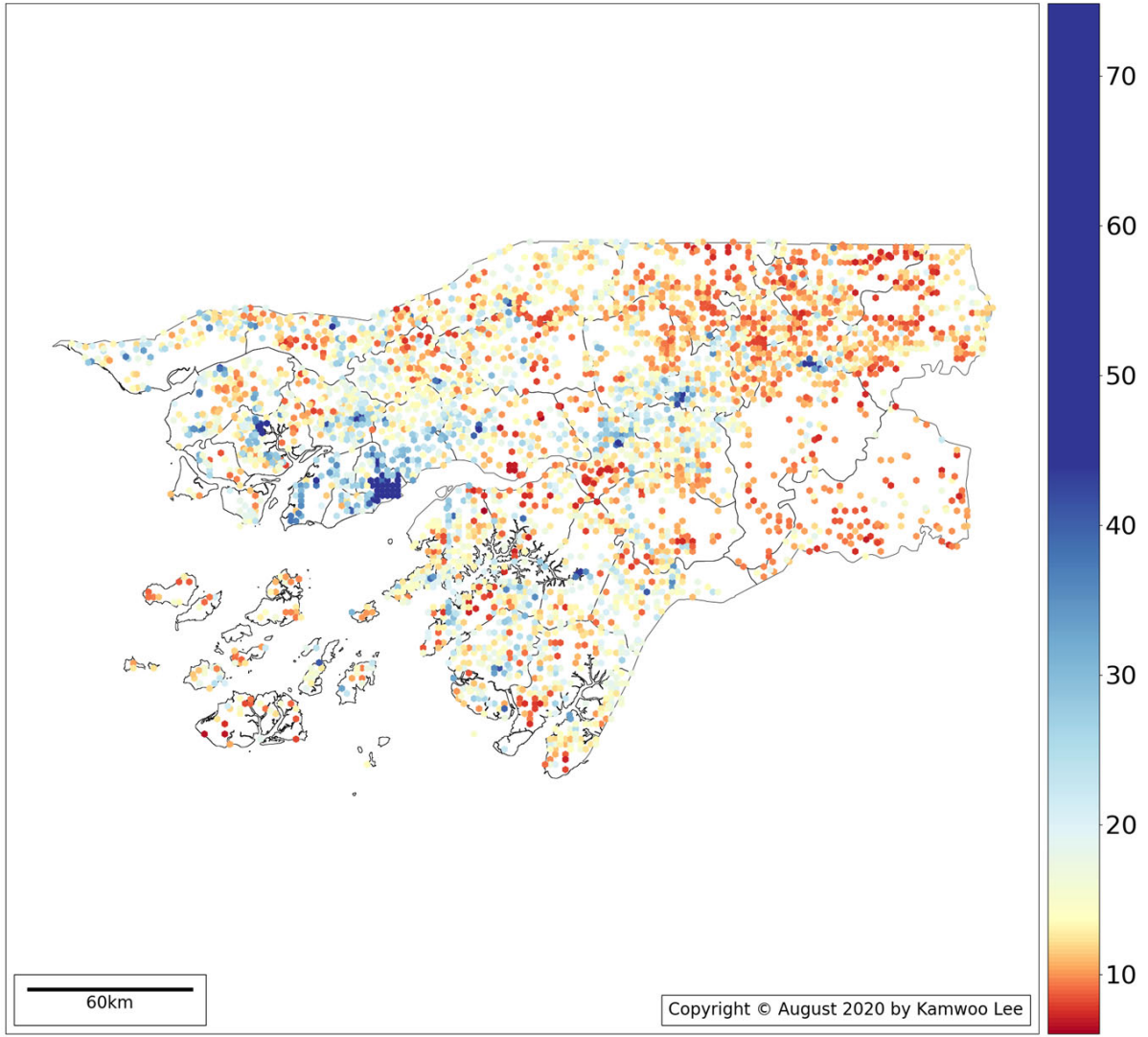

**Estimated Wealth Level (August 2020)**
This map displays estimated average wealth levels of households in 1 square-mile populated areas. The wealth level was estimated on the 0-100 International Wealth Index scale (color code: red-poor, yellow-median, blue-rich) using machine learning methods with geospatial information including OpenStreetMap, daytime satellite images, nighttime luminosity, and High-Resolution Population Densities. This is a cross-country estimation that is validated with the DHS data from 25 SSA countries ($R^2$: 0.91).

# High-Resolution Poverty Map
# Kenya

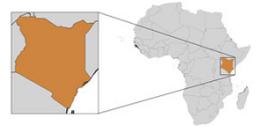

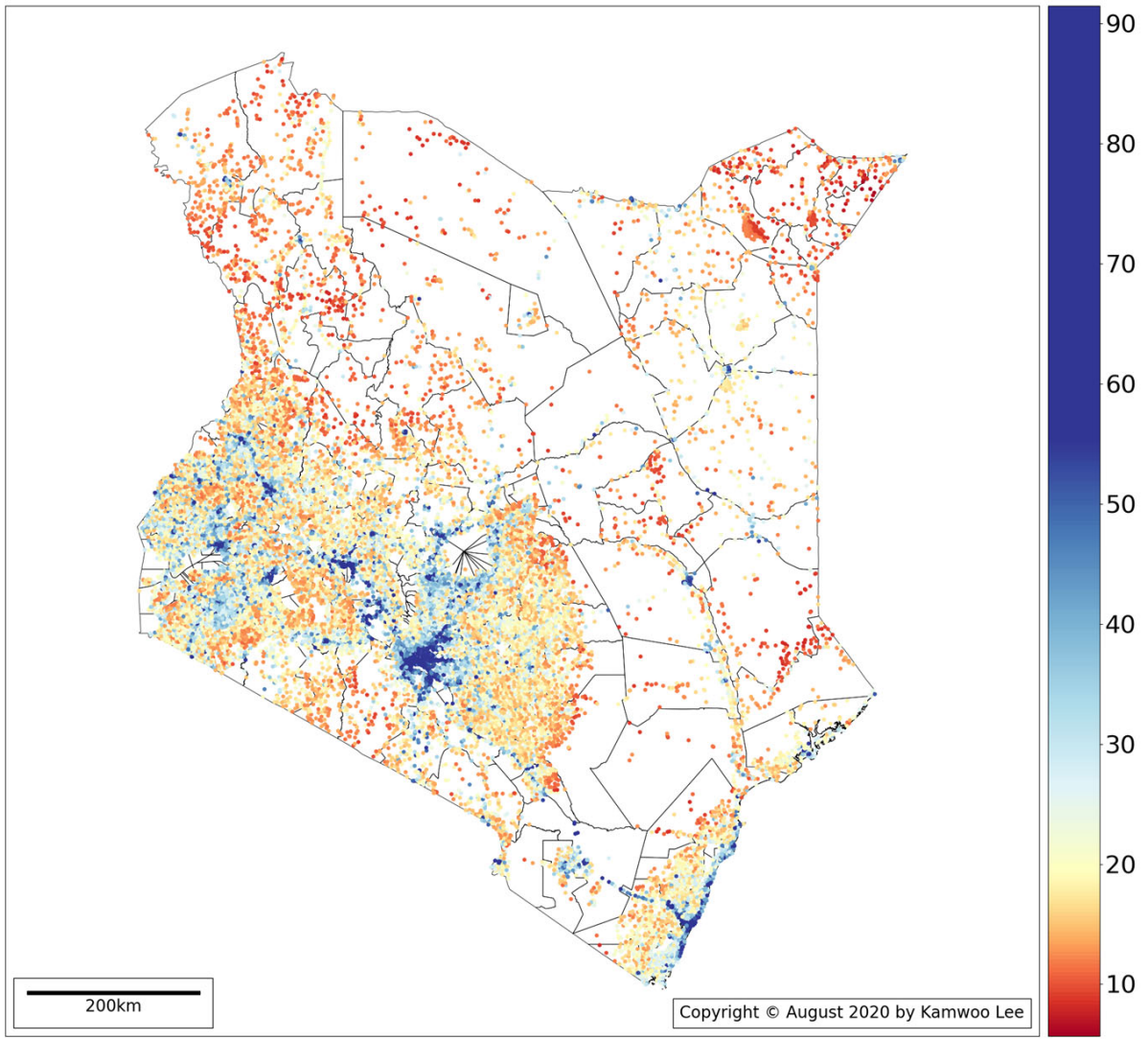

### Estimated Wealth Level (August 2020)

This map displays estimated average wealth levels of households in 1 square-mile populated areas. The wealth level was estimated on the 0-100 International Wealth Index scale (color code: red-poor, yellow-median, blue-rich) using machine learning methods with geospatial information including OpenStreetMap, daytime satellite images, nighttime luminosity, and High-Resolution Population Densities. The estimation was validated with 2015 MIS.

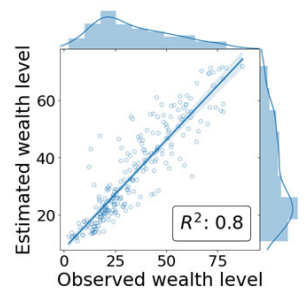

# High-Resolution Poverty Map
# Lesotho

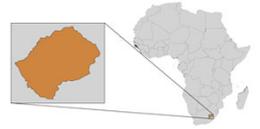

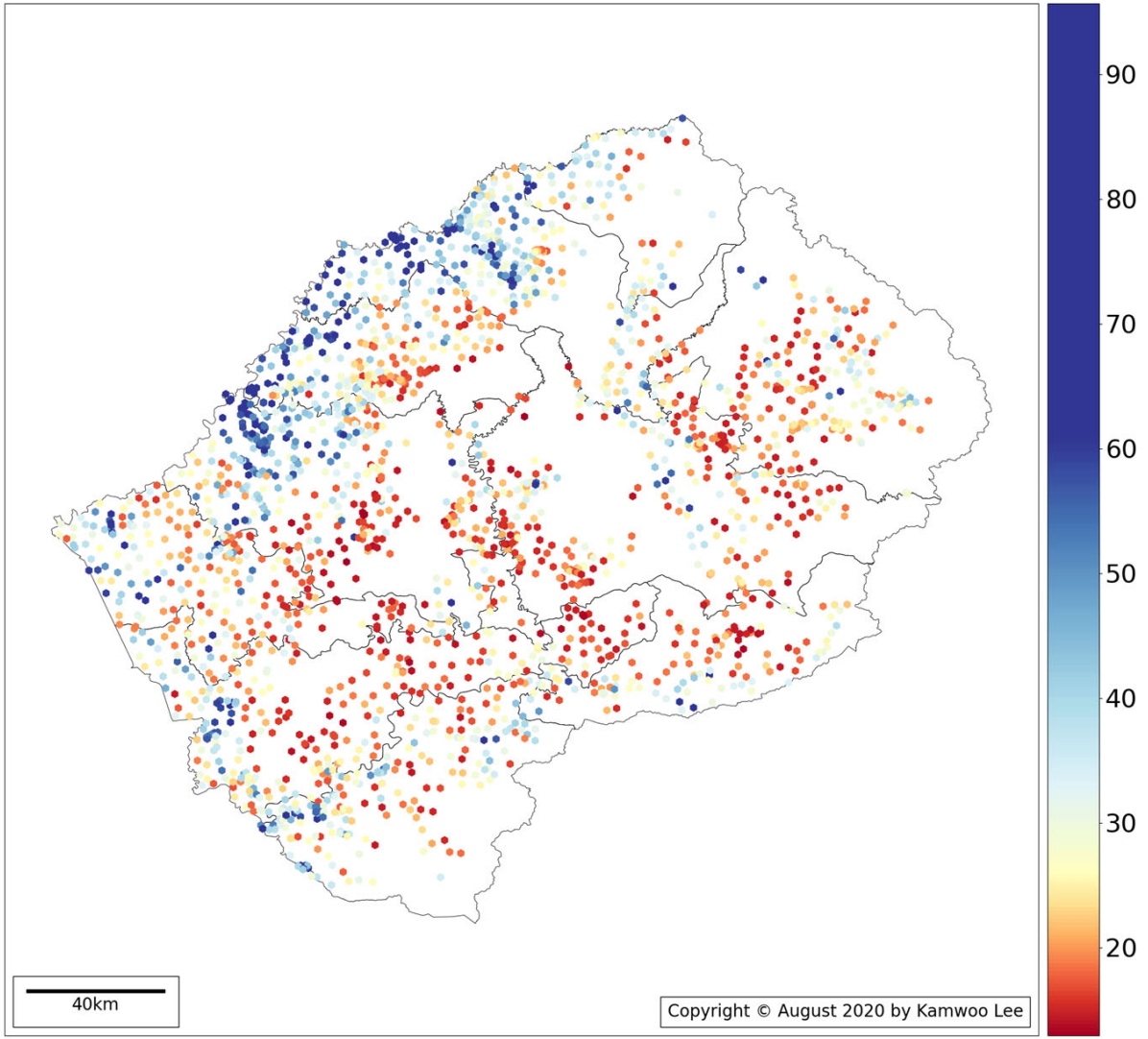

**Estimated Wealth Level (August 2020)**
This map displays estimated average wealth levels of households in 1 square-mile populated areas. The wealth level was estimated on the 0-100 International Wealth Index scale (color code: red-poor, yellow-median, blue-rich) using machine learning methods with geospatial information including OpenStreetMap, daytime satellite images, nighttime luminosity, and High-Resolution Population Densities. This is a cross-country estimation that is validated with the DHS data from 25 SSA countries ($R^2$: 0.91).

# High-Resolution Poverty Map
# Liberia

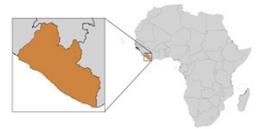

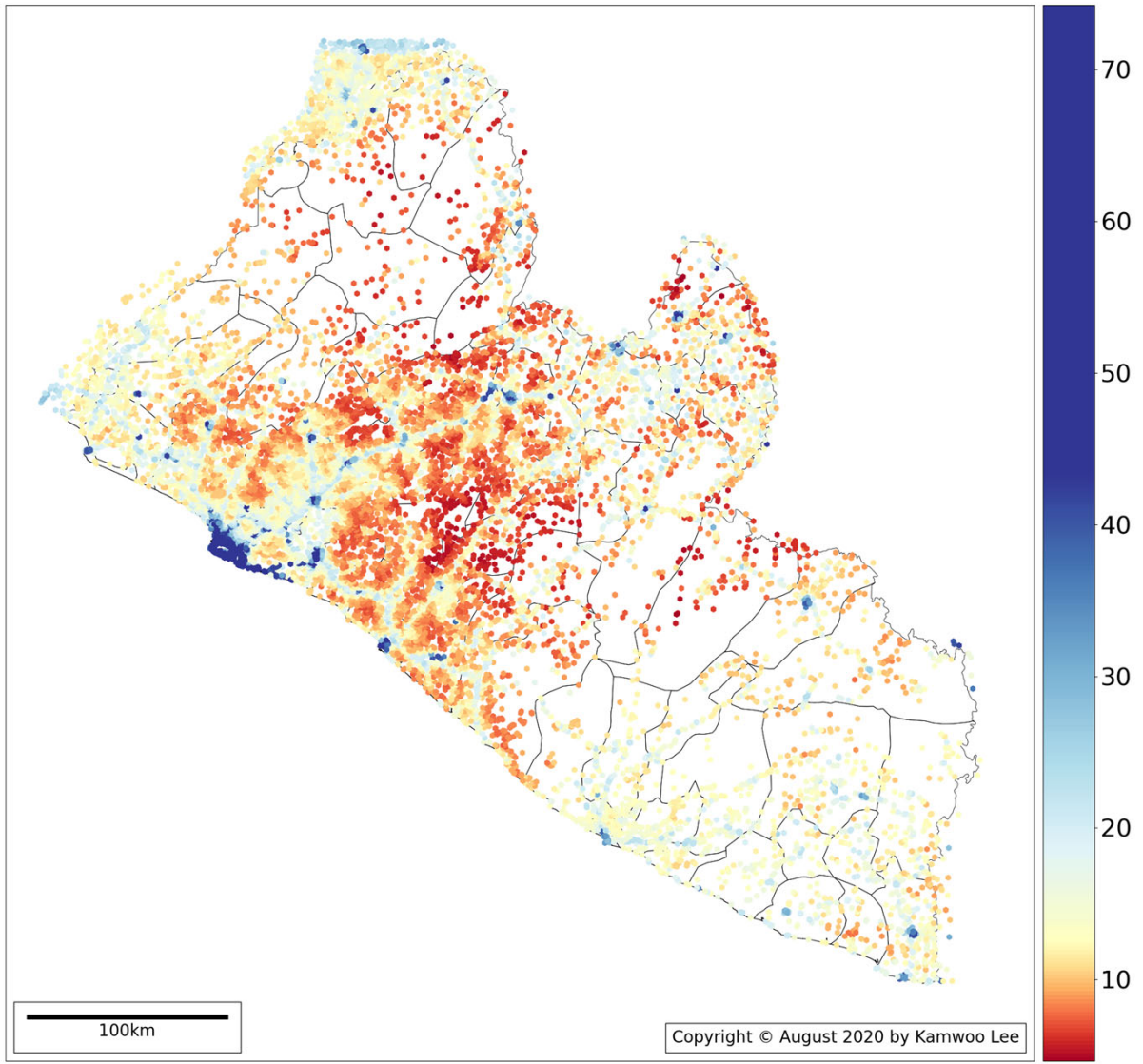

### Estimated Wealth Level (August 2020)

This map displays estimated average wealth levels of households in 1 square-mile populated areas. The wealth level was estimated on the 0-100 International Wealth Index scale (color code: red-poor, yellow-median, blue-rich) using machine learning methods with geospatial information including OpenStreetMap, daytime satellite images, nighttime luminosity, and High-Resolution Population Densities. The estimation was validated with 2016 MIS.

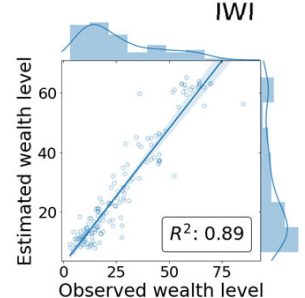

# High-Resolution Poverty Map
# Madagascar

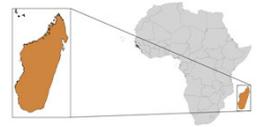

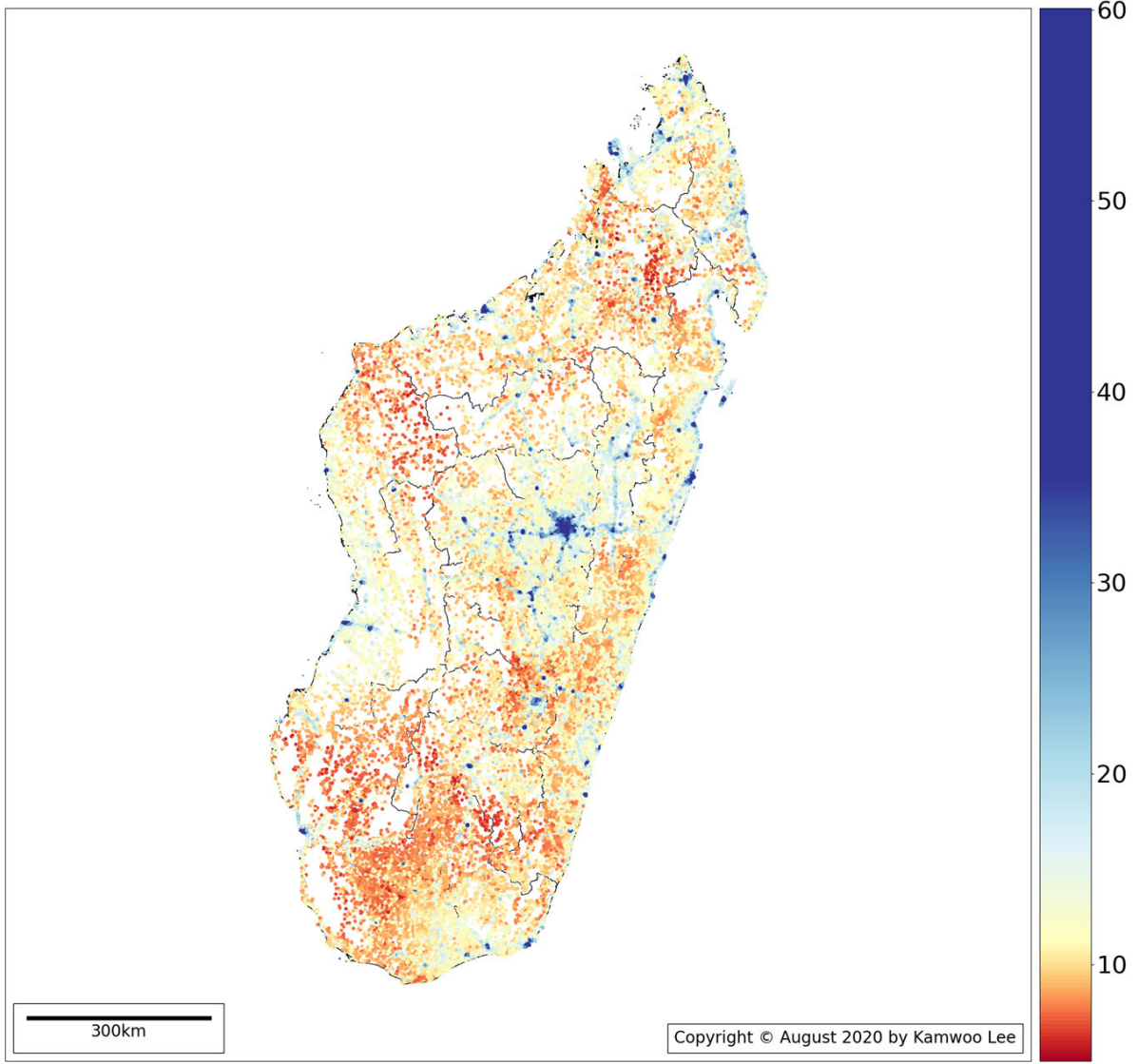

### Estimated Wealth Level (August 2020)
This map displays estimated average wealth levels of households in 1 square-mile populated areas. The wealth level was estimated on the 0-100 International Wealth Index scale (color code: red-poor, yellow-median, blue-rich) using machine learning methods with geospatial information including OpenStreetMap, daytime satellite images, nighttime luminosity, and High-Resolution Population Densities. The estimation was validated with 2016 MIS.

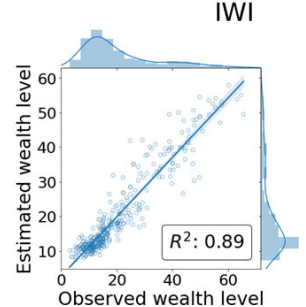

# High-Resolution Poverty Map
# Malawi

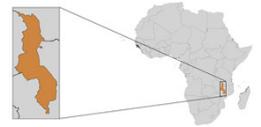

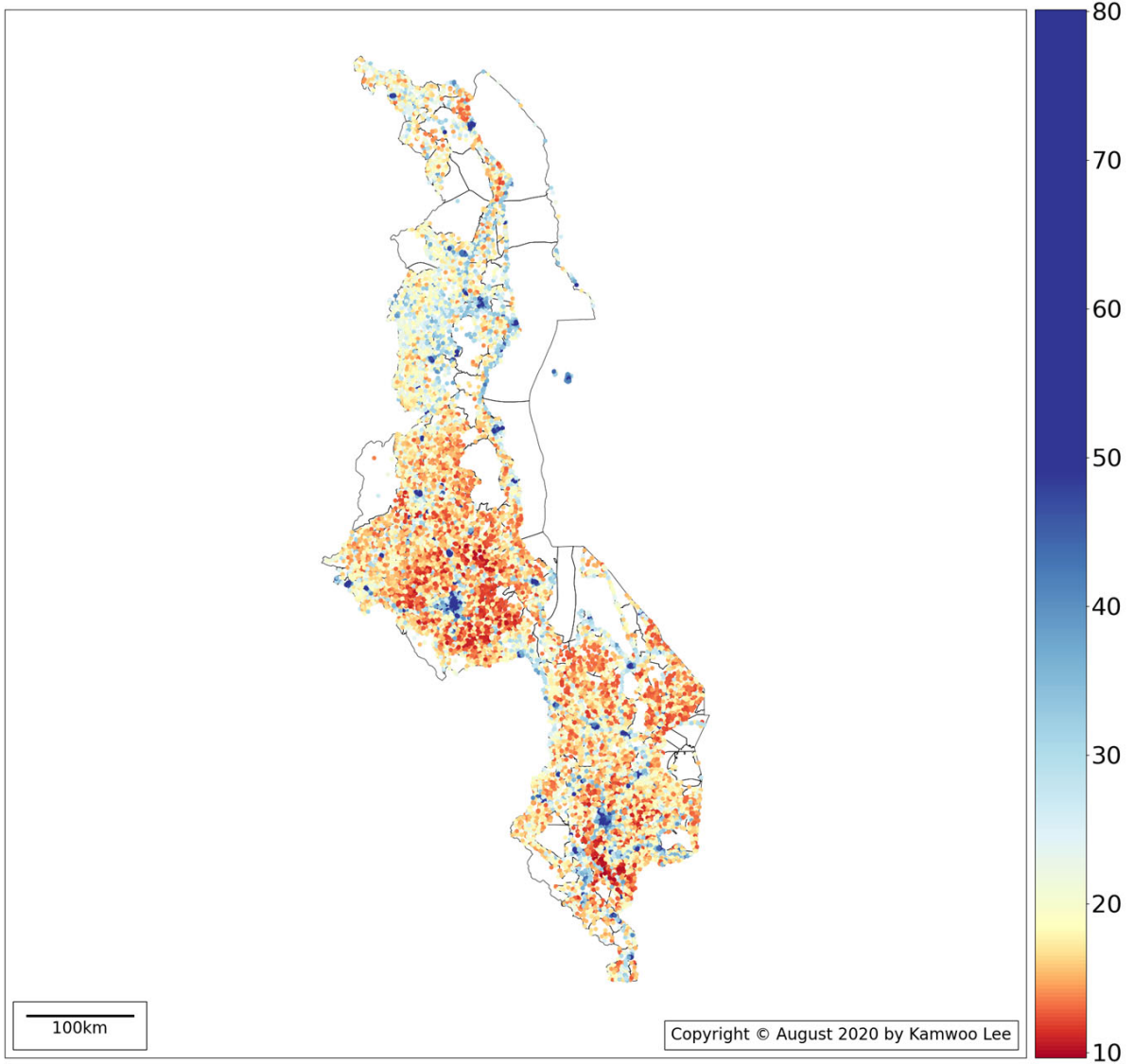

### Estimated Wealth Level (August 2020)
This map displays estimated average wealth levels of households in 1 square-mile populated areas. The wealth level was estimated on the 0-100 International Wealth Index scale (color code: red-poor, yellow-median, blue-rich) using machine learning methods with geospatial information including OpenStreetMap, daytime satellite images, nighttime luminosity, and High-Resolution Population Densities. The estimation was validated with 2015-2016 Standard DHS and 2017 MIS.

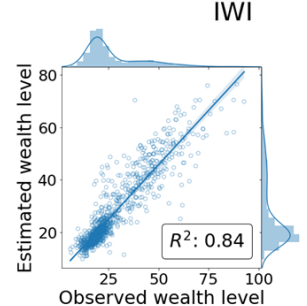

# High-Resolution Poverty Map
## Mali

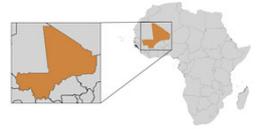

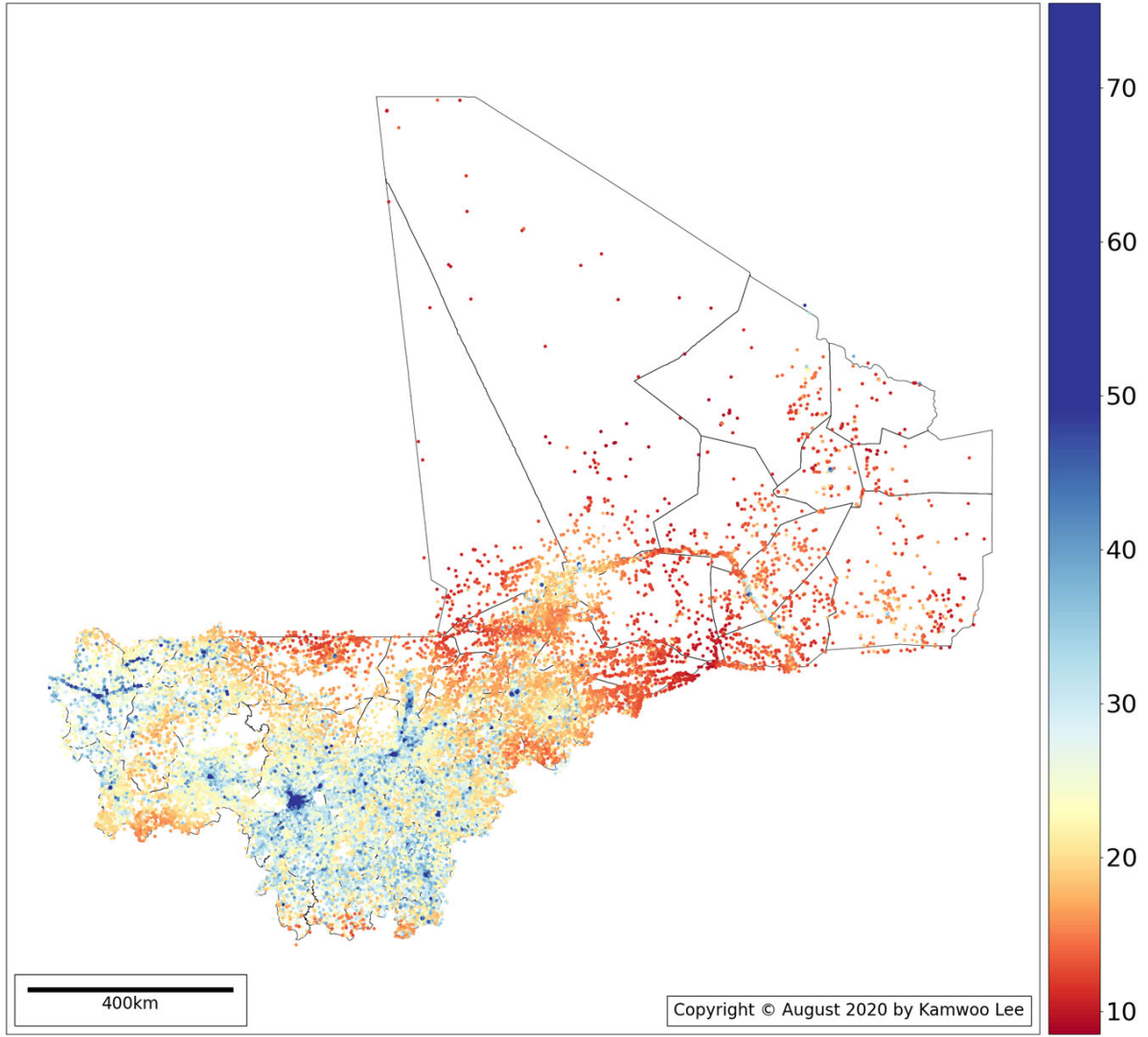

### Estimated Wealth Level (August 2020)

This map displays estimated average wealth levels of households in 1 square-mile populated areas. The wealth level was estimated on the 0-100 International Wealth Index scale (color code: red-poor, yellow-median, blue-rich) using machine learning methods with geospatial information including OpenStreetMap, daytime satellite images, nighttime luminosity, and High-Resolution Population Densities. The estimation was validated with 2015 MIS and 2018 Standard DHS.

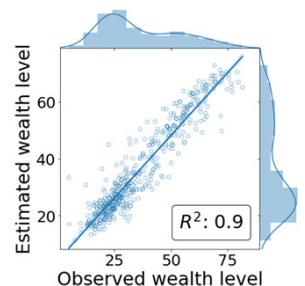

# High-Resolution Poverty Map

## Mauritania

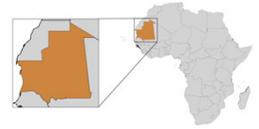

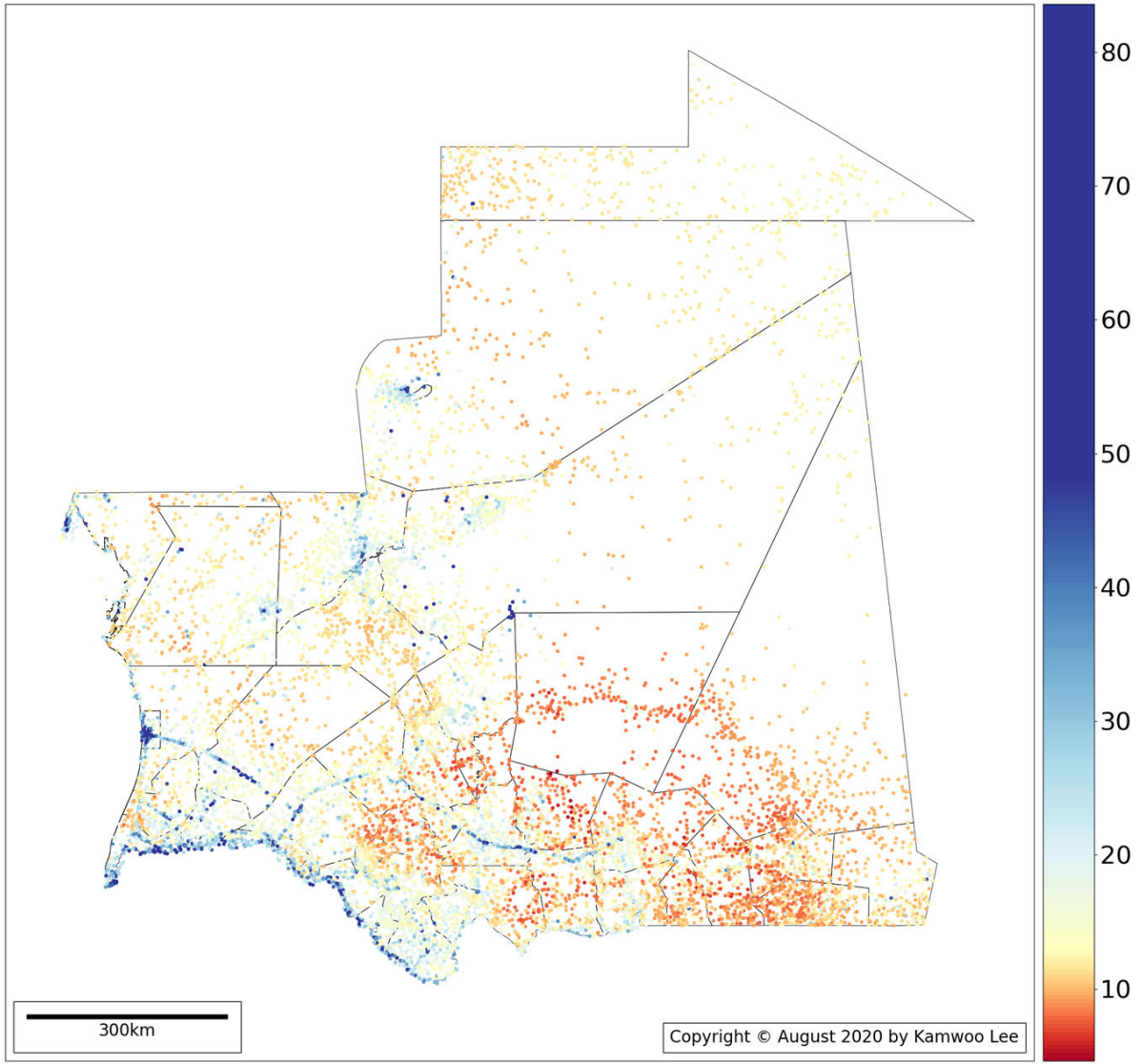

**Estimated Wealth Level (August 2020)**

This map displays estimated average wealth levels of households in 1 square-mile populated areas. The wealth level was estimated on the 0-100 International Wealth Index scale (color code: red-poor, yellow-median, blue-rich) using machine learning methods with geospatial information including OpenStreetMap, daytime satellite images, nighttime luminosity, and High-Resolution Population Densities. This is a cross-country estimation that is validated with the DHS data from 25 SSA countries ($R^2$: 0.91).

# High-Resolution Poverty Map
## Mozambique

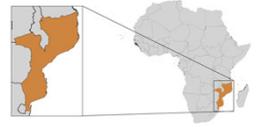

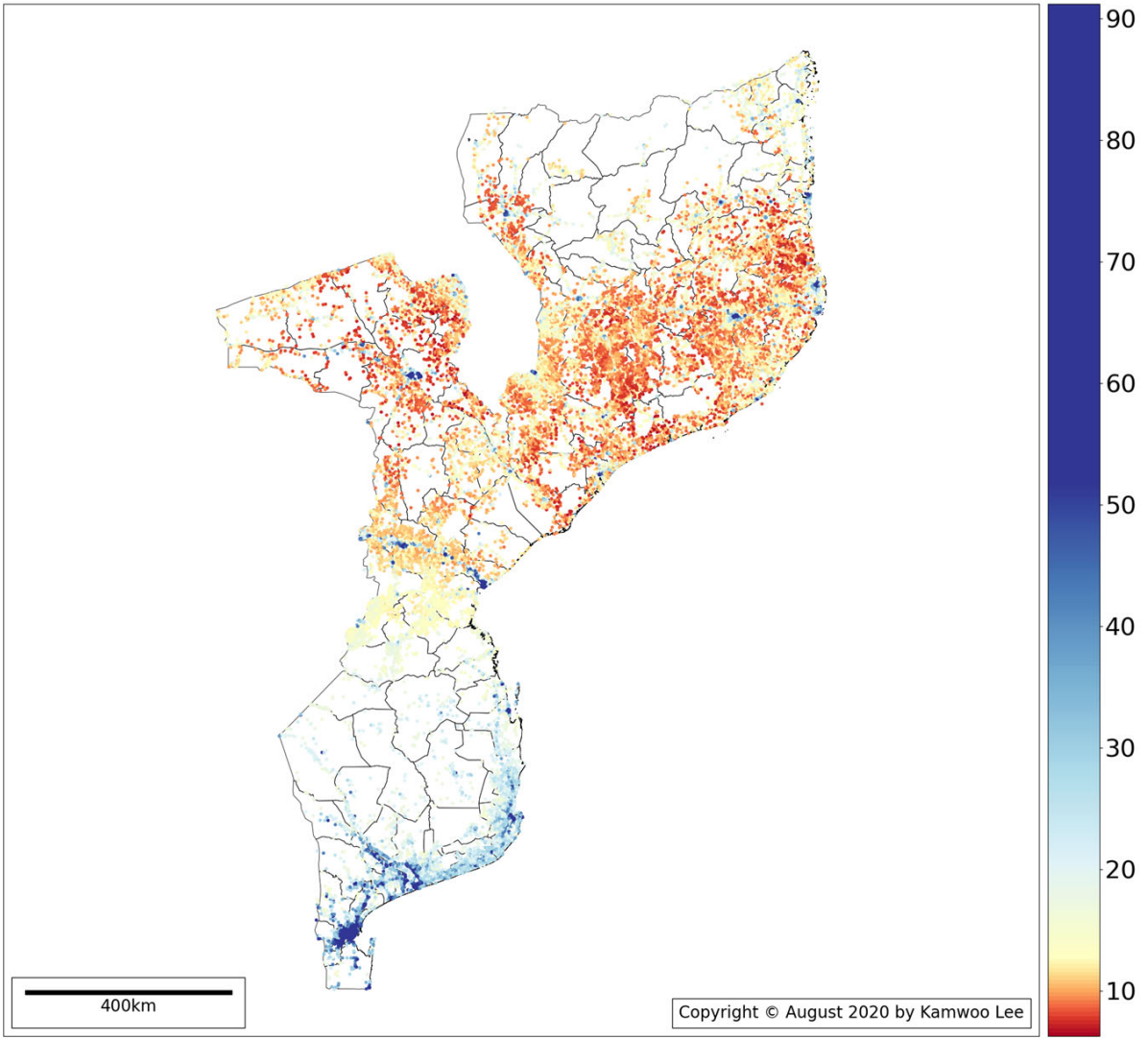

### Estimated Wealth Level (August 2020)

This map displays estimated average wealth levels of households in 1 square-mile populated areas. The wealth level was estimated on the 0-100 International Wealth Index scale (color code: red-poor, yellow-median, blue-rich) using machine learning methods with geospatial information including OpenStreetMap, daytime satellite images, nighttime luminosity, and High-Resolution Population Densities. The estimation was validated with 2015 and 2018 Standard DHS.

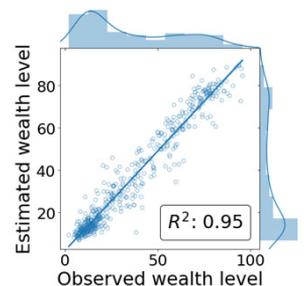

# High-Resolution Poverty Map
# Namibia

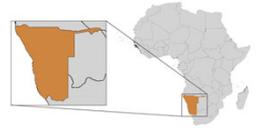

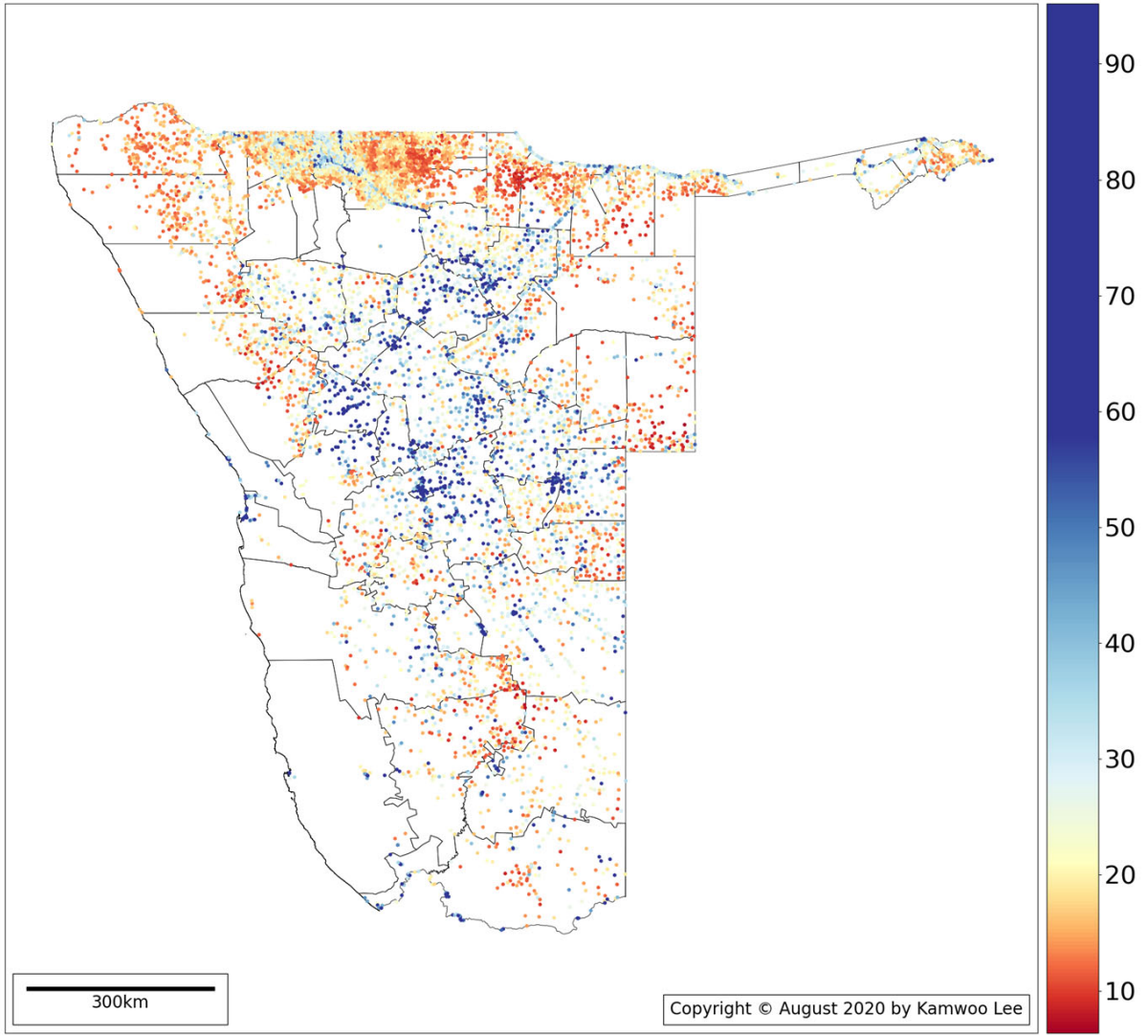

### Estimated Wealth Level (August 2020)
This map displays estimated average wealth levels of households in 1 square-mile populated areas. The wealth level was estimated on the 0-100 International Wealth Index scale (color code: red-poor, yellow-median, blue-rich) using machine learning methods with geospatial information including OpenStreetMap, daytime satellite images, nighttime luminosity, and High-Resolution Population Densities. This is a cross-country estimation that is validated with the DHS data from 25 SSA countries. ($R^2$: 0.91)

# High-Resolution Poverty Map
# Niger

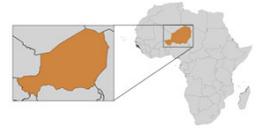

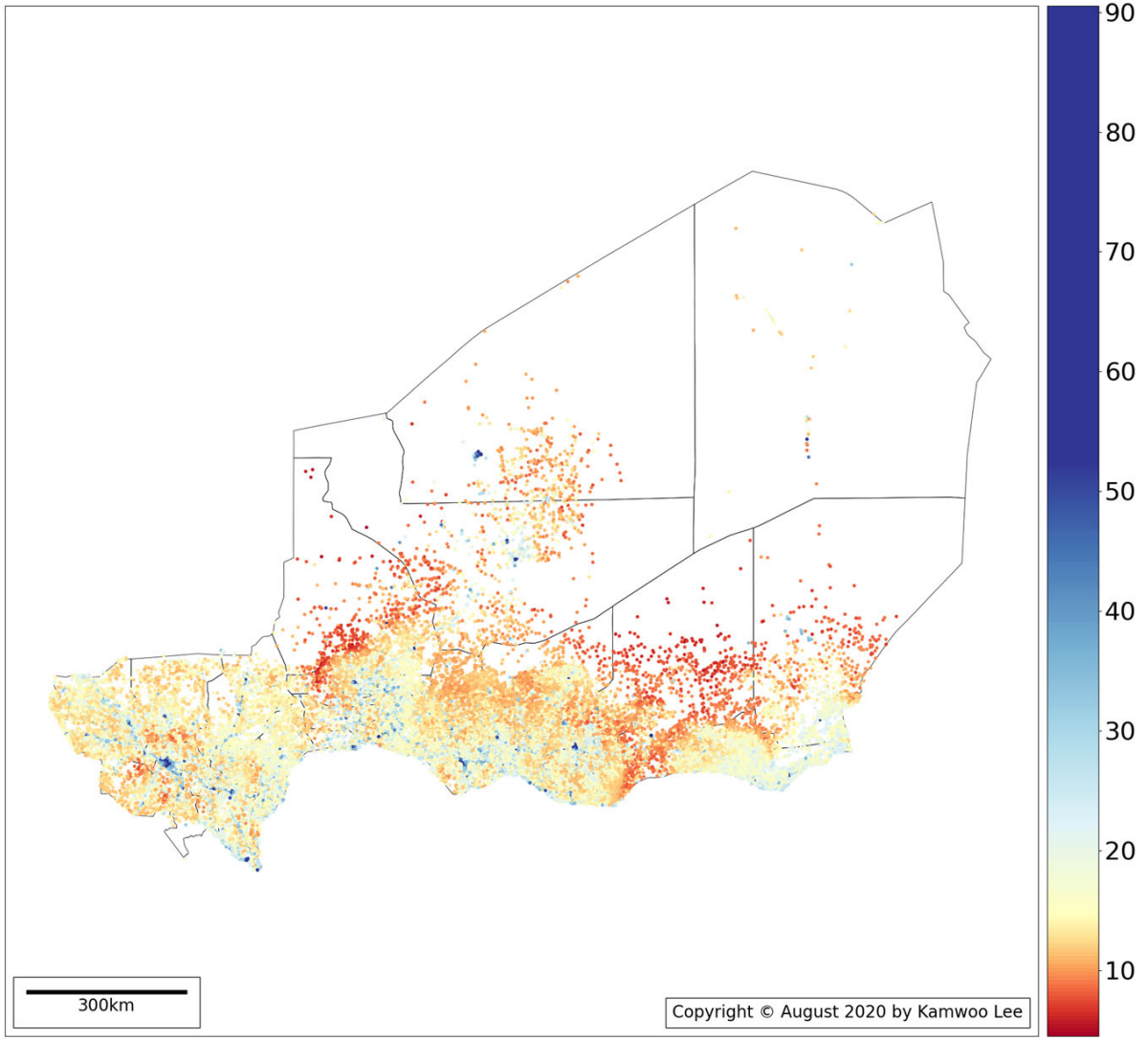

### Estimated Wealth Level (August 2020)
This map displays estimated average wealth levels of households in 1 square-mile populated areas. The wealth level was estimated on the 0-100 International Wealth Index scale (color code: red-poor, yellow-median, blue-rich) using machine learning methods with geospatial information including OpenStreetMap, daytime satellite images, nighttime luminosity, and High-Resolution Population Densities. This is a cross-country estimation that is validated with the DHS data from 25 SSA countries. ($R^2$: 0.91)

# High-Resolution Poverty Map
# Nigeria

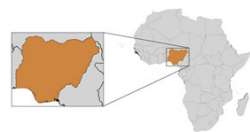

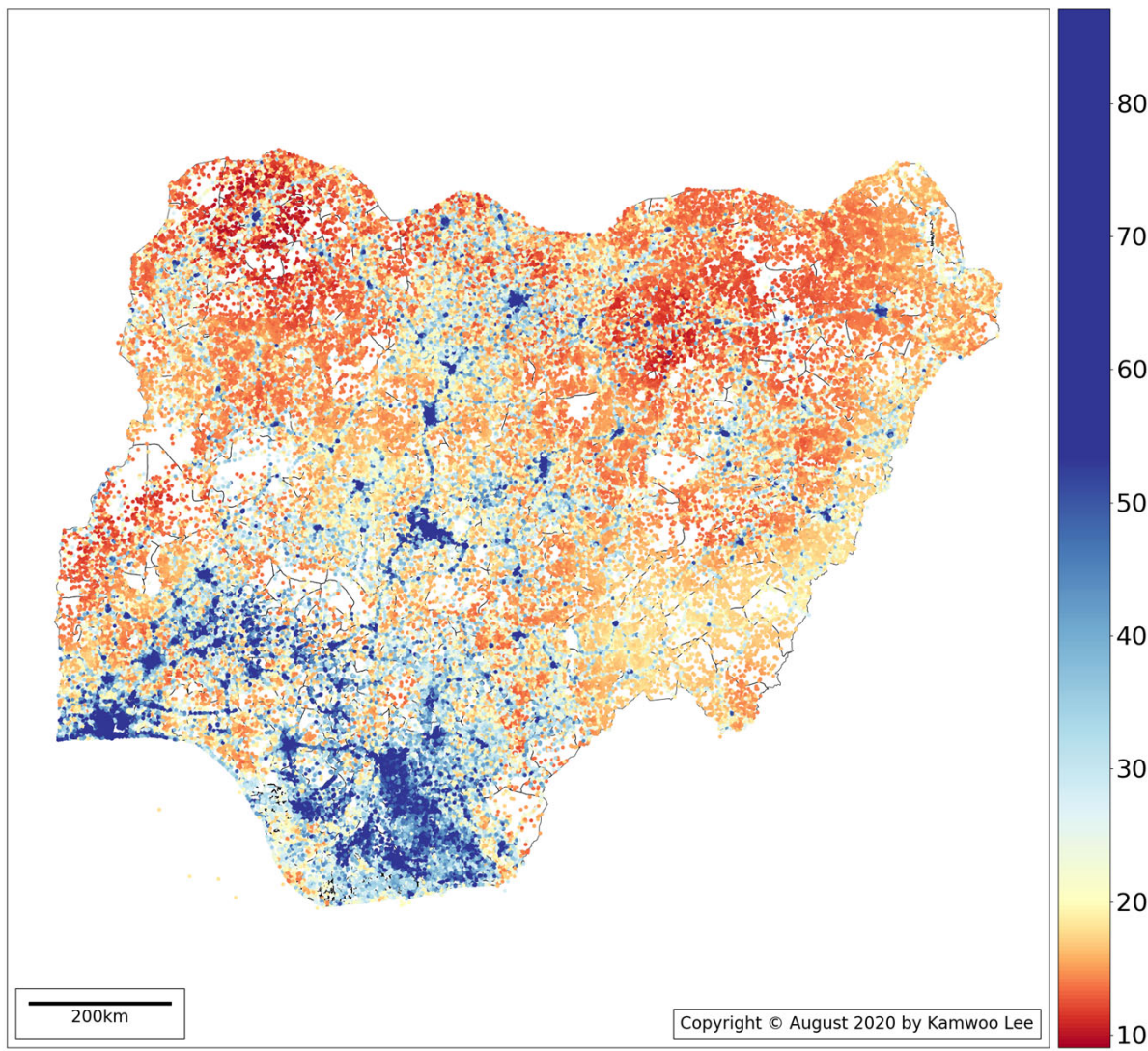

### Estimated Wealth Level (August 2020)

This map displays estimated average wealth levels of households in 1 square-mile populated areas. The wealth level was estimated on the 0-100 International Wealth Index scale (color code: red-poor, yellow-median, blue-rich) using machine learning methods with geospatial information including OpenStreetMap, daytime satellite images, nighttime luminosity, and High-Resolution Population Densities. The estimation was validated with 2015 MIS and 2018 Standard DHS.

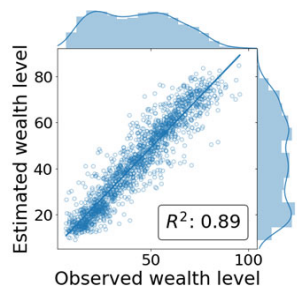

# High-Resolution Poverty Map
# Rwanda

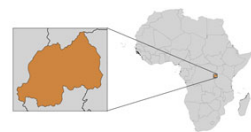

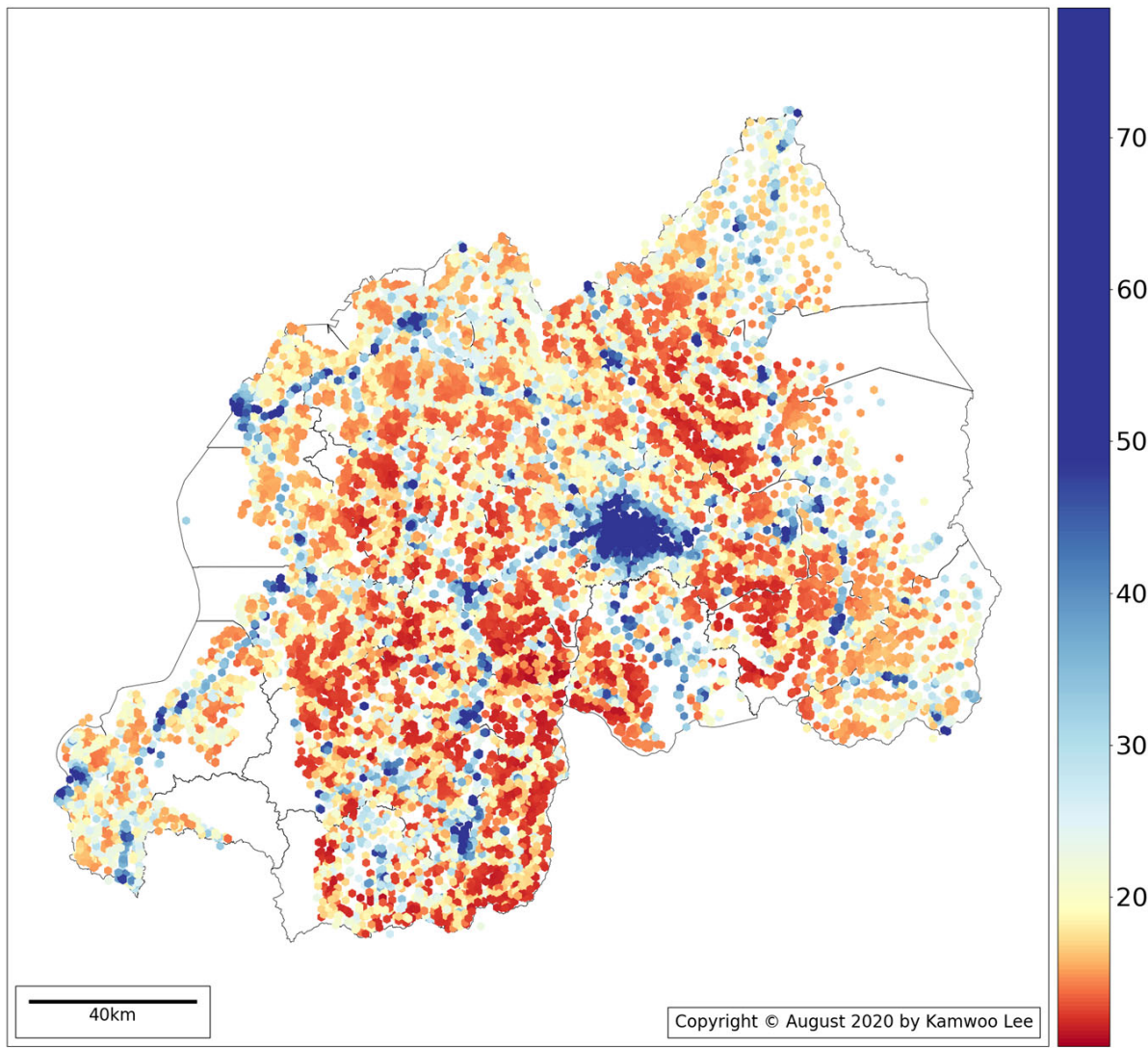

### Estimated Wealth Level (August 2020)

This map displays estimated average wealth levels of households in 1 square-mile populated areas. The wealth level was estimated on the 0-100 International Wealth Index scale (color code: red-poor, yellow-median, blue-rich) using machine learning methods with geospatial information including OpenStreetMap, daytime satellite images, nighttime luminosity, and High-Resolution Population Densities. The estimation was validated with 2014-2015 Standard DHS.

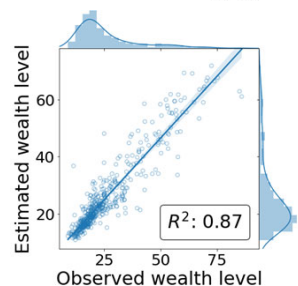

# High-Resolution Poverty Map
# Senegal

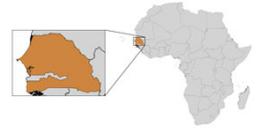

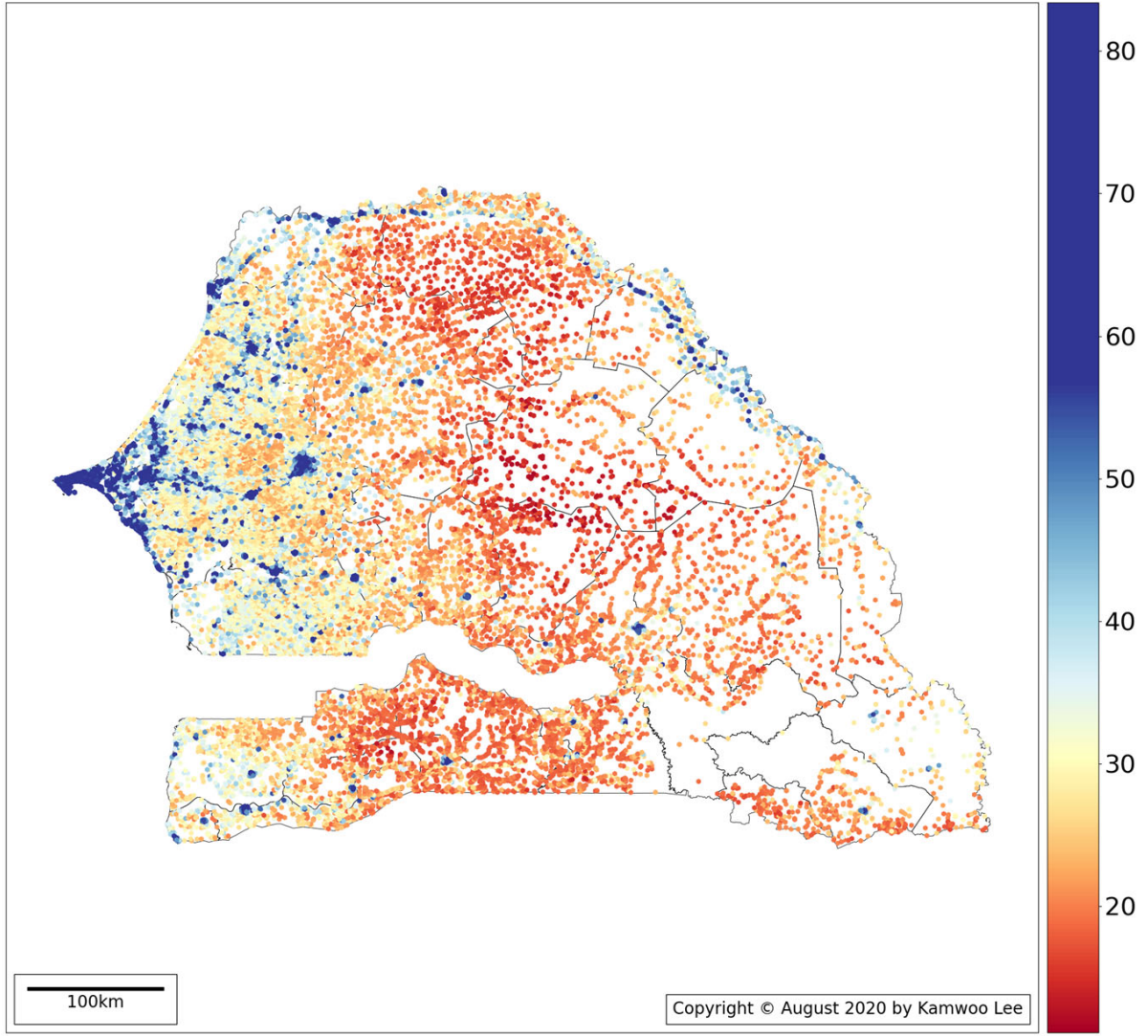

### Estimated Wealth Level (August 2020)

This map displays estimated average wealth levels of households in 1 square-mile populated areas. The wealth level was estimated on the 0-100 International Wealth Index scale (color code: red-poor, yellow-median, blue-rich) using machine learning methods with geospatial information including OpenStreetMap, daytime satellite images, nighttime luminosity, and High-Resolution Population Densities. The estimation was validated with 2015-2016 Continuous DHS.

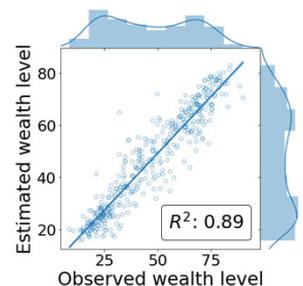

# High-Resolution Poverty Map
# Sierra Leone

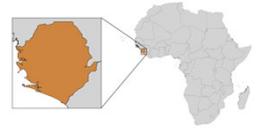

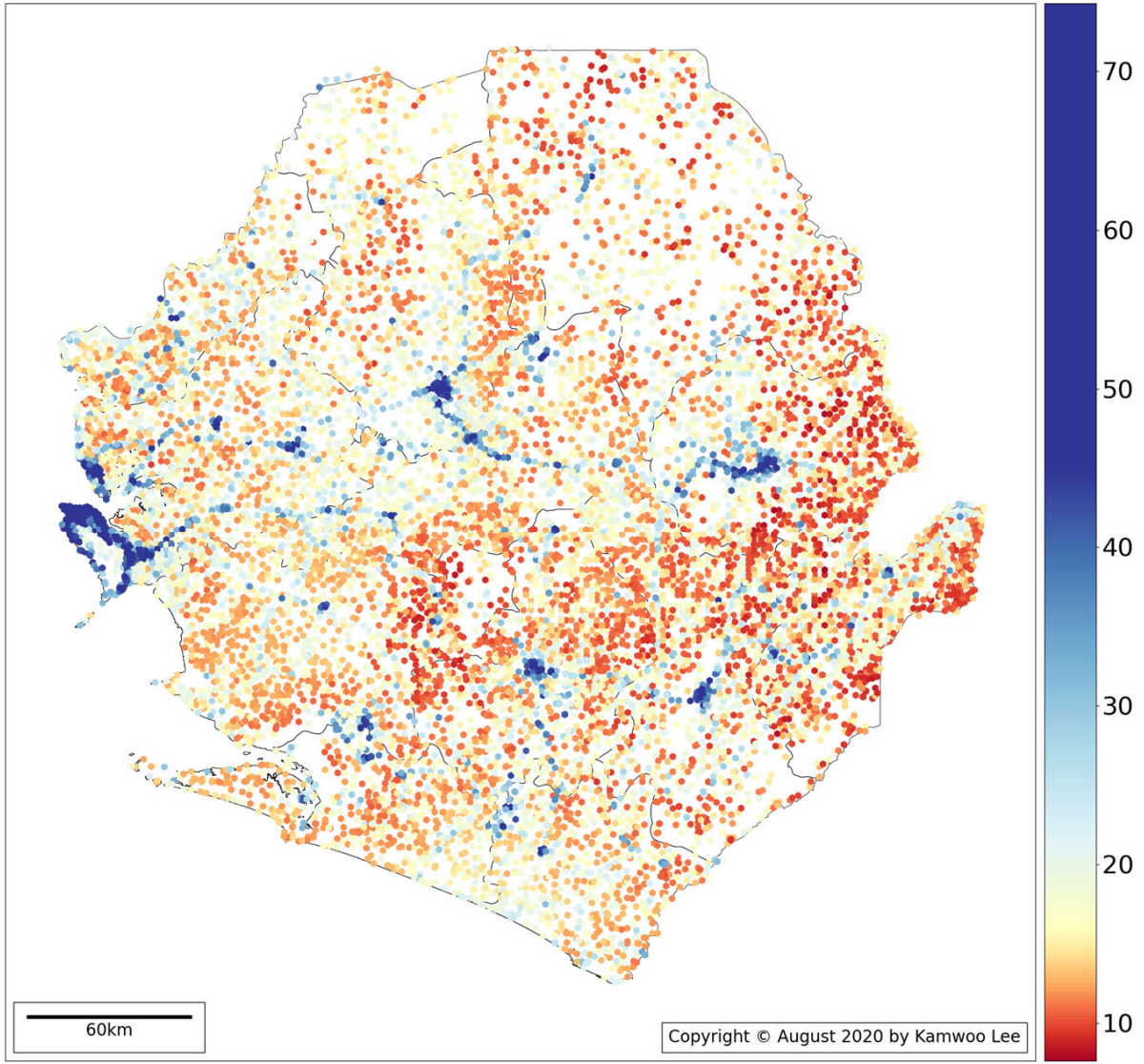

### Estimated Wealth Level (August 2020)
This map displays estimated average wealth levels of households in 1 square-mile populated areas. The wealth level was estimated on the 0-100 International Wealth Index scale (color code: red-poor, yellow-median, blue-rich) using machine learning methods with geospatial information including OpenStreetMap, daytime satellite images, nighttime luminosity, and High-Resolution Population Densities. The estimation was validated with 2016 MIS.

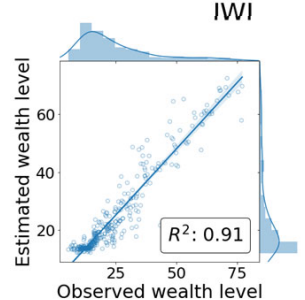

# High-Resolution Poverty Map
## Somalia

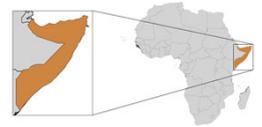

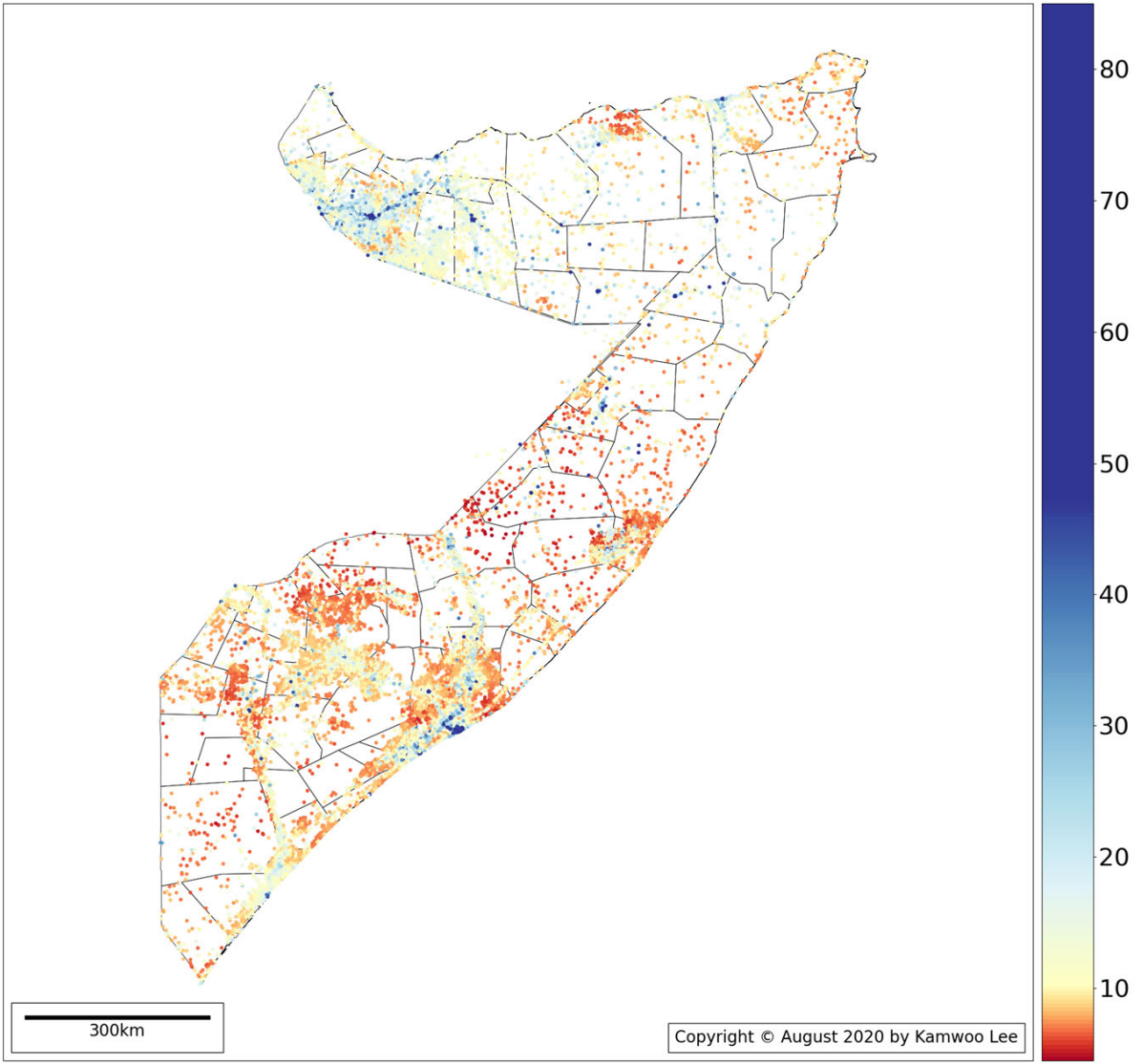

### Estimated Wealth Level (August 2020)
This map displays estimated average wealth levels of households in 1 square-mile populated areas. The wealth level was estimated on the 0-100 International Wealth Index scale (color code: red-poor, yellow-median, blue-rich) using machine learning methods with geospatial information including OpenStreetMap, daytime satellite images, nighttime luminosity, and High-Resolution Population Densities. This is a cross-country estimation that is validated with the DHS data from 25 SSA countries. ($R^2$: 0.91)

# High-Resolution Poverty Map
# South Africa

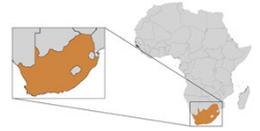

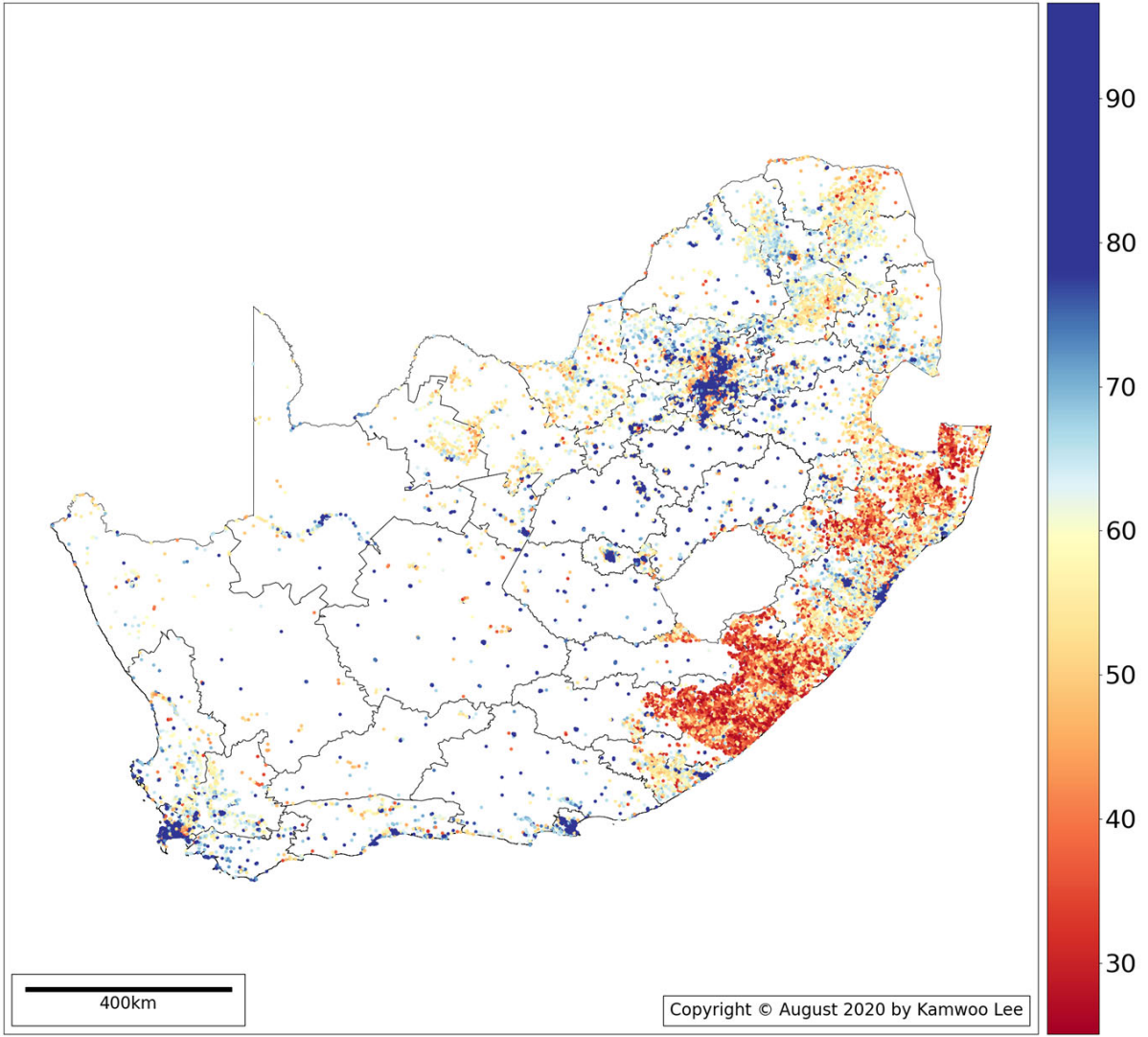

### Estimated Wealth Level (August 2020)
This map displays estimated average wealth levels of households in 1 square-mile populated areas. The wealth level was estimated on the 0-100 International Wealth Index scale (color code: red-poor, yellow-median, blue-rich) using machine learning methods with geospatial information including OpenStreetMap, daytime satellite images, nighttime luminosity, and High-Resolution Population Densities. The estimation was validated with 2016 Standard DHS.

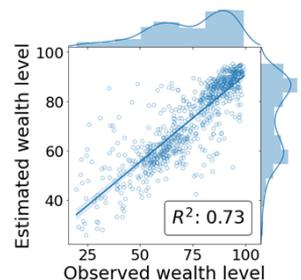

# High-Resolution Poverty Map
# South Sudan

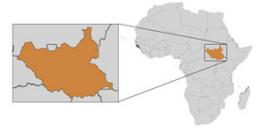

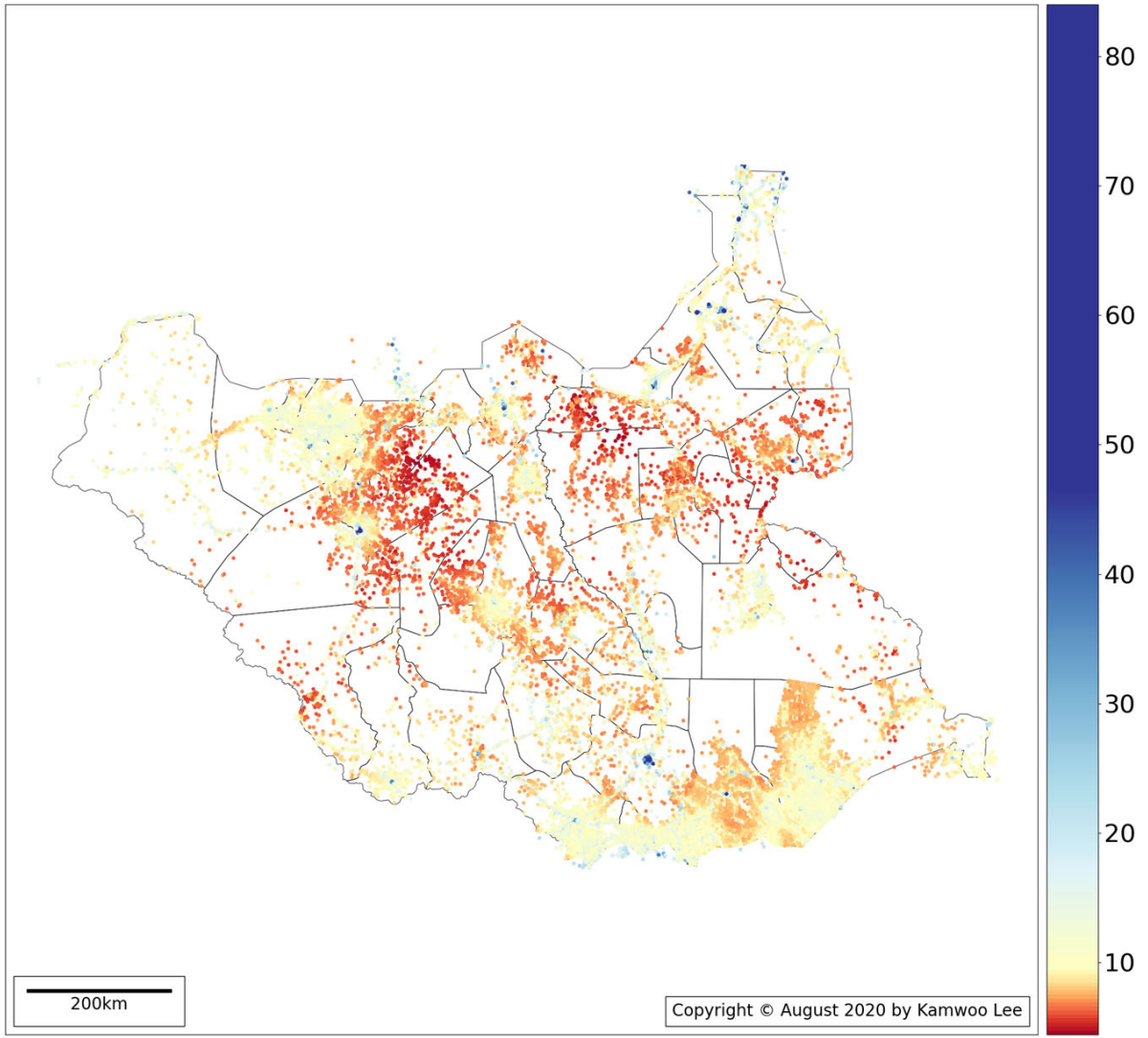

### Estimated Wealth Level (August 2020)
This map displays estimated average wealth levels of households in 1 square-mile populated areas. The wealth level was estimated on the 0-100 International Wealth Index scale (color code: red-poor, yellow-median, blue-rich) using machine learning methods with geospatial information including OpenStreetMap, daytime satellite images, nighttime luminosity, and High-Resolution Population Densities. This is a cross-country estimation that is validated with the DHS data from 25 SSA countries ($R^2$: 0.91).

# High-Resolution Poverty Map
## Sudan

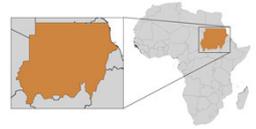

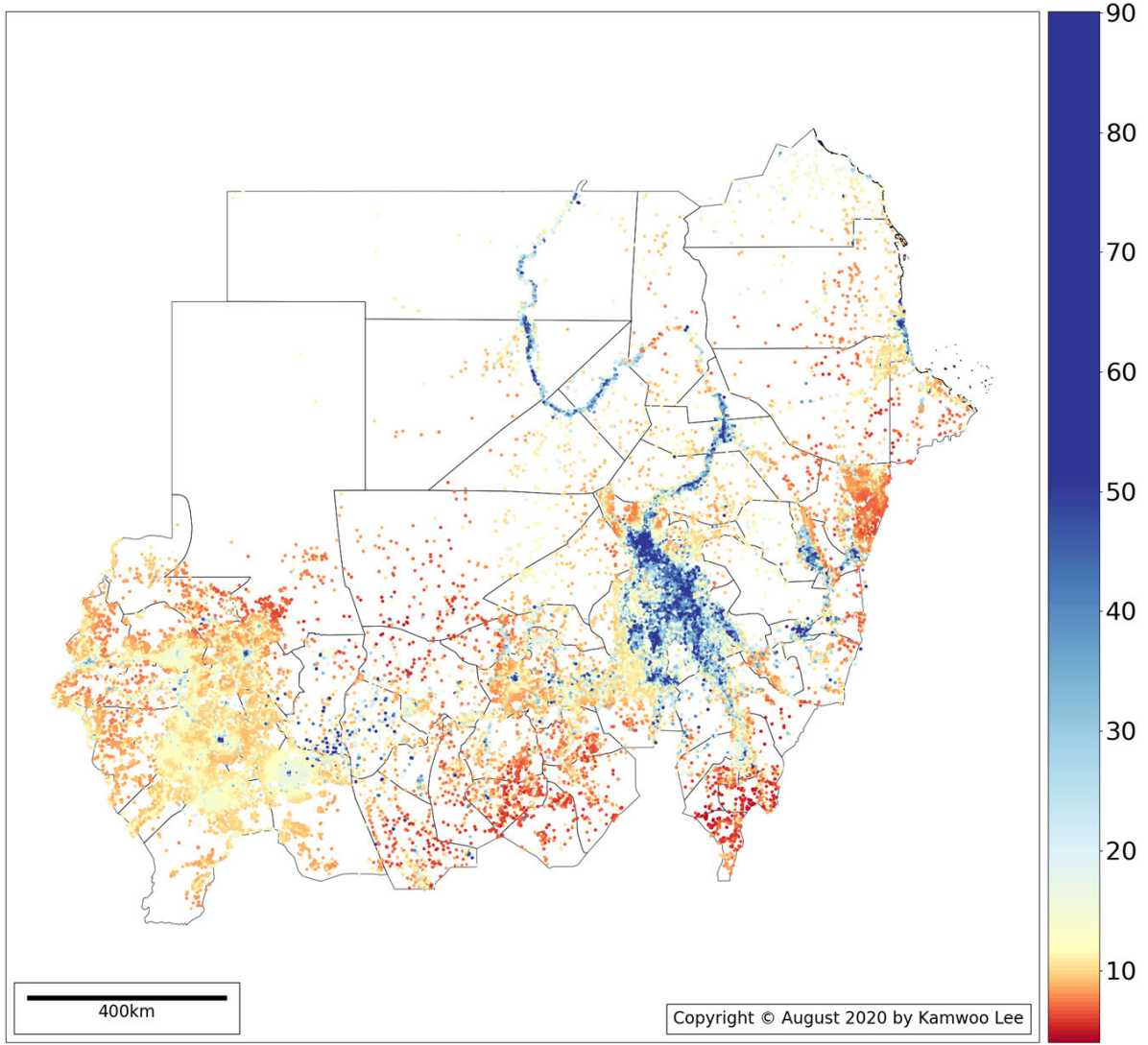

### Estimated Wealth Level (August 2020)
This map displays estimated average wealth levels of households in 1 square-mile populated areas. The wealth level was estimated on the 0-100 International Wealth Index scale (color code: red-poor, yellow-median, blue-rich) using machine learning methods with geospatial information including OpenStreetMap, daytime satellite images, nighttime luminosity, and High-Resolution Population Densities. This is a cross-country estimation that is validated with the DHS data from 25 SSA countries ($R^2$: 0.91).

# High-Resolution Poverty Map
# Tanzania

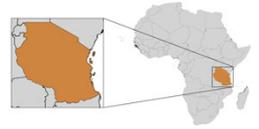

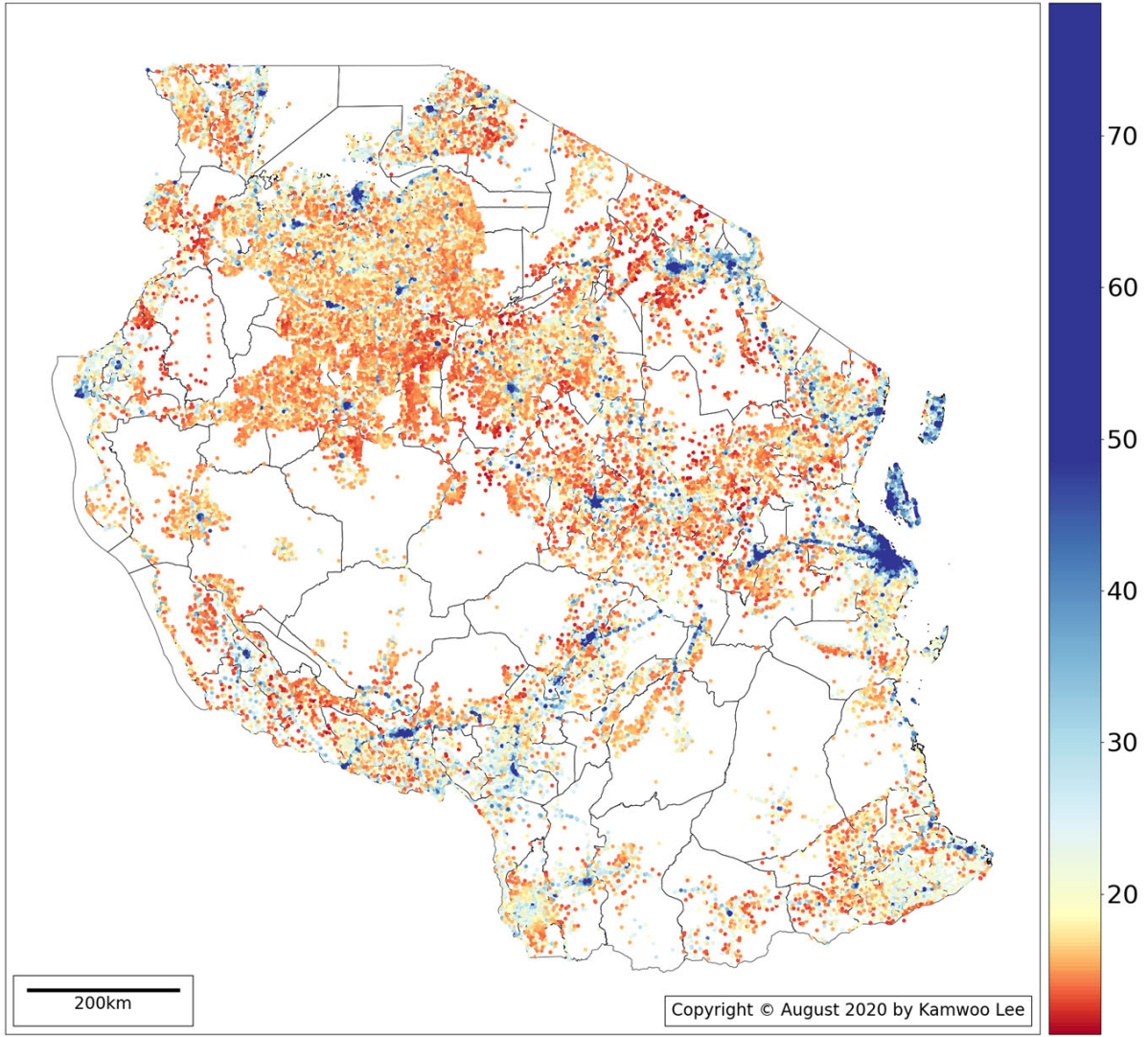

### Estimated Wealth Level (August 2020)
This map displays estimated average wealth levels of households in 1 square-mile populated areas. The wealth level was estimated on the 0-100 International Wealth Index scale (color code: red-poor, yellow-median, blue-rich) using machine learning methods with geospatial information including OpenStreetMap, daytime satellite images, nighttime luminosity, and High-Resolution Population Densities. The estimation was validated with 2015-2016 Standard DHS and 2017 MIS.

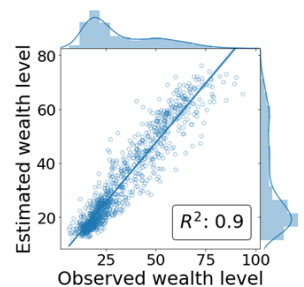

# High-Resolution Poverty Map
# Togo

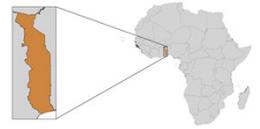

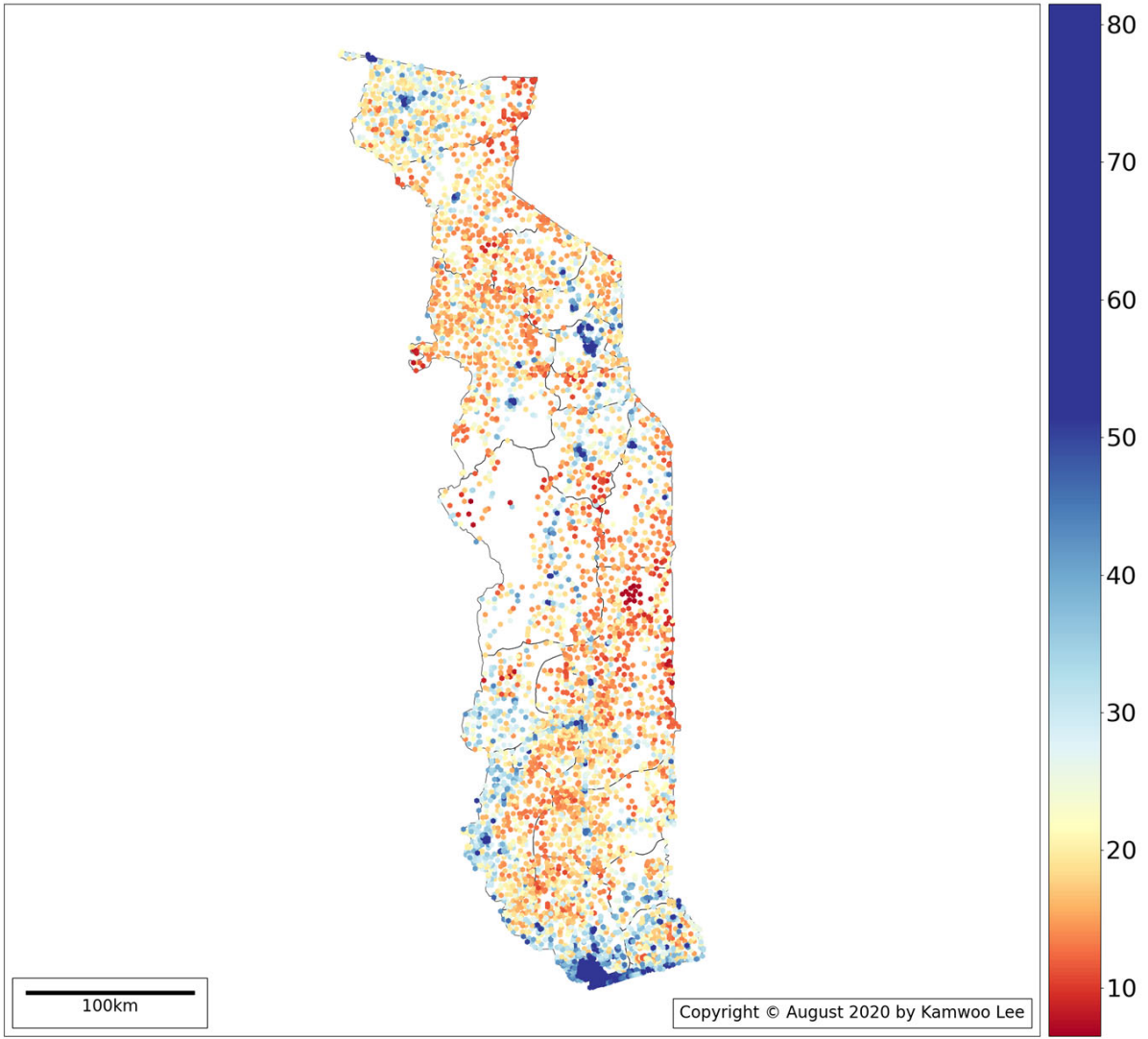

### Estimated Wealth Level (August 2020)
This map displays estimated average wealth levels of households in 1 square-mile populated areas. The wealth level was estimated on the 0-100 International Wealth Index scale (color code: red-poor, yellow-median, blue-rich) using machine learning methods with geospatial information including OpenStreetMap, daytime satellite images, nighttime luminosity, and High-Resolution Population Densities. The estimation was validated with 2017 MIS.

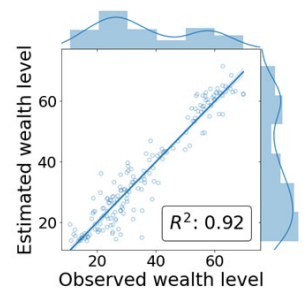

# High-Resolution Poverty Map
# Uganda

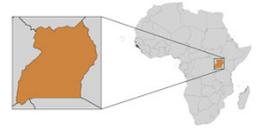

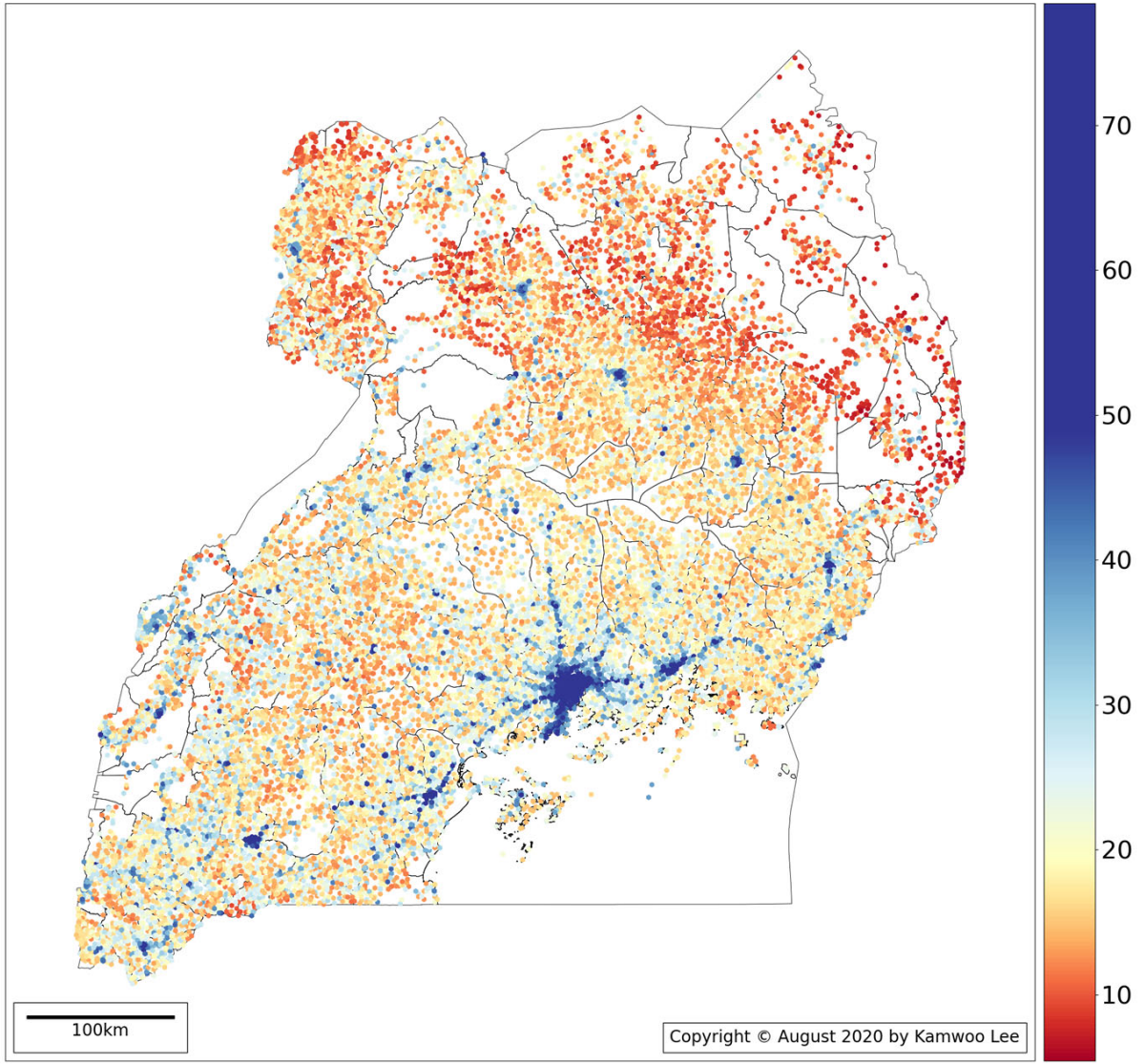

### Estimated Wealth Level (August 2020)
This map displays estimated average wealth levels of households in 1 square-mile populated areas. The wealth level was estimated on the 0-100 International Wealth Index scale (color code: red-poor, yellow-median, blue-rich) using machine learning methods with geospatial information including OpenStreetMap, daytime satellite images, nighttime luminosity, and High-Resolution Population Densities. The estimation was validated with 2016 Standard DHS and 2018-2019 MIS.

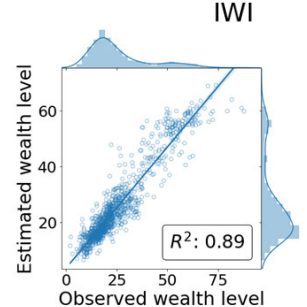

# High-Resolution Poverty Map
# Zambia

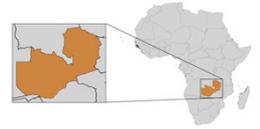

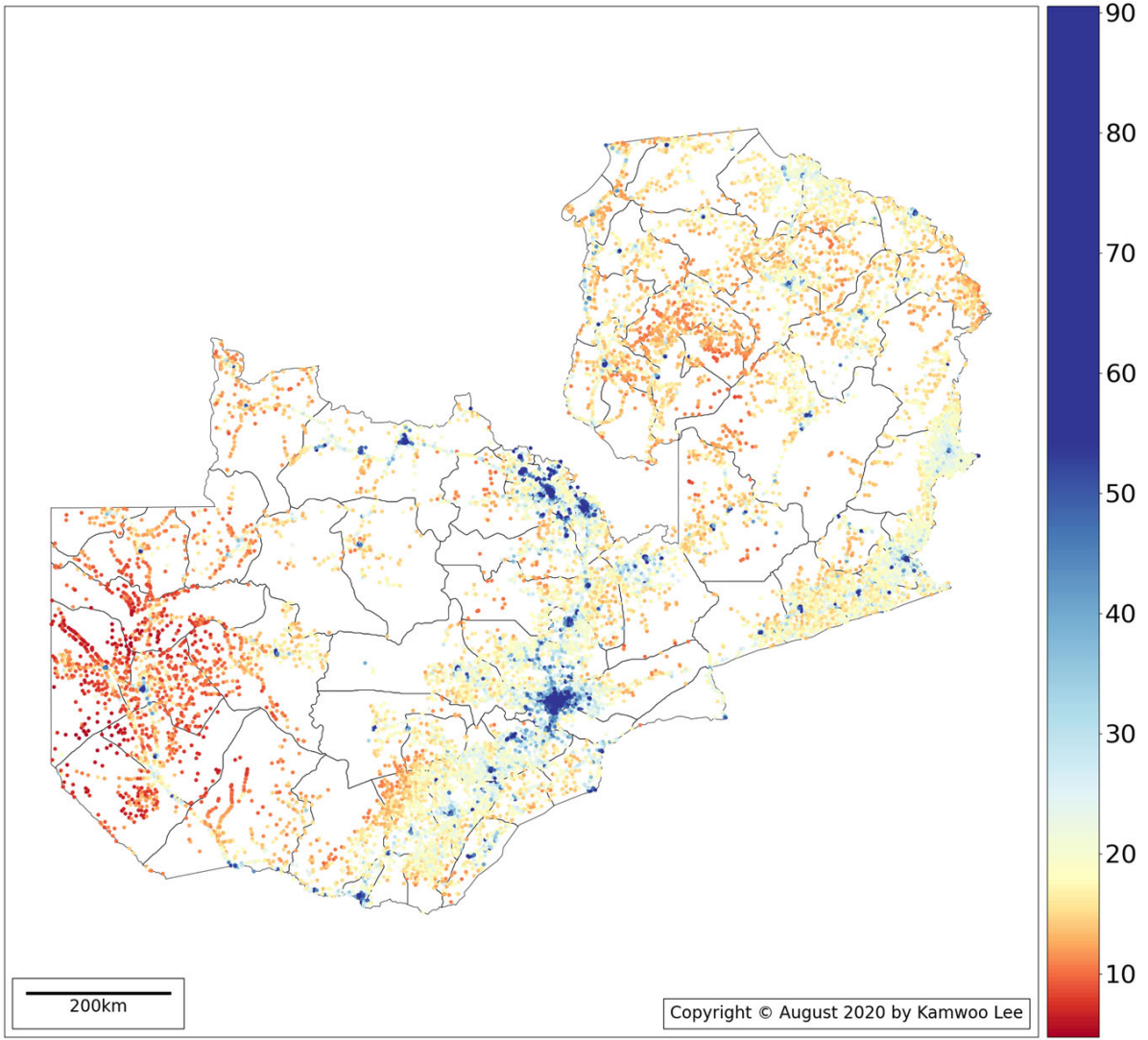

### Estimated Wealth Level (August 2020)
This map displays estimated average wealth levels of households in 1 square-mile populated areas. The wealth level was estimated on the 0-100 International Wealth Index scale (color code: red-poor, yellow-median, blue-rich) using machine learning methods with geospatial information including OpenStreetMap, daytime satellite images, nighttime luminosity, and High-Resolution Population Densities. The estimation was validated with 2018 Standard DHS.

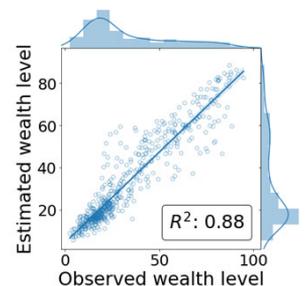

# High-Resolution Poverty Map
# Zimbabwe

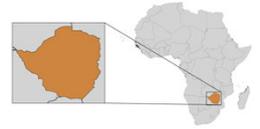

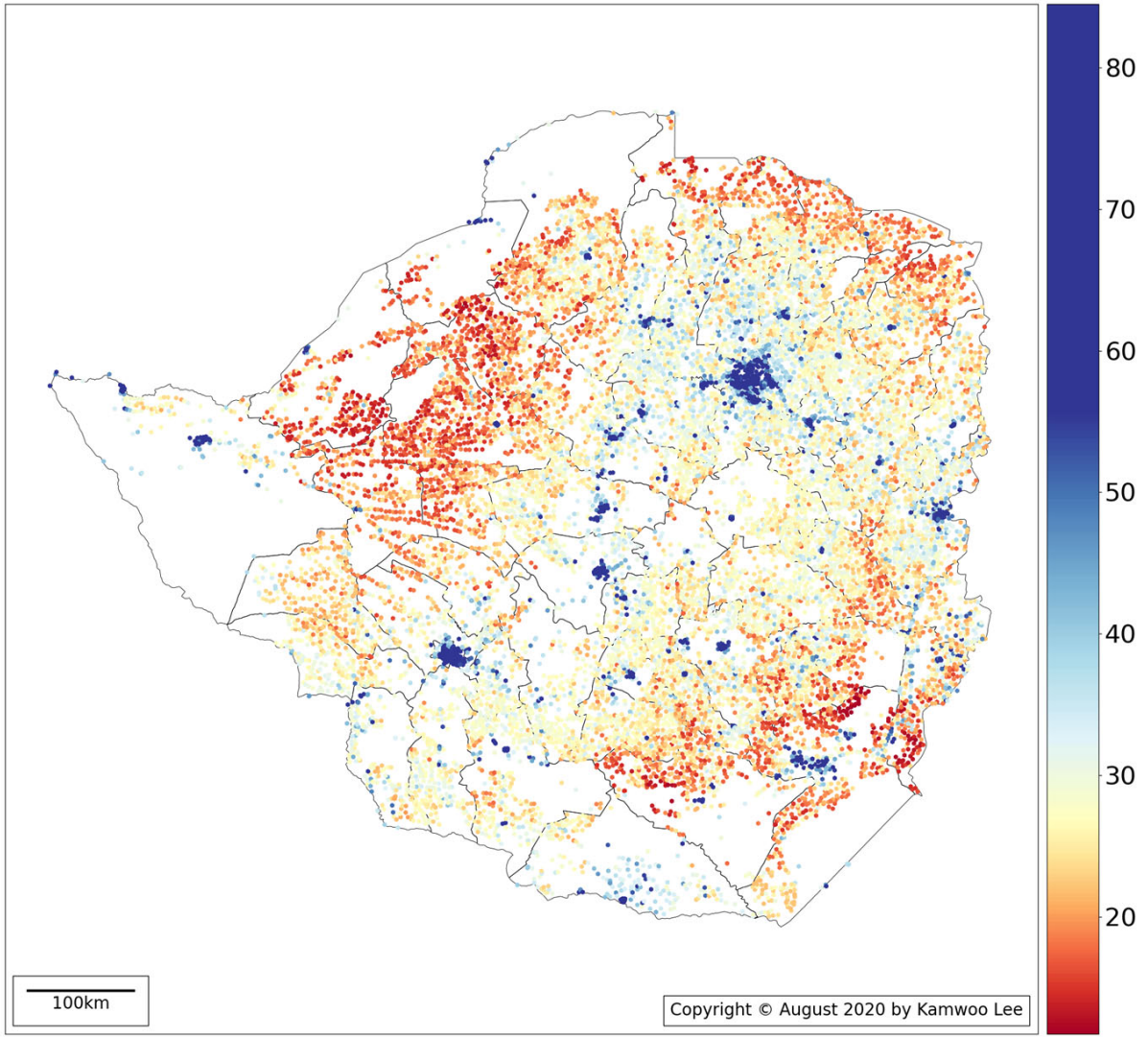

### Estimated Wealth Level (August 2020)
This map displays estimated average wealth levels of households in 1 square-mile populated areas. The wealth level was estimated on the 0-100 International Wealth Index scale (color code: red-poor, yellow-median, blue-rich) using machine learning methods with geospatial information including OpenStreetMap, daytime satellite images, nighttime luminosity, and High-Resolution Population Densities. The estimation was validated with 2015 Standard DHS.

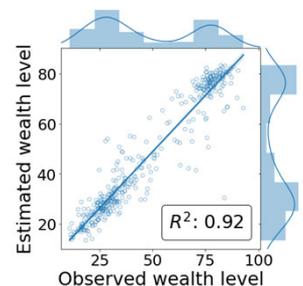

# Appendix B

# High-Resolution Poverty Maps

## Number of Places in the Maps

| Source<br>Country | UN OCHA settlement dataset | OSM populated places | HRSL | DHS clusters | Total populated places |
|---|---|---|---|---|---|
| Angola | - | 4,094 | 12,191 | 625 | 18,898 |
| Benin | 4,534 | 5,468 | 1,537 | 540 | 13,922 |
| Botswana | - | 694 | 3,947 | - | 4,641 |
| Burkina Faso | 10,243 | 12,530 | 4,205 | 224 | 22,995 |
| Burundi | - | 959 | 6,266 | 552 | 13,249 |
| Cameroon | 17,498 | 12,209 | 2,200 | 430 | 28,633 |
| Central African Republic | 7,427 | 11,973 | 3,900 | - | 18,344 |
| Chad | 16,962 | 29,548 | 3,473 | 624 | 43,565 |
| Congo, DR | - | 26,519 | 24,272 | - | 50,791 |
| Congo, Republic of | 5,774 | 1,697 | 1,484 | - | 8,028 |
| Côte d'Ivoire | 8,369 | 12,351 | 3,778 | - | 18,269 |
| Djibouti | 312 | 139 | 382 | - | 783 |
| Equatorial Guinea | 2,076 | 115 | 54 | - | 2,158 |
| Eritrea | 4,432 | 430 | 768 | - | 5,302 |
| Eswatini | - | 45 | 1,113 | - | 1,158 |
| Ethiopia | 2,840 | 3,853 | 37,900 | 622 | 47,037 |
| Gabon | 3,772 | 453 | 854 | - | 4,835 |
| Gambia | 1,628 | 1,029 | 9 | - | 1,827 |
| Ghana | 9,143 | 7,242 | 2,443 | 384 | 18,199 |
| Guinea | 2,223 | 13,674 | 3,266 | 401 | 19,797 |
| Guinea-Bissau | 3,356 | 1,509 | 74 | - | 3,869 |

| Country \ Source | UN OCHA settlement dataset | OSM populated places | HRSL | DHS clusters | Total populated places |
|---|---|---|---|---|---|
| Kenya | 2,675 | 3,336 | 14,450 | 245 | 20,167 |
| Lesotho | - | 1,209 | 962 | - | 2,171 |
| Liberia | 13,834 | 16,226 | 76 | 150 | 16,525 |
| Madagascar | 22,003 | 24,504 | 4,789 | 358 | 45,057 |
| Malawi | - | 1,512 | 8,999 | 998 | 17,618 |
| Mali | 18,423 | 19,154 | 2,493 | 505 | 31,897 |
| Mauritania | 8,636 | 3,215 | 1,342 | - | 11,774 |
| Mozambique | - | 2,171 | 17,183 | 527 | 21,623 |
| Namibia | 607 | 1,504 | 7,513 | - | 9,385 |
| Niger | 28,851 | 19,641 | 1,222 | - | 36,037 |
| Nigeria | 45,598 | 51,146 | 22,899 | 1,681 | 89,459 |
| Rwanda | - | 422 | 11,972 | 492 | 23,323 |
| Senegal | 23,700 | 9,848 | 238 | 428 | 30,131 |
| Sierra Leone | 7,892 | 9,689 | 318 | 336 | 13,040 |
| Somalia | 9,586 | 20,843 | 3,346 | - | 32,567 |
| South Africa | - | 2,770 | 16,421 | 746 | 22,690 |
| South Sudan | 18,538 | 6,001 | 2,715 | - | 23,491 |
| Sudan | 11,989 | 13,039 | 11,869 | - | 31,804 |
| Tanzania | - | 8,935 | 22,703 | 1,044 | 37,076 |
| Togo | 3,663 | 3,087 | 865 | 171 | 7,803 |
| Uganda | 5,163 | 10,700 | 10,228 | 1,001 | 27,482 |
| Zambia | - | 958 | 11,822 | 535 | 14,971 |
| Zimbabwe | 909 | 857 | 13,516 | 400 | 16,904 |
| **Sum** | **322,656** | **377,298** | **302,057** | **14,019** | **929,295** |